\documentclass[twocolumn,aps,rmp,floats]{revtex4}
\usepackage{graphicx}
\usepackage{amssymb}
\usepackage{epstopdf}
\usepackage{epsfig}
\begin{document}

\newcommand{\beq}[1]{\begin{equation}\label{#1}}
\newcommand{\eeq}{\end{equation}}
\newcommand{\beqa}{\begin{eqnarray}}
\newcommand{\eeqa}{\end{eqnarray}}

\newcommand{\eqref}[1]{Eq.~(\ref{#1})}
\newcommand{\eqsref}[1]{Eqs.~(\ref{#1})}
\newcommand{\thmref}[1]{Theorem~\ref{#1}}
\newcommand{\clmref}[1]{Claim~\ref{#1}}
\newcommand{\defref}[1]{Definition~\ref{#1}}
\newcommand{\figref}[1]{Figure~\ref{#1}}
\newcommand{\tblref}[1]{Table~\ref{#1}}
\newcommand{\partref}[1]{Part~\ref{#1}}
\newcommand{\chpref}[1]{Chapter~\ref{#1}}
\newcommand{\secref}[1]{Section~\ref{#1}}
\newcommand{\exref}[1]{Exercise~\ref{#1}}
\newcommand{\xamref}[1]{Example~\ref{#1}}
\newcommand{\chapref}[1]{Chapter~\ref{#1}}

\newcommand{\F}{{\cal{F}}}
\newcommand{\boldsymbol}[1]{{\mathbf{#1}}}
\newcommand{\bA}{\boldsymbol{A}}

\newcommand{\commentout}[1]{}
\newcommand{\ignore}[1]{}

\title{Information based clustering:   Supplementary material}

\author{Noam Slonim, Gurinder Singh Atwal, Ga\v{s}per Tka\v{c}ik, and William Bialek}

\affiliation{Joseph Henry Laboratories of Physics, and
Lewis--Sigler Institute for Integrative Genomics\\
Princeton University, Princeton, New Jersey 08544 USA}

\date{\today}

\begin{abstract}
This technical report provides the supplementary material 
for a paper entitled ``Information based clustering,'' to appear shortly in 
{\it Proceedings of the National Academy of Sciences (USA)\/}.
In \secref{sec:alg} we present in detail the iterative clustering algorithm
used in our experiments and in
\secref{sec:coh} we describe the validation scheme used to 
determine the statistical significance of our results.  Then in subsequent sections
we provide all the experimental results for  three very different applications: the response of gene expression in  yeast to different forms of environmental stress, the dynamics of stock prices in the 
Standard and Poor's 500, and viewer ratings of popular movies. 
In particular, we highlight some of the results that seem
to deserve special attention. 
All the experimental results and relevant code, 
including a freely available web application, 
can be found at {\it http://www.genomics.princeton.edu/biophysics-theory\/}.
\end{abstract}

\maketitle

\tableofcontents

\section{The {\it Iclust\/} algorithm}
\label{sec:alg}

Although clustering is a widely used method of data analysis and exploration, there is at present no unique or universal mathematical formulation of the clustering problem.  In practice, clustering a given data set  involves
many choices at different levels of the analysis.
In recent work we suggest that some generality can be achieved through the use of information theory \cite{pnas}.  Here we review this formulation briefly and then proceed to the technical details of its implementation that were left out of Ref \cite{pnas}.

We formulate clustering as a tradeoff between  maximizing the mean similarity of elements within a cluster and minimizing the complexity of the description provided by cluster membership.  Thus if we have some similarity measure $s({\rm i}, {\rm j})$ between elements ${\rm i}$ and ${\rm j}$, optimal clustering is a probabilistic assignment to clusters $C$ according to $P(C|{\rm i})$ such that we maximize 
\begin{equation}
{\cal F} = \langle s \rangle - T I(C;{\rm i}) ,
\end{equation}
where $\langle s \rangle$ is the mean similarity of elements chosen at random out of each cluster,
\begin{equation}
\langle s \rangle = \sum_C P(C) \sum_{\rm i,j} P({\rm i}|C) P({\rm j}|C) s({\rm i}, {\rm j}),
\end{equation}
and $I(C;{\rm i})$ is the   information that clusters provide about the identity of their elements, 
\begin{equation}
I(C;{\rm i}) = \sum_C \sum_{\rm i} P(C|{\rm i}) P({\rm i}) \log\left[{{P(C|{\rm i})}\over{P(C)}}\right] ;
\end{equation}
as usual we have
\begin{eqnarray}
P({\rm i}|C) &=& P(C|{\rm i}) P({\rm i})\cdot {1\over {P(C)}}\\
P(C) &=& \sum_{\rm i} P(C|{\rm i}) P({\rm i}) ,
\end{eqnarray}
and since in many cases all examples ${\rm i}$ occur with equal probability [$P({\rm i}) = 1/N$]
we consider this case for simplicity, although it is not essential. 
This formulation can be generalized to handle similarity measures defined on groups of more than two elements \cite{pnas}, but here we concentrate on the conventional case where only pairwise interactions 
are considered. 
Importantly, as opposed to using a problem specific similarity measure $s$, 
we use the generality of information theory once more, and take $s({\rm i},{\rm j})$ to be the 
pairwise mutual information $I_{\rm ij}$ between the observed patterns that correspond to data 
items ${\rm i}$ and ${\rm j}$ \cite{pnas}.\footnote{This report does not deal 
with the technical details of estimating mutual (and multi) information from empirical data.
The reader is referred to \cite{Slo+al04} for a complete description 
of the estimation procedure used in our experiments.} 

It is shown in Ref \cite{pnas} that any stationary point of our target functional, $\F$, must obey
\begin{equation}
P(C|{\rm i}) = {{P(C)} \over {Z({\rm i}; T)}}\exp\bigg{\{} {1\over T}[ 2 s(C;{\rm i} ) - s(C)]\bigg{\}} ,
\label{selfcon}
\end{equation}
where $Z({\rm i}; T)$ is a normalization function, 
$s(C;{\rm i})$ is the expected similarity between $\rm i$ and a member of cluster $C$,
\begin{equation}
s(C;{\rm i}) = \sum_{{\rm i}_1 =1}^N P({\rm i}_1 | C) s({\rm i}_1, {\rm i} ) ,
\end{equation}
and $s(C)$ is the average similarity among pairs chosen independently out of the cluster $C$,
\begin{equation}
s(C) = \sum_{{\rm i}_1 =1}^N \sum_{{\rm i}_2 =1}^N P({\rm i}_1 | C) P({\rm i}_2 | C) s({\rm i}_1, {\rm i}_2) .
\end{equation}
\eqref{selfcon} defines an implicit set of equations 
since the right hand side depends on $P({\rm i}|C)$ and $P(C)$.
This is a common situation in variational methods, also present, for example,
in conventional rate--distortion clustering \cite{Cover91},
in maximum likelihood estimation with hidden variables \cite{Dempster+al77}
and in the Information Bottleneck framework \cite{TPB99}.
The standard strategy is to turn the self--consistency condition into an 
iterative algorithm. 
Specifically, let us denote the intermediate solution of the algorithm
at the $m^{\rm th}$ iteration by $P^{(m)}(C|{\rm i})$. Then, 
at the $m+1^{\rm st}$ iteration, the algorithm applies  
the following update rule:
\ignore{
\begin{widetext}
\beq{update}
P^{(m+1)}(C|{\rm i}) \leftarrow {P^{(m)}(C)} \exp\bigg{\{} {1\over T}[ 2 s^{(m)}(C;{\rm i} ) - s^{(m)}(C)]\bigg{\}},~\forall~C=1:N_c\;,
\eeq
\end{widetext}
}
\beq{update}
P^{(m+1)}(C|{\rm i}) \leftarrow {P^{(m)}(C)} 
\exp{\bigg \{} 
{1\over T}[ 2 s^{(m)}(C;{\rm i} ) - s^{(m)}(C)] {\bigg \}}
\label{iter}
\eeq
followed by a normalization step.  
Notice that the terms $\{P^{(m)}(C),\;s^{(m)}(C;{\rm i} ),\;s^{(m)}(C)\}$ all are calculated
using $P^{(m)}(C|{\rm i})$.
Pseudo--code for this algorithm is given in \figref{algorithm}.
It is easy to verify that with a straightforward implementation, the complexity 
of this algorithm is $O(N^3\cdot N_c)$ for a single pass over the entire data set,
where $N_c$ is the number of clusters.
We will refer to this algorithm as the {\it Iclust\/} algorithm.

To gain some intuition let us consider a typical situation where
${\rm i}$ is relatively similar to elements in $C$,
but very different from elements in $C'$. 
Thus, the exponent in Eq (\ref{iter}) will be positive for ${\rm i}$ and $C$,
but might be negative for ${\rm i}$ and $C'$. Consequently, while applying the update step 
the weight of assignment of $\rm i$ to $C$ $[P(C|{\rm i})]$ will be boosted while its assignment
to $C'$ will be decreased. This is clearly a desirable outcome,
which in particular should increase $\F$. 
Thus, since $\F$ is upper bounded (as a sum of information terms), 
after a finite number of such updates the algorithm 
is expected to converge to a fixed point which corresponds
to a (possibly local) maximum of $\F$. 

This example also illustrates one of the differences between
our algorithm and previous approaches. While in 
the Blahut--Arimoto algorithm in rate--distortion theory \cite{Cover91},
in the iterative Information Bottleneck algorithm \cite{TPB99}, 
and typically also in EM for maximum likelihood \cite{Dempster+al77}, 
the sign of the exponent is constant (for a given $\rm i$), this is not true
in our case. In principle, such a non--constant exponent sign should imply
faster convergence to a local stationary point, but might also imply
higher sensitivity to the random initialization of $P(C|{\rm i})$. 
Thus, as  in other work, we typically perform several
runs with different random initializations of $P(C|{\rm i})$
from which we choose the best solution, {\it i.e.\/}, the one that maximizes $\F$.

The {\it Iclust\/} algorithm presented here uses a sequential,
or incremental iterative procedure in which the updates for some $\rm i$
incorporate the implications of the updates for preceding elements,
${\rm i'} \neq {\rm i}$. 
As a simple example, consider the case where we have three elements 
($N=3)$ and two clusters ($N_c=2$).
We start from some random conditional distribution matrix, $P^{(0)}(C|{\rm i})$,
which in particular defines $s^{(0)}(C),\;\forall\;C=1:2$.
At the first iteration we find a new distribution for the first element (${\rm i}=1$)
over the two clusters. Thus, we now have a new conditional 
distribution matrix, $P^{(1)}(C|{\rm i})$ which differs from the previous $P^{(0)}(C|{\rm i})$
only by its first row. This distribution is used
to define $s^{(1)}(C),\;\forall\;C=1:2$. 
Now, in the next iteration, we find a new distribution
for the second element (${\rm i}=2$) over the two clusters. 
This yields another new conditional distribution matrix, $P^{(2)}(C|{\rm i})$ 
which differs from the previous $P^{(1)}(C|{\rm i})$ only by its middle row, 
and so on. This process is somewhat in the spirit of the 
incremental EM \cite{NealHinton98} and the sequential Information Bottleneck algorithm \cite{sIB}.
An alternative optimization routine, which seems less natural 
in our case, would be parallel optimization, used {\it e.g.\/}, in standard EM \cite{Dempster+al77}.
In this case, if we continue our example, at the first iteration
we will update {\it all\/} the rows in the conditional distribution matrix,
$P^{(0)}(C|{\rm i})$ using $s^{(0)}(C)$, to find the new $P^{(1)}(C|{\rm i})$.

In some extreme cases the above algorithm might produce a non--monotonic behavior in $\F$. 
That is, some of the updates might reduce $\F$, 
suggesting that obtaining a general proof of convergence
is a challenging goal. 
Nonetheless, even in these extreme cases, and more generally in all
our experiments (which included more than $1000$ runs over
real world problems with different $T$ values and different numbers of clusters),
the algorithm always converged to a stationary point. 
Moreover, for the regime $T \geq {\rm max}_{{\rm i}_1,{\rm i}_2}s({\rm i}_1,{\rm i}_2)$ 
it is possible to prove this convergence analytically 
(the details will be presented elsewhere).

%\newpage
\section{Evaluating clusters' coherence}
\label{sec:coh}

The central question in clustering is whether an essentially unsupervised analysis of a data set can recover categories that have ``meaning.''  In practice we assess this by comparison with some set of labels for the data that were generated by human intervention.  To get started, then, we need a set of annotations (or labels) provided for every data item we clustered.
Importantly, these annotations are not used during the 
clustering process but rather are exposed only for the post--clustering
validation. Every data item might be assigned  more than one
annotation via different sources of information. The assumption is that these annotations
reflect to some extent the ``real'' structure of the data that 
one wishes to reveal through the clustering process. 

To be more concrete, let us assume that we clustered $N$ elements,  where each one of these elements is assigned some set of annotations. 
Formally, this could be represented through an annotation matrix,
denoted as $\bA$, with $N$ rows and $R$ columns, where $R$ is the number
of distinct annotations in our data.
Thus, $\bA({\rm i},{\rm j})=1$ if and only if the ${\rm i}$-th element is assigned 
annotation $a_{\rm j}$, and zero otherwise. A simple example
is given in \tblref{Tbl:AnnotMatrix}.

When we examine a single cluster, consisting of $n < N$ 
elements, the first question we might ask is whether some annotations 
occur in this cluster with a ``suspiciously'' high frequency. 
Let us consider a specific annotation $a_{\rm j}$ that is assigned to $K \leq N$
elements in the entire population and to $x \leq n$ elements 
in the cluster. The probability of this event, under the null hypothesis
that elements are assigned to clusters at random,  is given by the hypergeometric distribution:
\beq{Eq:HyperP}
P_{\rm hyper} (x\,|\,n,K,N) = \displaystyle{\frac{ {K \choose x} {N-K \choose n-x} }{ {N \choose n} }}~.
\eeq
The corresponding $P$-value is defined as the tail of this distribution:
\beq{Eq:Pval}
Pval(x\,|\,n,K,N) = \sum_{x'=x}^{{\rm min}(K,n)} P_{\rm hyper} (x'\,|\,n,K,N)~.
\eeq
In words, it is the probability of observing $x$ or more elements
in the cluster with annotation $a_{\rm j}$ where the members of the cluster are chosen independently
of this annotation. Alternatively, it is the probability of {\it wrongly\/} rejecting 
the hypothesis that the cluster has nothing to do with the
annotation $a_{\rm j}$. The smaller the $P$-value the more unlikely this null 
hypothesis becomes. To gain some intuition, several examples
are presented in \tblref{Tbl:PvalDefinitions}.

Having defined the statistical significance of a single event we need
to bear in mind that in a single cluster one typically observes 
several (perhaps many) different annotations.
Naturally, the more hypotheses one tests the less surprising
it is to find one with a small $P$-value, even in a randomly chosen cluster. 
The simplest and most conservative approach to correct 
for this multiple hypotheses testing effect is to apply the 
Bonferroni correction (see,{\it e.g.\/}, \cite{Bonf}).
Specifically, if the statistical significance level is $q\;$ ({\it e.g.\/}, $q=0.05$),
an event is considered significant if and only if its $P$-value satisfies:
\beq{Eq:Bonf}
Pval < \frac{q}{H}~,
\eeq
where $H$ is the number of hypotheses being tested. 
We will say that a cluster is {\it enriched\/} with the annotation
if the corresponding $P$-value satisfies \eqref{Eq:Bonf}.

Finally, while the above procedure determines the significance
of every annotation that occurs in the cluster, it also is useful
to have a single score that roughly summarizes how homogeneous
the cluster is with respect to all annotations. 
Different alternatives have been proposed to this end
and here we use the {\it coherence\/} score, 
suggested by Segal {\it et al.} \cite{Segal+al03},
\beq{Eq:coh}
coh (C) \equiv 100 \cdot {n_{enriched} \over n},
\eeq
where $n$ is the number of items in the cluster $C$, 
and $n_{enriched}$ is the number of items in $C$ with 
an annotation that was found to be significantly enriched in $C$.
In other words, the coherence of a cluster is simply the percentage
of the cluster's elements covered by some annotation that was found to be enriched in that cluster.
In particular, a coherence value above zero means that at least one annotation 
is enriched in the cluster, namely that there is at least a single hint 
regarding the reason for forming this cluster.

\section{First application: The yeast ESR data}
\label{sec:ESR}

\subsection{Description of the data}

We considered experiments on the response of gene expression levels
in yeast to various forms of environmental stress \cite{Gasch00}.
Previous analysis of expression patterns from all $\sim 6000$ genes identified a 
group of $283$ stress--induced and $585$ stress--repressed genes 
that had apparently ``nearly identical but opposite'' expression profiles \cite{Gasch00}.
This collection of $868$ genes was thus termed the 
yeast environmental stress response (ESR) module.  
As seen in \figref{ESR_data},
differences in expression profiles within the 
ESR module indeed are relatively subtle.
More recent manual analysis with attention 
to background biological data suggests that some 
of these differences are biologically significant \cite{Gasch02}.
Thus, it seems a good challenge for our approach to ask if we can discover 
automatically any meaningful substructure in these data.

Each of the $868$ ESR genes was represented by its log--ratio expression profile
in the $173$ microarray 
experiments \cite{Gasch00},
available at {\it http://genome-www.stanford.edu/yeast\_stress/data.shtml\/}.\footnote{This log transformation has no effect on our analysis 
since the mutual information is invariant to such transformations \cite{Slo+al04}.}
From these data we estimated all the $\sim 376,000$ pairwise
mutual information relations $I_{\rm ij}$, as described in \cite{Slo+al04},
ending up with a $868\times 868$ matrix   which defined the input to our clustering procedure. 
For convenience, we provide here some statistics of the estimated mutual information values. 
For a complete description, including different verification schemes
that support the reliability of our estimates, 
the reader is referred to \cite{Slo+al04}.

Across all pairs of genes, the average estimated mutual information was $0.48\;{\rm bits}$ with a variance of $0.0425\;{\rm bits}^2$. 
This relatively high average value corresponds
to the strong positive/negative linear correlations known
to be present in these data. 
Almost $7,000$ pairs had a mutual information greater than $1\;{\rm bit}$, and 
the maximal estimated mutual information was $1.58\;{\rm bits}$.
All the pairwise mutual information relations are presented in \figref{ESR_MIs},
where the genes are sorted according to the clustering partition
into $N_c=20$ clusters that we analyze in detail (see below).
The diagonal elements of this matrix, or self--information were set to 
$I_{\rm ii} = \log_2(5)$,   the maximal 
possible information under a quantization into five bins \cite{Slo+al04}.

\subsection{Quality--complexity trade--off curves}

Given the pairwise mutual information matrix we applied the {\it Iclust\/} algorithm
described in \secref{sec:alg}.
Recall that our target functional, $\F$, is given by:
\beq{equation}
\label{F}
{\cal F} = \langle s \rangle - T I (C; {\rm i}) ,
\eeq
where $T$ is a (temperature) trade--off parameter,
\begin{equation}
\langle s \rangle = \sum_{C =1}^{N_c} P(C) \sum_{\rm ij} P({\rm i}|C) P({\rm j}|C) I_{\rm ij} 
\end{equation}
measures
the quality of the clusters,
and $I (C; {\rm i})$ measures the cost of coding cluster identity.

For a fixed number of clusters, $N_c$, 
the term $\langle s \rangle$ gradually saturates
as the temperature $T$ is lowered, while $I(C;{\rm i})$ increases accordingly. 
We explored this trade--off for different numbers of clusters:
$N_c=5,10,15,20$. 
For each of these values we tried several values of $T$; 
we found that $1/T = \{5,10,15,20,25\}$ typically
suffices to obtain a relatively clear saturation of $\langle s \rangle$, hence
we present the results for these $T$ values. 

For each $\{N_c\;,T\}$ pair we performed $10$ different random initializations,
ending up with $10$ (possibly) different
local maxima of $\F$, from which we chose the best one.
The resulting trade--off curves are presented in the left
panel of \figref{Fig:tradeoff}. 
For a given $N_c$, as $T$ is lowered, $\langle s \rangle$ increases
but so does $I(C;{\rm i})$. In addition, the solutions become more
deterministic. For example, for $N_c=20$
and $1/T=15$, only $\sim\;44\%$ of the genes
have nearly deterministic assignment 
[{\it i.e.\/}, $P(C|{\rm i}) > 0.9$ for a particular $C$]. 
For $1/T=25$ this percentage grows to $\sim\;85\%$. 

The entire continuum of solutions, 
represented by the trade--off  curves, may encompass a lot of insights
about the data. Nonetheless, for brevity, we  focus
our analysis on solutions for which the saturation
of $\langle s \rangle$ is relatively clear, {\it i.e.\/},
on the four solutions with $N_c=\{5,10,15,20\}$ and $1/T=25$.
In all these partitions most of the genes (between $75\%$ to $85\%$)
had a nearly deterministic assignment 
[$P(C|{\rm i}) > 0.9$ for a particular $C$]. In the rest of the analysis we treat these solutions 
as hard ({\it i.e.\/}, deterministic) partitions where every gene is assigned solely to its most
probable cluster. 
In the next section we explore the possible hierarchical relations
between these four solutions. In later sections we analyze in detail
the specific solution with $\{N_c=20,\;1/T=25\}$ that obtained the highest
value of $\langle s \rangle$.

\subsection{Comparing solutions at different numbers of clusters}
\label{sec:ESR_Hierarchy}

A common dichotomy in the cluster analysis literature
is between hierarchical and non--hierarchical, or partitional
clustering algorithms [see,{\it e.g.\/}, Ref \cite{JaneRev99}].
What is often missed, though, is the fact that applying 
a hierarchical clustering algorithm  typically
enforces the output to be of a hierarchical nature, 
regardless of whether the data  indeed call for this view. 
For example, applying an agglomerative clustering algorithm 
to  the ESR data will produce, by definition, 
a nested tree--like hierarchy of partitions, 
although {\it a priori\/} it is not obvious whether the functional classification
of these genes should be of a hierarchical nature.

Because our approach is not constrained to hierarchical structures, the emergence of even an approximate hierarchy would be a genuine result.
To test for this, we start with
 several solutions {\it that were found independently\/}
at different numbers of clusters and ask to what extent
these solutions form a hierarchy. This is done in two steps. 
First, for every cluster we identify its best parent
in the next (less detailed) level. 
Specifically, if $C$ is some cluster at a partition with $N_c$ clusters,
then its best parent in a less detailed partition with $N_c' < N_c$
clusters will be the one that includes the maximal number of $C$ members. 
Second, we measure how well this parent includes
its child and represent the result through the type of the edge that we draw
between the two clusters. 

The hierarchical graph produced by this scheme is different
from the standard output of hierarchical clustering algorithms in several aspects.
To start, a cluster might have more than one parent
if its members are equally distributed among
several clusters in the less detailed partition. 
Next, a cluster might have no children if it is not the best parent
of any cluster at the more detailed partition. 
Last, the characteristics of the edges convey further information
regarding how well the independent solutions form a hierarchy. 
In particular, a graph with many high quality inclusion edges
is a good indication that the   data are hierarchical in nature.
In contrast, a graph in which many of the inclusions
from one level to the other are only partial 
suggests that a hierarchical view of the data is somewhat misleading.

We applied this scheme to the four solutions we obtained
independently for $N_c=\{5,10,15,20\}$ with $1/T=25$.
The results are presented in \figref{Fig:ESR_Hierarchy}. 
As can be seen in the figure, the independent solutions
form an approximately  hierarchical structure.
Interestingly, some functional modules are better preserved than others across
the different levels. 
For example, the ribosome cluster, $c18$, 
clearly is identified at all the independent solutions.

\subsection{Coherence results}
\label{sec:ESR_Coh}

\subsubsection{Constructing the annotation matrices}
\label{sec:ESR_annotation}

As already mentioned, clusters' coherence is estimated
with respect to a given annotation matrix. 
For yeast genes, different sources of information may
provide these data. One  important resource
is the Gene Ontology (GO) database \cite{GO},
which is the one that we use in this work; specifically, 
we used the December 2003 version.

The GO database consists of three structured ontologies (controlled vocabularies)
that describe gene products in terms of their associated 
Biological Processes ($GO_{BP}$), 
Molecular Functions ($GO_{MF}$), 
and Cellular Components ($GO_{CC}$).
For each of these three ontologies we constructed a corresponding 
annotation matrix. Thus, for example, if $\bA_{BP}$ is the
matrix constructed for the $GO_{BP}$ ontology then 
$\bA_{BP}({\rm i},{\rm j})=1$ if and only if the ${\rm i}$-th gene in our data
is assigned to the ${\rm j}$-th biological process.
A small subset of this annotation matrix is presented
in \tblref{Tbl:AnnotMatrixBP}.

Each of the GO ontologies is organized in a hierarchical manner
where more specific annotations correspond to nodes which are 
more distant from the ontology root.
This might yield evaluation difficulties if one considers 
only the particular GO terms with which a gene is annotated 
\cite{Troyanskaya+al03}. An example is given in 
\tblref{Tbl:c15example}. Here, several genes that were all 
assigned to the same cluster are annotated with different specific $GO_{BP}$ terms,
and their functional relationship becomes evident 
only if one notices that all these annotations have
a common (more general) ancestor in the ontology. 
We applied a standard routine to overcome this difficulty, in which 
every gene was assigned not only  its direct GO annotations
but also all the ancestors of these annotations in the GO hierarchy.
This is consistent with the GO organization, in which
if a GO term describes some gene product
then all its parent terms in the ontology also apply to that gene product.

Last, while estimating clusters' coherence we removed
annotations that were assigned to less than two genes in our data,
since these annotations obviously can not be enriched in any cluster.
We also removed from the analysis genes that were 
not assigned any annotations, or assigned the {\it unknown\/} annotation.
The details of the resulting annotation matrices 
are given in \tblref{Tbl:ESR_AnnotDetails}.

\subsubsection{Coherence results and comparisons}

We estimated the statistical coherence of the clusters obtained
at the low--temperature end of the trade--off curves where $1/T=25$. 
This coherence was estimated with respect to each of the three Gene Ontologies.
To gain some perspective, we applied similar analysis with 
a recent release of the {\it Cluster\/} software, called {\it Cluster 3.0\/} \cite{Cluster}.
This software is considered to be a state--of--the--art (and quite popular) 
tool for cluster analysis of gene expression data. 
We experimented extensively with all the basic algorithms available in this package.
These include two different variants of iterative $K$--means clustering
({\it $K$--means\/} and {\it $K$--medians\/}) and four different variants of
hierarchical clustering ({\it Complete linkage, Average linkage,
Centroid linkage\/}, and {\it Single linkage\/}).
With each of these algorithms we tried three standard similarity measures:
the Pearson correlation (``centered correlation'') \cite{Pearson},
the absolute value of the Pearson correlation, 
and the Euclidean distance. 
Thus, altogether we compared our performance to $18$ 
different configurations of this software
which are probably the most commonly used configurations. 
For the six $K$--means variants we tried $100$ 
different random initializations in every run, from which the best
solution (the one with the smallest sum of within--cluster distances) was chosen. 
The comparison was undertaken at all the different numbers of clusters,
$N_c=5,10,15,20$. The results are summarized in
\tblref{Tbl:ESR_Nc20} to \tblref{Tbl:ESR_Nc5}.
The average results are given \figref{Fig:ESR_Coh}.

In all cases the {\it Iclust\/} algorithm was clearly superior
to all of the $12$ hierarchical algorithms we tried.
It should be stressed that these algorithms are considered
a powerful tool for analyzing genomic datasets,
and many published applications are based on this type of hierarchical analysis.
Nonetheless, standard hierarchical clustering typically failed to
see a significant substructure in the ESR module.
In most cases {\it Iclust\/} was also superior to the average performance of
the six $K$--means variants, and in some cases ({\it e.g.\/}, $N_c=5$)
it was in fact superior to all the $K$--means variants.
Averaging over all three Gene Ontologies and over all four $N_c$ values,
Iclust obtains a coherence of $\sim 56\%$ while
the average $K$--means coherence is $\sim 42\%$ and
the average Hierarchical coherence is $\sim 12\%$. 

We further repeated this comparison with all the competing algorithms
while considering the $\log_2$ ratios of expression profiles as input, 
instead of the raw ratios.
Even with this preprocessing (to which our approach is invariant),
the {\it Iclust\/} average performance is superior to almost all the $18$ alternatives,
typically by a significant margin. 
Specifically, when averaging over all three Gene Ontologies 
and over all four $N_c$ values,
the average $K$--means coherence is $\sim 52\%$ and
the average Hierarchical coherence is $\sim 19\%$. 
While there exists some intuitive motivation for the $\log_2$ preprocessing, there is no formal justification. 
Clearly,  from a principled point of view, 
an approach which is invariant to such
transformations is preferable, even if it were to generate only comparably good results.

\subsection{Detailed results for $N_c=20$ clusters}
\label{sec:ESR_fulldetails}

In \tblref{Tbl:ESR_Coh} we present all enriched annotations
for the {\it Iclust\/} partition with $N_c=20$ clusters and $1/T=25$. 
Further examination of these clusters yields several observations that allow us to see in more detail what makes these clusters meaningful solutions to our problem.

First, in several cases the extracted clusters
consist of genes from both the nominal induced and repressed groups.
For example, $c5$ consists of $26$ induced
genes (enriched with {\it oxidoreductase activity\/})
and $6$ repressed ones. In \figref{Fig:ESR_Centroids}A we see that the genes in
this cluster have a relatively augmented response under Menadione exposure
and a relatively reduced response in a stationary phase, 
as opposed to genes not in this cluster.

In \figref{Fig:ESR_Centroids}B we display the average behavior of the $22$ induced
genes in $c8$ versus the $49$ induced genes in 
$c19$ in two opposing temperature shifts. 
Although all are induced by heat, the genes in $c19$ are more sensitive to this
treatment which is consistent with the enrichment
of {\it heat shock protein activity\/} in this cluster. 

Cluster $c18$ consists of $122$ repressed genes which were mainly 
ribosomal proteins. 
In \figref{Fig:ESR_Centroids}C we see that the genes in this clusters 
exhibit a distinguished transient expression pattern
under,{\it e.g.\/}, Diamide treatment, a fact that was already mentioned in \cite{Gasch02}.
On the other hand, cluster $c16$ consists of $87$ repressed genes 
and is enriched for {\it ribosome biogenesis\/} and other
related annotations. In the same figure we see
that this cluster exhibits another distinctive behavior
with respect to the rest of the repressed genes. 

In \figref{Fig:ESR_Centroids}D we consider again two clusters,
$c2$ and $c7$, which seem to involve ribosomal proteins 
and ribosome biogenesis, respectively. 
As seen in the figure, when the cells converge to a quiescent state
under Nitrogen depletion, these two clusters exhibit
quite different behaviors.

In \figref{Fig:ESR_Centroids}E we see another intriguing behavior of two clusters, 
$c15$ and $c17$, 
under steady--state growth at different temperatures.
From the GO annotations we find that $c15$, which includes
$12$ repressed genes, is enriched for tRNA aminoacylation,
while $c17$ which includes $7$ repressed genes is enriched
with cell cycle related annotations. 
\figref{Fig:ESR_Centroids}F demonstrates that the distinction between these two clusters
is not spurious, as they display different behaviors, {\it e.g.\/},  
in response to hyper--osmotic shock.

As two complementary validation schemes we used the regulator--promoter region 
interactions reported in \cite{Lee+al02}\footnote{
In these data, every gene is ``annotated'' with $106$
``$P$-value'' scores that determine the probability of
this gene being regulated by each of $106$
yeast transcriptional regulators. By considering only
interactions with a $P$-value lower than $0.005$ 
we constructed out of these data an annotation matrix
with $868$ (gene) rows, $106$ (regulator) columns
and a total $1307$ predicted interactions.}
and the DNA--binding sequence motifs provided in \cite{Pilpel+al01}.\footnote{Here, again, one can construct an annotation matrix 
where $A({\rm i},{\rm j})=1$ if and only if the $1,000$ base--pair promoter sequence 
of the ${\rm i}$-th gene includes the ${\rm j}$-th motif.
After considering only the $100$ most frequent motifs we ended
up with an annotation matrix, with $868$ (gene) rows,
$100$ (motif) columns and $19,517$ predicted interactions.}
In most of our clusters 
we found enrichment of regulatory interactions
and/or known sequence motif in the corresponding upstream sequences
($Pval < 0.05$, Bonferroni corrected). 
For example, $c5,\,c19,$ and $c17$ were enriched for YAP1,
HSF1, and MBP1, respectively. 
As YAP1 is known to be involved with response to oxidative stress,
HSF1 with response to heat, and MBP1 with cell cycle regulation,
these enrichments are clearly in consistent with the GO enrichments
for the same clusters. 
$c18$ and $c2$ (Figs. 4C,D) were enriched with FHL1 
which is required for rRNA processing, and $c18$ was further
enriched with RAP1 -- known to be involved in regulating
ribosomal proteins, 
and with four other regulators (GAT3, YAP5, PDR1, and RGM1),
suggesting similar, yet not identical regulatory
programs for these two functionally related clusters. 
$c16$ was enriched for ABF1 
and both $c7$ and $c16$ were enriched with several motifs
which are known to be related to rRNA processing and synthesis, 
consistently with the GO annotations enriched for these clusters.

\subsection{A cluster enriched with uncharacterized genes}

In the statistical validation of our clusters (\secref{sec:ESR_Coh})
we removed from the analysis uncharacterized genes.
Nonetheless, the distribution of the uncharacterized genes among our 
clusters yields an intriguing result.
One might have suspected that almost every process in the cell has a few components 
that have not been identified, and hence that as these processes are regulated 
there would be a handful of unknown genes that are regulated in concert with 
many genes of known function.  
For  at least one of our clusters, 
our results reveal a different picture.

Given the fraction of uncharacterized genes
in a cluster and the corresponding fraction at the entire population,
one can use the hypergeometric distribution to calculate a $P$-value
for this event (see \secref{sec:coh}). 
Applying this to our partition into $N_c=20$ clusters
we find that $c7$ is significantly enriched with 
genes that are uncharacterized in the $GO_{BP}$ and $GO_{MF}$ ontologies.

Specifically, out of the $123$ genes in $c7$,
$\;72$ have an unknown molecular function.
This level of concentration has a ($P$-value)
probability of $\sim 10^{-8}$ to have arisen by chance. 
Moreover, if we consider only the repressed genes in the ESR module
(since $c7$ consists mainly of such genes),
we see that $69$ out of the $114$ repressed genes in $c7$
are uncharacterized in the $GO_{MF}$ ontology,
which has a $P$-value of $\sim 10^{-15}$.  

Closer examination of the $GO_{BP}$ {\it characterized\/} genes in the same cluster 
reveals several enriched annotations (see \tblref{Tbl:ESR_Coh})
related to ribosome biogenesis and ribosomal RNA processing,
suggesting that most of the previously unannotated genes in this cluster are
involved in these processes as well.
Nonetheless, the extremely high concentration of uncharacterized genes
in this cluster suggests that these genes are involved with biological
processes which are harder to detect and characterize with the current technologies. 

\ignore{
Clearly, it is very unlikely that the high concentration
of uncharacterized genes in this cluster is a pure coincidence
and one should look for alternative explanations. 

A possible lead is given by the cellular--component ($GO_{CC}$) 
annotations in this cluster.
Specifically, out of $43$ repressed genes in $c7$ 
which are uncharacterized both in $GO_{MF}$ and in $GO_{BP}$,
$31$ has a $GO_{CC}$ annotation. 
$11$ out of these $31$ are 
annotated {\it simultaneously\/} with two $GO_{CC}$ annotations:
``nucleus'' and ``nucleolus'', 
or ``nucleus'' and ``cytoplasm'' ($P$-value $<\;0.0005$ with 
respect to the entire repressed genes population).
Moreover, almost all the rest of these $31$ genes
are directly annotated as either 
``nucleus'', ``nucleolus'', or ``cytoplasm''
($P$-value $<\;10^{-6}$ with respect to the entire repressed genes population).
Since the current GO annotations are sometimes only partial
it seems reasonable to conclude that further genes out of the above--mentioned
$43$ will be identified in the future as annotated simultaneously
with more than one of the three annotations, 
``nucleus'', ``nucleolus'', and ``cytoplasm''.

Thus, we see that a suspiciously high fraction
of the $GO_{BP}$ and $GO_{MF}$ uncharacterized genes in $c7$
are annotated simultaneously as either ``nucleus'' and ``nucleolus'', 
or ``nucleus'' and ``cytoplasm''. 
This, together with the fact that $c7$ as a whole
seem to be involved in ribosomal biogenesis and processing, 
suggests that its uncharacterized genes are involved in
transporting ribosomal proteins among these different cell components.
In particular, it might be that this ``non--static'' activity
is somewhat harder to characterize (as opposed to well located activities,
{\it e.g.\/}, a protein which is a structural constituent of the ribosome).
This might explain why such a large fraction of this module
remains uncharacterized to date.
}

Finally, it is also worthwhile to point out that the cluster $c7$ 
is extremely conserved when one tries to find partitions
with a smaller number of clusters,
as demonstrated in \figref{Fig:ESR_Hierarchy}.
In fact, all the parent clusters of this $c7$ cluster
(for $N_c=5,10,15$) were similarly enriched for
$GO_{BP}$ and $GO_{MF}$ uncharacterized genes. 

\section{Second application: The SP500 data}
\label{sec:SP500}

\subsection{Description of the data}

In our second application we consider a very different data set, 
the companies in the Standard and Poor's $500$ list. 
We used the May 2004 listing of the $500$ companies,
available at 
{\it http://www.standardandpoors.com\/}.
For these companies we examine the day--to--day fractional changes in stock price 
during the trading days between December 2, 2002, and December 31, 2003,
(a total of $273$ trading days), as seen in \figref{SP_data}.\footnote{
These data are available at 
{\it http://wrds.wharton.upenn.edu\/}. We identified the different companies
by their ticker symbols as reported in {\it http://www.standardandpoors.com\/}.
However, these symbols are not unique, and as a result the database at
{\it http://wrds.wharton.upenn.edu\/} returned slightly more than 500 companies; for $501$ of these the data were available for the entire 2003 year, hence these 
are the companies we consider in our analysis.}

From these data we estimated all the $\sim 125,000$ 
mutual information relations, as described in \cite{Slo+al04},
ending up with a $500 \times 500$ matrix $I_{\rm ij}$ which, as before,
defines the input to our clustering procedure. 
For convenience, we provide here some statistics of the estimated mutual information values. 
For a complete description, including different verification schemes
that support the reliability of our estimates, 
the reader is referred to \cite{Slo+al04}.

Across all pairs of companies, the average estimated mutual information was $0.10\;{\rm bits}$ with a variance of $0.0054\;{\rm bits}^2$, and
the maximal estimated mutual information was $0.97\;{\rm bits}$.
All the pairwise mutual information relations are presented in \figref{SP_MIs},
where the companies are sorted according to the clustering partition
into $N_c=20$ clusters that we analyze in detail (see below).
The self--information relations were set to 
$I_{\rm ii} = \log_2(5)$ which corresponds to the maximal 
possible information under a quantization into five bins \cite{Slo+al04}.

\subsection{Quality--complexity trade--off curves}

Given the pairwise mutual information matrix we applied the {\it Iclust\/} algorithm
described in \secref{sec:alg}.
As in the first application, we explored the trade--off between $\langle s \rangle$ 
and $I(C;{\rm i})$ for different numbers of clusters:
$N_c=5,10,15,20$ and for different values
of the trade--off parameter, $T$.
Specifically, we found that $1/T = \{15,20,25,30,35\}$ were
typically sufficient to obtain a relatively clear saturation of $\langle s \rangle$, hence
we present the results for these $T$ values. 
For each $\{N_c\;,T\}$ pair we performed $10$ different random initializations
ending up with $10$ (possibly) different
local maxima of $\F$, from which we chose the best one.
The resulting trade--off curves are presented in the middle
panel of \figref{Fig:tradeoff}. 

As before, as $T$ is lowered, $\langle s \rangle$ increases
but so does $I(C;{\rm i})$. In addition, the solutions become more
deterministic. For example, for $N_c=20$
and $1/T=25$, only $\sim\;36\%$ of the companies
have nearly deterministic assignment 
[$P(C|{\rm i}) > 0.9$ for a particular $C$].
On the other hand, for  $1/T=35$, all the assignments
are nearly deterministic [$P(C|{\rm i}) > 0.9$].

For brevity, we focus
our analysis on solutions for which the saturation
of $\langle s \rangle$ is relatively clear, {\it i.e.\/},
on the four solutions with $N_c=\{5,10,15,20\}$ and $1/T=35$.
In all these partitions almost all of the companies
had a nearly deterministic assignment 
[$P(C|{\rm i}) > 0.9$ for a particular $C$],
so we treat these solutions 
as hard partitions where every company is assigned solely to its most
probable cluster.

\subsection{Comparing solutions at different numbers of clusters}
\label{sec:SP500_Hierarchy} 

We examine directly how well our independent solutions
form a hierarchical structure. 
Accordingly, we apply exactly the same scheme as described 
in \secref{sec:ESR_Hierarchy} to the four solutions we obtained
independently for $N_c=\{5,10,15,20\}$ with $1/T=35$.
The results are presented in \figref{Fig:SP500_Hierarchy}. 
Again, the independent solutions
form only an approximate hierarchy.
Nonetheless, this approximation seems more suitable in this case,
as demonstrated,{\it e.g.\/}, by the larger percentage of nearly perfect
inclusion relations (solid bold edges in the figure).
It should be noted that indeed the standard classification
of these companies is hierarchical in nature (see \secref{sec:SP500_annotation}).

Again, it is worthwhile to point out that some of the clusters
are better preserved than others across the different levels. 
For example, the {\it Semiconductor Equipment\/} cluster, $c11$,
is clearly identified in all the independent solutions.

\subsection{Coherence results}

\subsubsection{Constructing the annotation matrices}
\label{sec:SP500_annotation}

We used the Global Industry Classification Standard (GICS),
which classifies companies at four different levels:
sector, industry group, industry, and sub-industry
(see {\it http://www.standardandpoors.com\/}).
These four levels are arranged in a well defined, tree--like hierarchy. 
The bottom (sector) level consists of $10$ different annotations:
{\it Consumer Discretionary,
Consumer Staples,
Energy,
Financials,
Health Care,
Industrials,
Information Technology,
Materials,
Telecommunication Services,\/} 
and {\it Utilities\/}.
The next (industry group) level consists of $24$ distinct annotations. 
The next (industry) level consists of $59$ distinct annotations. 
The last (sub--industry) level consists of $114$ distinct annotations. 
Thus, altogether there are $207$ different annotations where
every company is assigned exactly four annotations, 
one at every level of the hierarchy. 

As in the first application, while estimating clusters' coherence we removed
annotations that were assigned to less than two companies in our data,
ending up with a total of $178$ distinct annotations.

\subsubsection{Coherence results and comparisons}

We estimated the coherence of the clusters obtained
at the low--temperature end of the trade--off curves where $1/T=35$. 
To gain some perspective, we applied a similar analysis to the results obtained with the
{\it Cluster 3.0\/} software \cite{Cluster}.
We experimented with the same $18$ basic configurations
as in the previous application 
($K$-means variants, again with $100$ different initializations),
and applied the comparison 
to all the different numbers of clusters we examined,
$N_c=5,10,15,20$. The results are summarized in
\tblref{Tbl:SP_Nc20} to \tblref{Tbl:SP_Nc5} and in \figref{Fig:SP500_Coh}.

In all cases, {\it Iclust\/} was superior to the average performance
of the $K$--means and the hierarchical {\it Cluster 3.0\/} variants.
In fact, except for the $K$--medians configurations, 
none of the other algorithms came even close to the {\it Iclust\/} performance. 
Averaging over all four $N_c$ values,
Iclust obtains an average coherence of $\sim 90\%$ while
the average $K$--means coherence is $\sim 79\%$ and
the average Hierarchical coherence is only $\sim 19\%$. 

It is interesting to point out that although the annotations
for these data are arranged in a relatively simple
and clear hierarchical structure, the performance
of the hierarchical algorithms are still relatively poor,
perhaps due to the greedy nature of these optimization
routines, which typically yield suboptimal solutions.

\subsection{Detailed results for $N_c=20$ clusters}
\label{sec:SP500_fulldetails}

In \tblref{Tbl:SP_Coh} we present all enriched annotations
for the {\it Iclust\/} partition with $N_c=20$ clusters and $1/T=35$. 
Several specific results are noted in the following. 

First, $8$ out of the $20$ clusters are perfectly ($100\%$) coherent.
For example, $c11$ consists of $18$ companies which are all 
{\it Information Technology\/} companies, mainly sub--classified as 
{\it Semiconductors \& Semiconductor Equipment\/} companies
such as {\it Intel\/} and {\it Texas Instruments\/}.  In contrast, 
$c17$ consists mainly  of different types of retail stores:
{\it Department Stores\/} like {\it Sears\/}, {\it General Merchandise Stores\/}
like {\it Target\/}, {\it Speciality Stores\/} like {\it Staples\/}, and so on.

Perhaps more interesting is the relatively subtle distinction
between $c7$ and $c1$, both of which are perfectly coherent.
The former includes mainly companies which are classified
as {\it Investment Banking \& Brokerage\/} ({\it e.g.\/}, {\it Merrill Lynch\/})
or {\it Asset Management \& Custody Banks\/}, while the latter
corresponds to {\it Commercial (Regional) Banks\/} like {\it PNC\/}. 
Indeed, in \figref{Fig:SP500_Hierarchy} we see that these two clusters
nicely merge with each other at the independent solution found for
$N_c=15$ clusters. A similar relatively subtle distinction also is
captured between $c6$ and $c20$ (again, both are perfectly coherent),
where both clusters correspond to different sub--classifications
of the {\it Oil \& Gas\/} category. As for the previous pair, these two
clusters also merge for $N_c=15$.

Even in clusters with non--perfect coherence we typically see a clear
reasoning behind the automatically recovered structure. 
For example, $c16$ is  
enriched only for three {\it Hotels Resorts \& Cruise Lines'\/} companies,
with a coherence level of only $30\%$. Nonetheless, 
it further contains two banks ({\it MBNA\/} and {\it Capital One Financial\/})
which specialise in credit card issuing and therefore consumer spending, 
a company ({\it CINTAS\/}) which is a builder of corporate identity, 
and another company ({\it Paychex\/}) which handles payroll and human resource services for employees. 
In addition, the {\it Walt Disney Co.\/} is also in this cluster, presumably
due to its parks and resorts division. 

\section{Third application: The EachMovie data}
\label{sec:EachMovie}

\subsection{Description of the data}

In our third test case we consider the $EachMovie$ dataset, 
movie ratings provided by more than $70,000$ viewers.\footnote{See
{\it http://www.research.digital.com/SRC/eachmovie/\/}.}
These data are inherently quantized as only six discrete possible ratings were used. 
Indeed, many real life clustering problems involve such 
categorical data. 
In these cases the issue of what similarity measure to use seems 
even more obscure, especially if the descriptive attributes are 
not naturally ordered, and our general information theoretic approach seems especially promising.

We represented each movie by its ratings from different viewers
and focused on the $500$ movies that got the maximal number of votes. 
These data are presented in \figref{EachMovie_data}.
We estimated all the $\sim 125,000$ 
mutual information relations as in the previous applications; again see Ref \cite{Slo+al04} for details.
Notice that in estimating the mutual information for a pair of movies, only 
viewers who voted for both movies can be  considered. 

Across all pairs of movies, the average estimated mutual information was $0.052\;{\rm bits}$, with a variance of $0.0026\;{\rm bits}^2$, and
the maximal estimated mutual information was $0.89\;{\rm bits}$.
All the pairwise mutual information relations are presented in \figref{EachMovie_MIs},
where the movies are sorted according to the clustering partition
into $N_c=20$ clusters that we analyze in detail (see below).
The self-information relations were set to 
$I_{\rm ii} = \log_2(6)$ which corresponds to the maximal 
possible information under a quantization into six bins.

\subsection{Quality--complexity trade--off curves}

Given the pairwise mutual information matrix we applied the {\it Iclust\/} algorithm
described in \secref{sec:alg}.
As in the previous applications, we explored the trade--off between $\langle s \rangle$ 
and $I(C;{\rm i})$ for different numbers of clusters:
$N_c=5,10,15,20$ and for different values
of the trade--off parameter, $T$.
Specifically, we found that $1/T = \{20,25,30,35,40\}$ were
typically sufficient to obtain a relatively clear saturation of $\langle s \rangle$, hence
we present the results for these $T$ values. 
For each $\{N_c\;,T\}$ pair we performed $10$ different random initializations
ending up with $10$ (possibly) different
local maxima of $\F$, from which we chose the best one.
The resulting trade--off curves are presented in the right
panel of \figref{Fig:tradeoff}. 

As before, as $T$ is lowered, $\langle s \rangle$ and $I(C;{\rm i})$ increase
and the solutions become more
deterministic. For example, for $N_c=20$
and $1/T=30$, only $\sim\;32\%$ of the movies
have nearly deterministic assignment, 
while for $1/T=40$ almost all the movie assignments
are nearly deterministic [$P(C|{\rm i}) > 0.9$ for a particular $C$].

For brevity, we   focus
our analysis on solutions for which the saturation
of $\langle s \rangle$ is relatively clear, {\it i.e.\/},
on the four solutions with $N_c=\{5,10,15,20\}$ and $1/T=40$,
and  we treat these solutions 
as hard partitions where every movie is assigned solely to its most
probable cluster.

\subsection{Comparing solutions at different numbers of clusters}
\label{sec:EachMovie_Hierarchy} 

We examine directly how well our independent solutions
form a hierarchical structure by applying the same
scheme as in \secref{sec:ESR_Hierarchy}. 
The results are presented in \figref{Fig:EachMovie_Hierarchy}. 
Clearly, the relations between solutions at different numbers
of clusters are relatively weak, suggesting that the data really do not support a robust hierarchical structure.
Only a few clusters are somewhat preserved as we vary $N_c$, like
the {\it Family--Animation--Classic\/} cluster, $c12$,
or the {\it Action\/} cluster, $c9$.

\subsection{Coherence results}

\subsubsection{Constructing the annotation matrices}
\label{sec:EachMovie_annotation}

We used the genre labels provided for these data
to construct the annotation matrix. 
Specifically, these labels are:
{\it Action\/} ($110$ movies), 
{\it Animation\/} ($25$ movies), 
{\it Art-Foreign\/} ($45$ movies), 
{\it Classic\/} ($44$ movies), 
{\it Comedy\/} ($149$ movies),  
{\it Drama\/} ($160$ movies),  
{\it Family\/} ($67$ movies),  
{\it Horror\/} ($33$ movies),  
{\it Romance\/} ($61$ movies), 
and {\it Thriller\/} ($90$ movies).
Almost half of the movies 
were annotated with more than one genre and the average number
of genre annotations per movie was $1.6$, with a maximal number
of $4$ different genres for a single movie. 

It is important to notice that these annotations are broad, providing a somewhat simplistic view of 
the structure in these data. For example, it is quite reasonable
that more subtle distinctions like the movie director and/or
main actors are reflected in the viewer preferences that
were used to cluster the movies.
Nonetheless, for practical reasons we used these broad
labels as a first--order approximation for our evaluation.

\subsubsection{Coherence results and comparisons}

We estimated the statistical coherence of the clusters obtained
at the low--temperature end of the trade--off curves where $1/T=40$. 
As before, to gain some perspective, we also used the
{\it Cluster 3.0\/} software \cite{Cluster}.
We experimented with the same $18$ basic configurations
as in the previous applications 
($K$-means variants, again with $100$ different initializations),
and applied the comparison 
to all the different numbers of clusters we examined,
$N_c=5,10,15,20$. The results are summarized in
\tblref{Tbl:EachMovie_Nc20} to \tblref{Tbl:EachMovie_Nc5} 
and in \figref{Fig:EachMovie_Coh}.

In all cases, {\it Iclust\/} was clearly superior to the average performance
of the $K$--means and the Hierarchical {\it Cluster 3.0\/} variants.
In fact, except for the hierarchical complete-linkage configuration
with the Pearson correlation as the similarity measure,
none of the other algorithms came even close to the {\it Iclust\/} performance. 
Averaging over all four $N_c$ values,
Iclust obtains an average coherence of $\sim 53\%$ while
the average $K$--means coherence is only $\sim 12\%$ and
the average Hierarchical coherence is $\sim 24\%$. 

Notice that, in contrast to the previous
applications, here the $K$--means algorithms
are inferior to some of the hierarchical algorithms
(and both are inferior to {\it Iclust\/}).
These results demonstrate that while standard clustering algorithms
might work well in certain circumstances and fail completely in others,
our principled and model--independent 
approach maintains a high and robust performance across
a wide variety of applications.

\subsection{Detailed results for  $N_c=20$ clusters}
\label{sec:EachMovie_fulldetails}

In \tblref{Tbl:EachMovie_Coh} we present all enriched annotations
for the {\it Iclust\/} partition with $N_c=20$ clusters and $1/T=40$. 
Several results should be noted specifically.

For example, $c12$ consisted solely of $14$ classic family movies
such as {\it The Wizard of Oz\/} and {\it Snow White\/}.
$c8$ consisted mainly of {\it Art--Foreign\/} movies, including
all the {\it Three Colors\/} trilogy by 
Kieslowski. 
$c15$ included all seven {\it Star Trek\/} movies. 
Moreover, some of the obtained clusters
reflect more subtle distinctions than the 
broad genre definitions. 
For example, both $c4$ and $c6$ were enriched for {\it Comedy\/},
but while $c4$ was further enriched for {\it Romance\/}
$c6$ consisted mainly of Jim Carrey and Adam Sandler movies.
Both $c7$ and $c17$ were enriched for {\it Action\/},
but $c7$ was further enriched for {\it Classic\/} with some
emphasis on Science Fiction movies such as
the {\it Star Wars\/} trilogy, the {\it Terminator\/} movies, 
{\it Alien\/}, and {\it Back to the Future\/}.
In contrast $c17$ consisted mainly of movies starring
Sylvester Stallone, Jean-Claude Van Damme etc. 

\ignore{
In a separate text file we provide complete details of
the specific partition obtained by {\it Iclust\/} for $N_c=20$ clusters.
Similar detailed results for $N_c=15,10,5$ (including the analogous
tables of \tblref{Tbl:EachMovie_Coh}) are available at request
and will be posted online in the corresponding web site. 
}

\begin{acknowledgments}
We thank O Elemento and E Segal for their help in connection with the analysis of the ESR data, and C Callan, D Botstein, N Friedman, R Schapire and S Tavazoie for their helpful comments on early versions of the manuscript.  This work was supported in part by the National Institutes of Health Grant P50 GM071508.  GT was supported in part by the Burroughs--Wellcome Graduate Training Program in Biological Dynamics.
\end{acknowledgments}

\clearpage

\begin{widetext}
\section{Figures and tables}

\clearpage

\begin{figure}
\centerline{\fbox{
\begin{minipage}{\columnwidth} 
\small
\begin{tabbing}
xxx\=xxx\=xxx\=xxx\=xxx\=xxx\=xxx\=\kill
\underline{\large\bf Input:}\\\\
\>Pairwise similarity matrix, $s({\rm i}_1,{\rm i}_2),\;\forall\;{\rm i}_1=1,...,N,\;{\rm i}_2=1,...,N\;$.\\
\>trade--off parameter, $T\;$.\\
\>Requested number of clusters, $N_c\;$.\\
\>Convergence parameter, $\epsilon\;$.\\\\
\underline{\large\bf Output:}\\\\
\>A (typically ``soft'') partition of the $N$ elements into $N_c$ clusters.\\\\
\underline{\large\bf Initialization:}\\\\
\>$m = 0\;$.\\
\>$P^{(m)}(C|{\rm i}) \leftarrow\;$ A random (normalized) distribution $\;\forall\;{\rm i}=1,...,N\;$.\\\\
\underline{\large\bf While True}\\\\
\>For every ${\rm i}=1,...,N\;$:\\\\
\>\>$\bullet~~P^{(m+1)}(C|{\rm i}) \leftarrow {P^{(m)}(C)} \exp\bigg{\{} {1\over T}[ 2 s^{(m)}(C;{\rm i} ) - s^{(m)}(C)]\bigg{\}},\;\forall\;C=1,...,N_c\;$.\\\\
\>\>$\bullet~~P^{(m+1)}(C|{\rm i}) \leftarrow \frac{P^{(m+1)}(C|{\rm i})}{\sum_{C'=1}^{N_c} P^{(m+1)}(C'|{\rm i})},\;\forall\;C=1,...,N_c\;$.\\\\
\>\>$\bullet~~m \leftarrow m+1\;$.\\\\
\>If $\;\forall\;{\rm i}=1,...,N,\;\forall\;C=1,...,N_c$ we have
$|P^{(m+1)}(C|{\rm i})-P^{(m)}(C|{\rm i})| \leq \epsilon\;$,\\
\>\>Break.\\\\
\end{tabbing}
\end{minipage}}}
\caption{\small Pseudo-code of the {\it Iclust\/} algorithm.
Extending the algorithm for the general case (of more than
pairwise relations) is straightforward. 
In principle we repeat this procedure for different initializations
and choose the solution which maximizes 
$\F = \langle s \rangle - T I (C; {\rm i})\;$.}
\label{algorithm} 
\end{figure}

\clearpage

\begin{figure}[] 
\begin{center}
\begin{tabular}{c}
    \psfig{figure=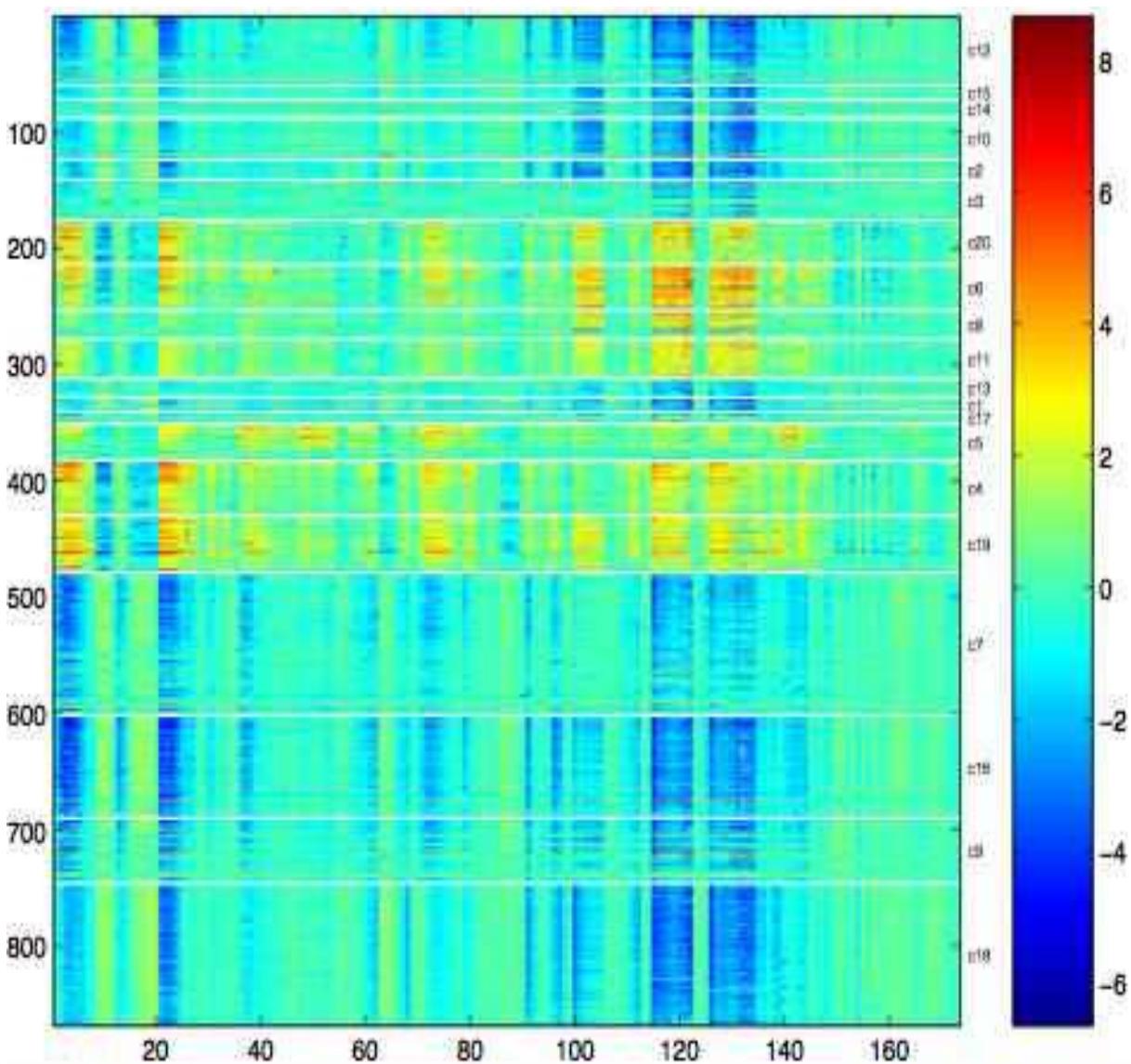,width=.9\columnwidth,height=6.0in} 
\end{tabular}
\end{center}
\caption{\small Expression profiles of the $868$ genes in the ESR data
across $173$ microarray experiments. 
Data taken from Gasch {\it et. al\/} \cite{Gasch00}.
Missing values are set to zero.
The genes are sorted according to the clustering partition 
into $20$ clusters that we 
analyze in detail later on. Inside each cluster, 
genes are sorted according to the average mutual information relation
with other cluster members.}
\label{ESR_data}
\end{figure}

\clearpage

\begin{figure}[] 
\begin{center}
\begin{tabular}{c}
    \psfig{figure=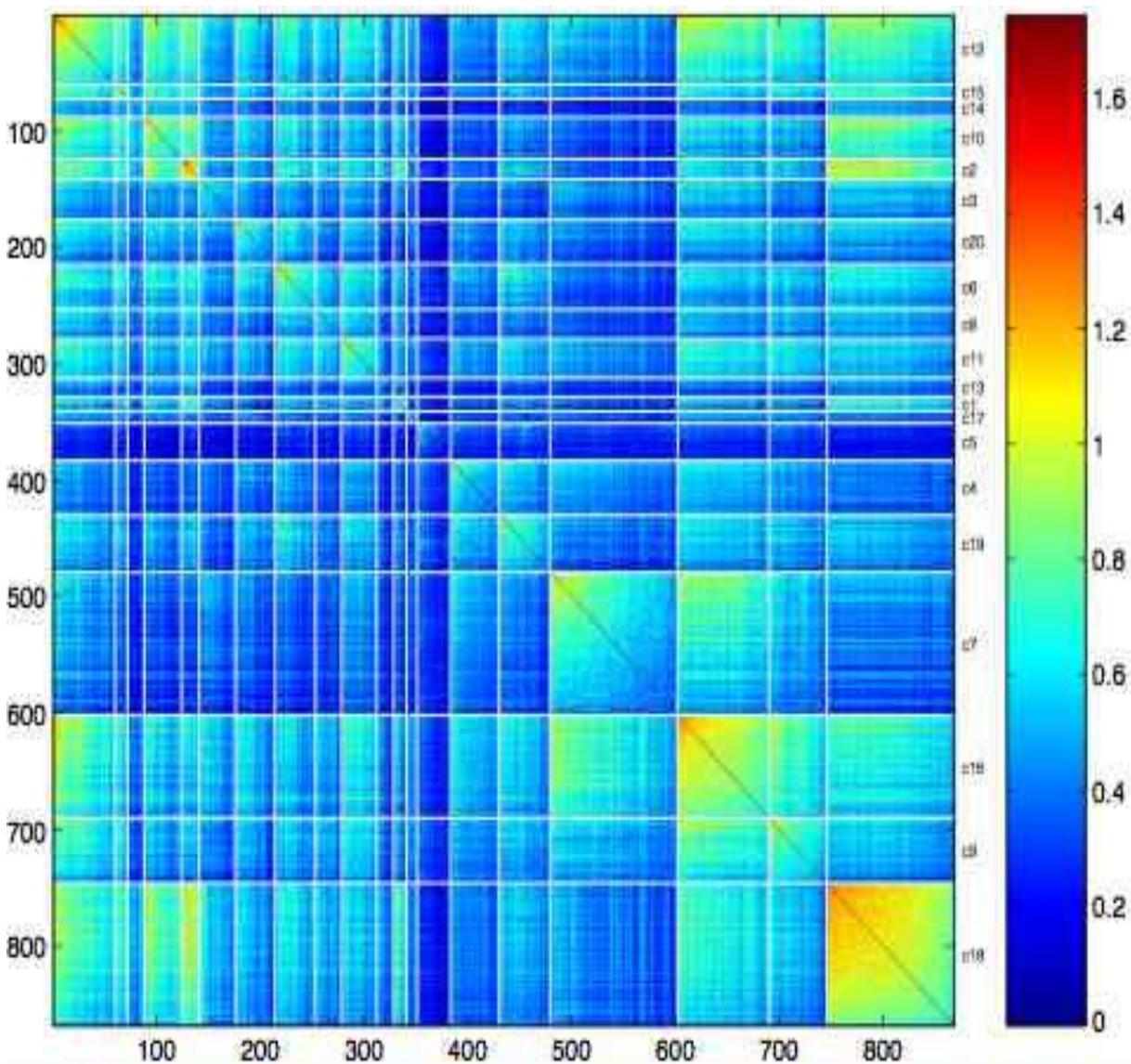,width=.9\columnwidth,height=6.0in} 
\end{tabular}
\end{center}
\caption{\small Pairwise mutual information relations for the $868$ genes in the ESR data.
The genes are sorted according to the clustering partition 
into $20$ clusters that we analyze in detail later on. Inside each cluster, 
genes are sorted according to the average mutual information relation
with other cluster members.}
\label{ESR_MIs}
\end{figure}

\clearpage

\begin{figure}[h] 
\label{Fig:TradeOffcurve}
\begin{center}
\begin{tabular}{ccc}
    \psfig{figure=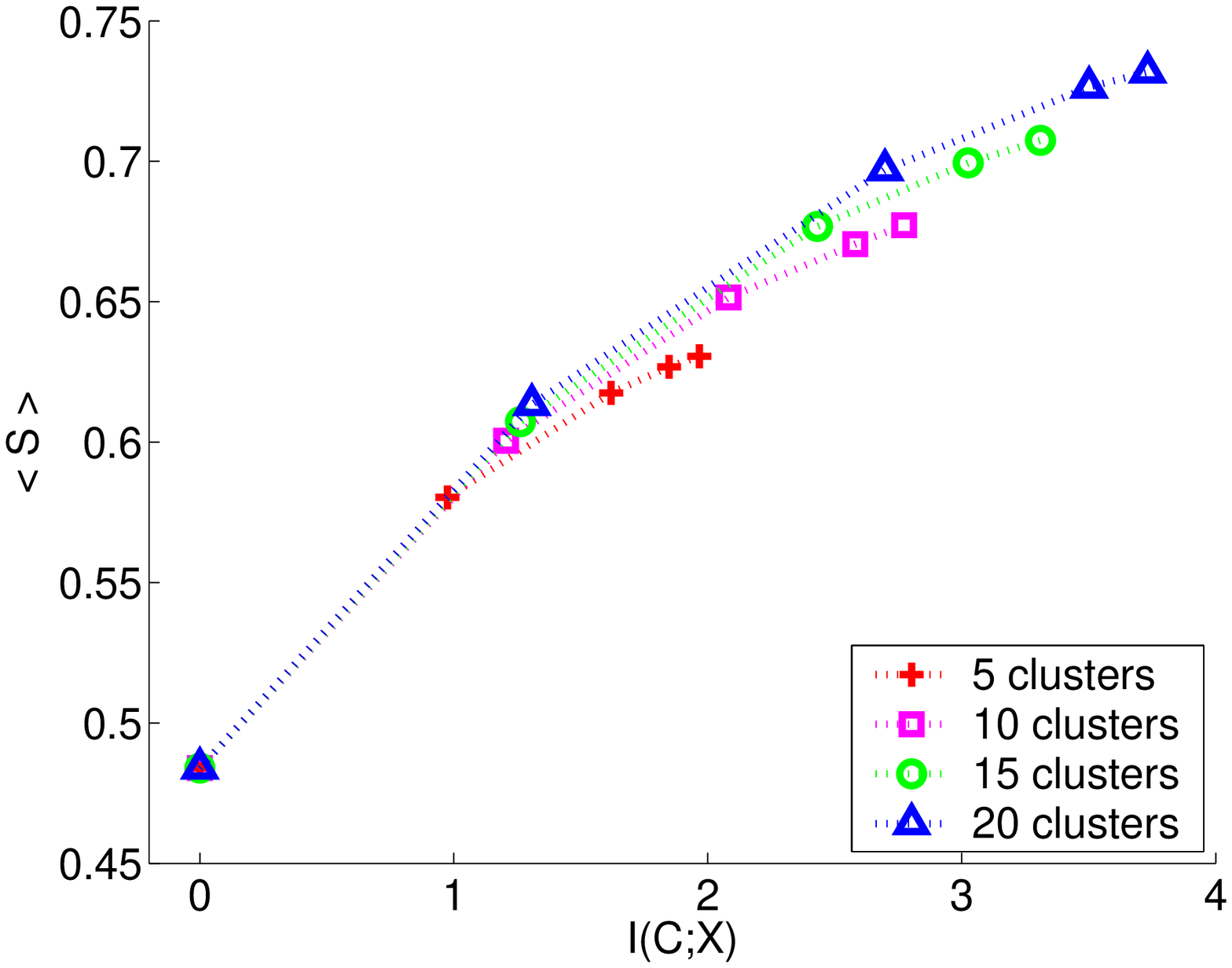,width=.31\columnwidth,height=2.0in} &
    \psfig{figure=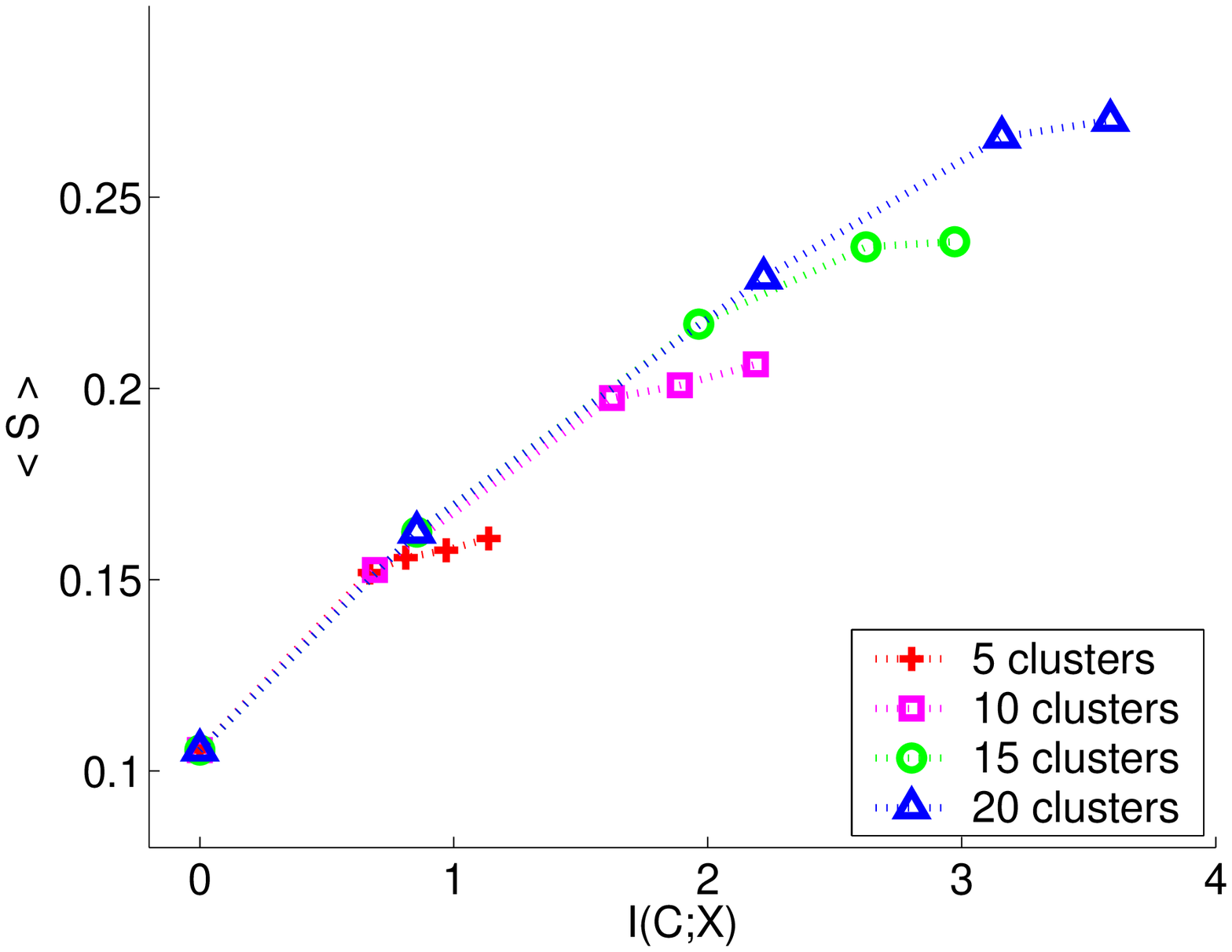,width=.31\columnwidth,height=2.0in} &
    \psfig{figure=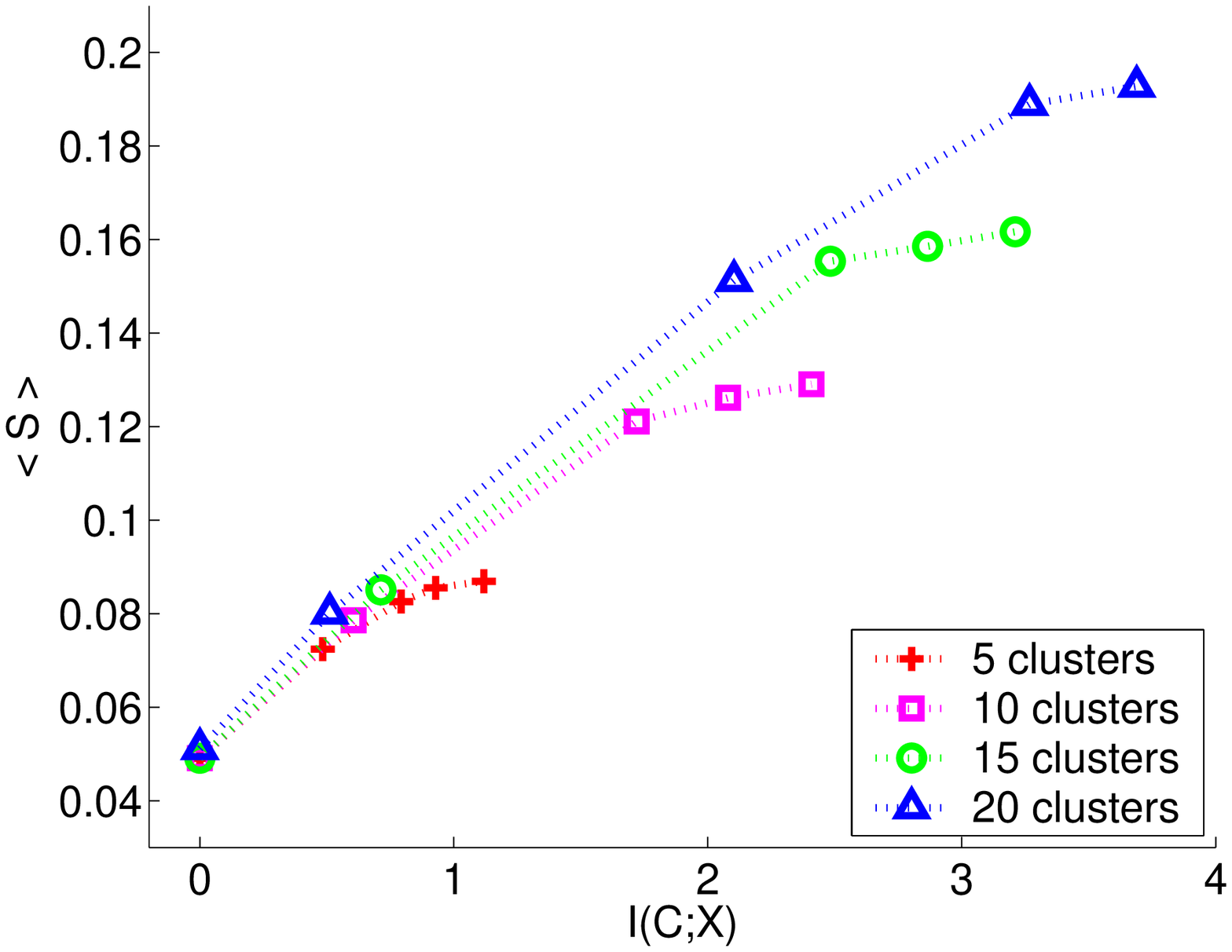,width=.31\columnwidth,height=2.0in} 
\end{tabular}
\end{center}
\caption{\small {\bf (Left)} Tradeoff curves obtained for the ESR data. 
Each curve describes the solutions obtained for a particular $N_c$ value,
{\it i.e.\/}, for a fixed number of clusters. 
Different points along each curve correspond to different local maxima
of $\F$ at different $T$ values. The results are presented 
for $\frac{1}{T}=\{5,10,15,20,25\}$ which 
suffices to obtain a relatively clear saturation of the average
mutual information per cluster, $\langle s \rangle$.
In \secref{sec:ESR_Hierarchy} we explore the possible hierarchical relations between 
the four saturated solutions at $\frac{1}{T}=25$ and $N_c=\{5,10,15,20\}\;$.
Further detailed analysis refers to the solution with $N_c=20$ and $\frac{1}{T}=25$
that obtained the highest $\langle s \rangle$ value. 
{\bf (Middle)} Similar tradeoff curves that were obtained for the SP500 data. 
The results are presented for $\frac{1}{T}=\{15,20,25,30,35\}$ which 
suffices to obtain a relatively clear saturation of $\langle s \rangle$.
Notice, that due to the lower average mutual information relations in these data
(with respect to the ESR example), 
one must apply lower $T$ values to obtain a clear saturation. 
In \secref{sec:SP500_Hierarchy} we explore the possible hierarchical relations between 
the four saturated solutions at $\frac{1}{T}=35$ and $N_c=\{5,10,15,20\}\;$.
Further detailed analysis refers to the solution with $N_c=20$ and $\frac{1}{T}=35$. 
{\bf (Right)} Similar tradeoff curves that were obtained for the EachMovie data. 
The results are presented for $\frac{1}{T}=\{20,25,30,35,40\}$ which 
suffices to obtain a relatively clear saturation of $\langle s \rangle$.
In \secref{sec:EachMovie_Hierarchy} we explore the possible hierarchical relations between 
the four saturated solutions at $\frac{1}{T}=40$ and $N_c=\{5,10,15,20\}\;$.
Further detailed analysis refers to the solution with $N_c=20$ and $\frac{1}{T}=40$.}
\label{Fig:tradeoff}
\end{figure}

\clearpage

\begin{figure}[] 
\begin{center}
\begin{tabular}{c}
   \psfig{figure=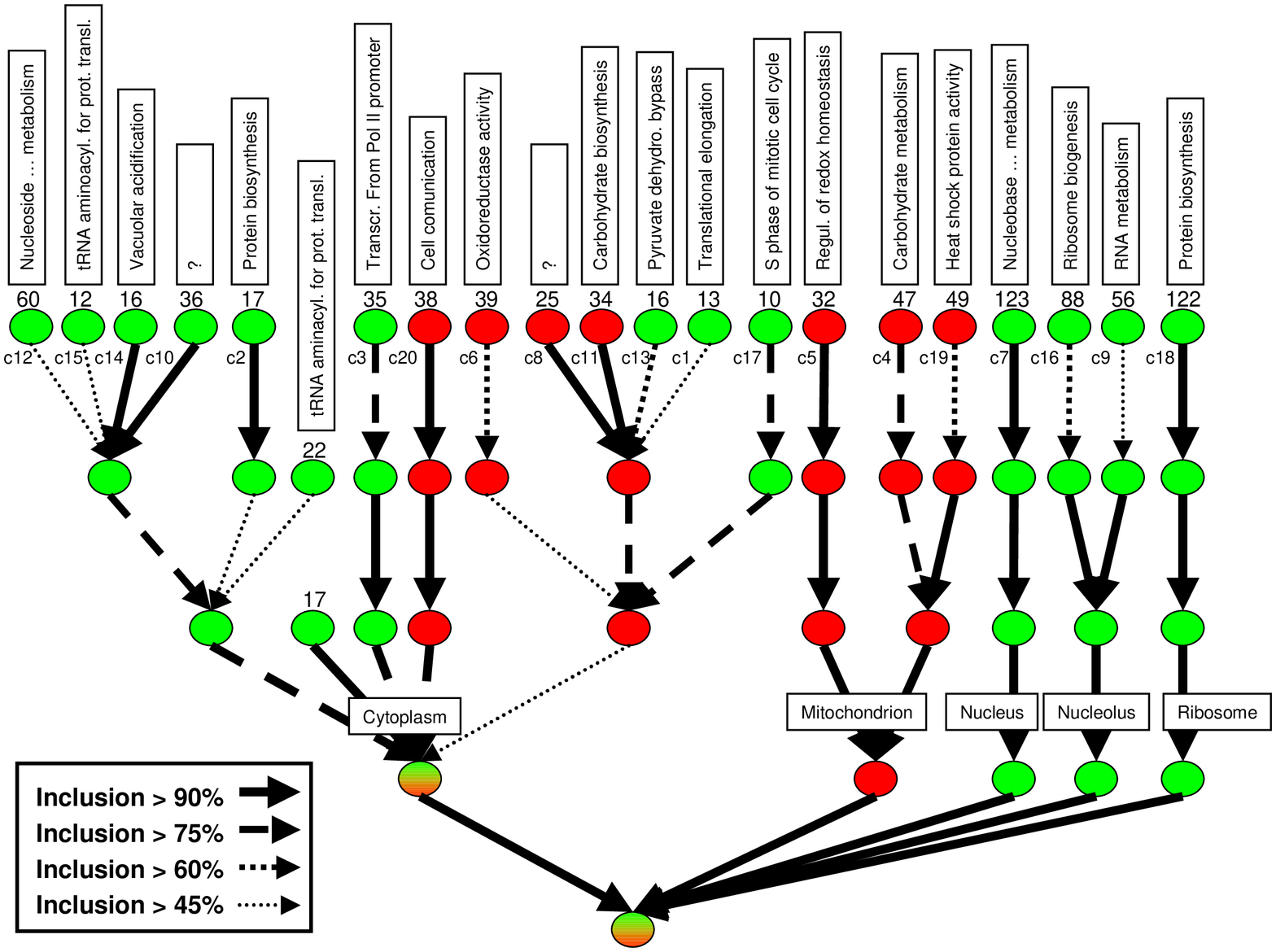,width=.9\columnwidth,height=5.5in} 
\end{tabular}
\end{center}
\caption{\small Relations between the optimal solutions with $N_c=\{5,10,15,20\}$
at $\frac{1}{T}=25$ for the ESR data.
At the upper level, $N_c=20$ clusters, 
and the clusters are sorted as in \figref{ESR_data} and \figref{ESR_MIs}.
The numbers above every cluster indicate the number of genes in this cluster. 
The title of each cluster 
correspond to the most enriched $GO_{BP}$ (biological process) annotation in the cluster,
{\it i.e.\/}, to the $GO_{BP}$ annotation with the smallest $P$-value in the cluster
(see \secref{sec:ESR_annotation}).
The only exceptions are $c6$, not enriched in $GO_{BP}$,
and $c19$, enriched with a non-informative annotation
({\it response to stress\/}). For these two clusters we use their
most enriched $GO_{MF}$ (molecular function) annotation as a title.
The titles of the five clusters at the lower level ($N_c=5$) are by
their most enriched $GO_{CC}$ (cellular component) annotation.
Notice, that most clusters were enriched with more than
one annotation, hence the short titles might be too concise 
in some cases (see \secref{sec:ESR_fulldetails} for a detailed
description of every cluster at the top level).
Red and green clusters represent clusters with a clear majority of 
stress--induced or stress--repressed genes, respectively.
In the {\it cytoplasm\/} cluster we had a relatively balanced
mixture of stress-repressed ($58\%$) and
stress-induced ($42\%$) genes.}
\label{Fig:ESR_Hierarchy}
\end{figure}

\clearpage

\begin{figure}[] 
\begin{center}
\begin{tabular}{c}
    \psfig{figure=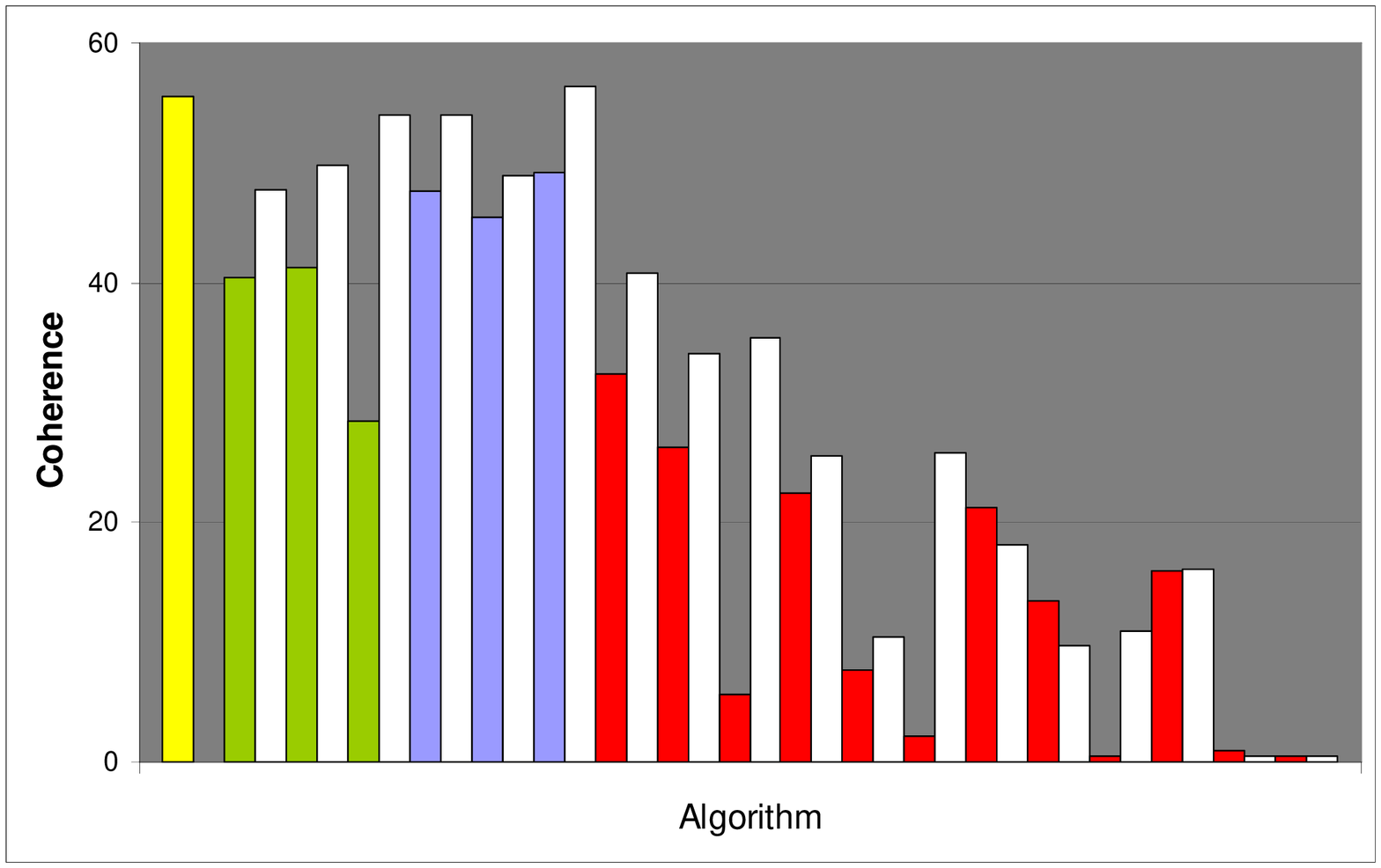,width=.9\columnwidth,height=3.0in} 
\end{tabular}
\end{center}
\caption{{\bf ESR data}:
Comparison of average coherence results of the {\it Iclust\/} algorithm 
(yellow) with conventional clustering algorithms \cite{Cluster}:
$K$--means (green); $K$--medians (blue); Hierarchical (red).
For the hierarchical algorithms, four different variants are tried:
complete, average, centroid, and single linkage, respectively
from left to right.
For every algorithm, three different similarity measures are applied:
Pearson correlation (left); absolute value of Pearson correlation
(middle); Euclidean distance (right).
The white bars correspond to applying the algorithm
to the logarithmically transformed  expression ratios. 
In all cases, the results are averaged over all the different numbers of clusters that 
we tried: $N_c=5,10,15,20$, and over the three Gene Ontologies.}
\label{Fig:ESR_Coh}.
\end{figure}

\begin{figure}[t] 
\begin{center}
\begin{tabular}{ccc}
    \psfig{figure=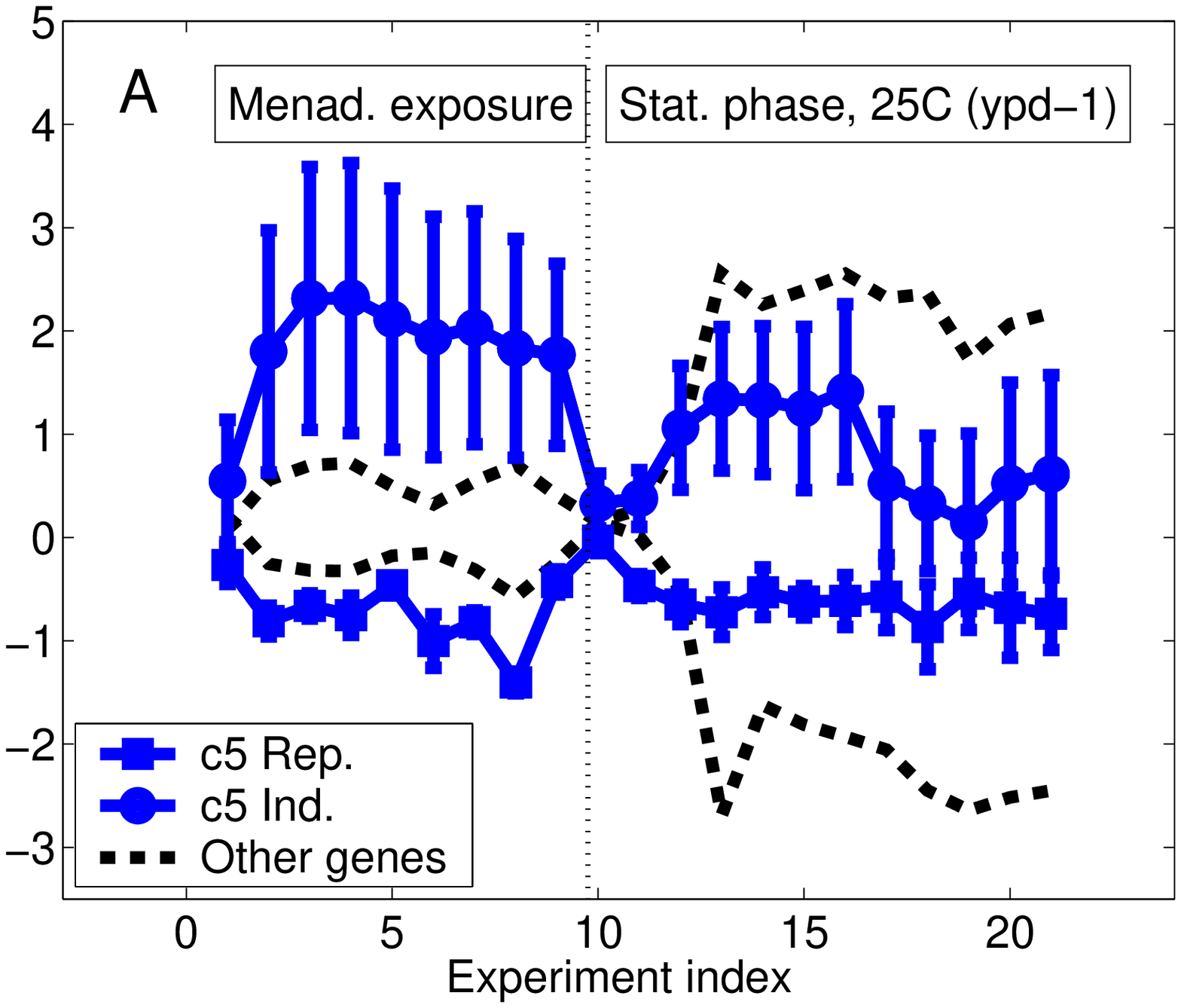,width=.31\columnwidth,height=1.8in} &
    \psfig{figure=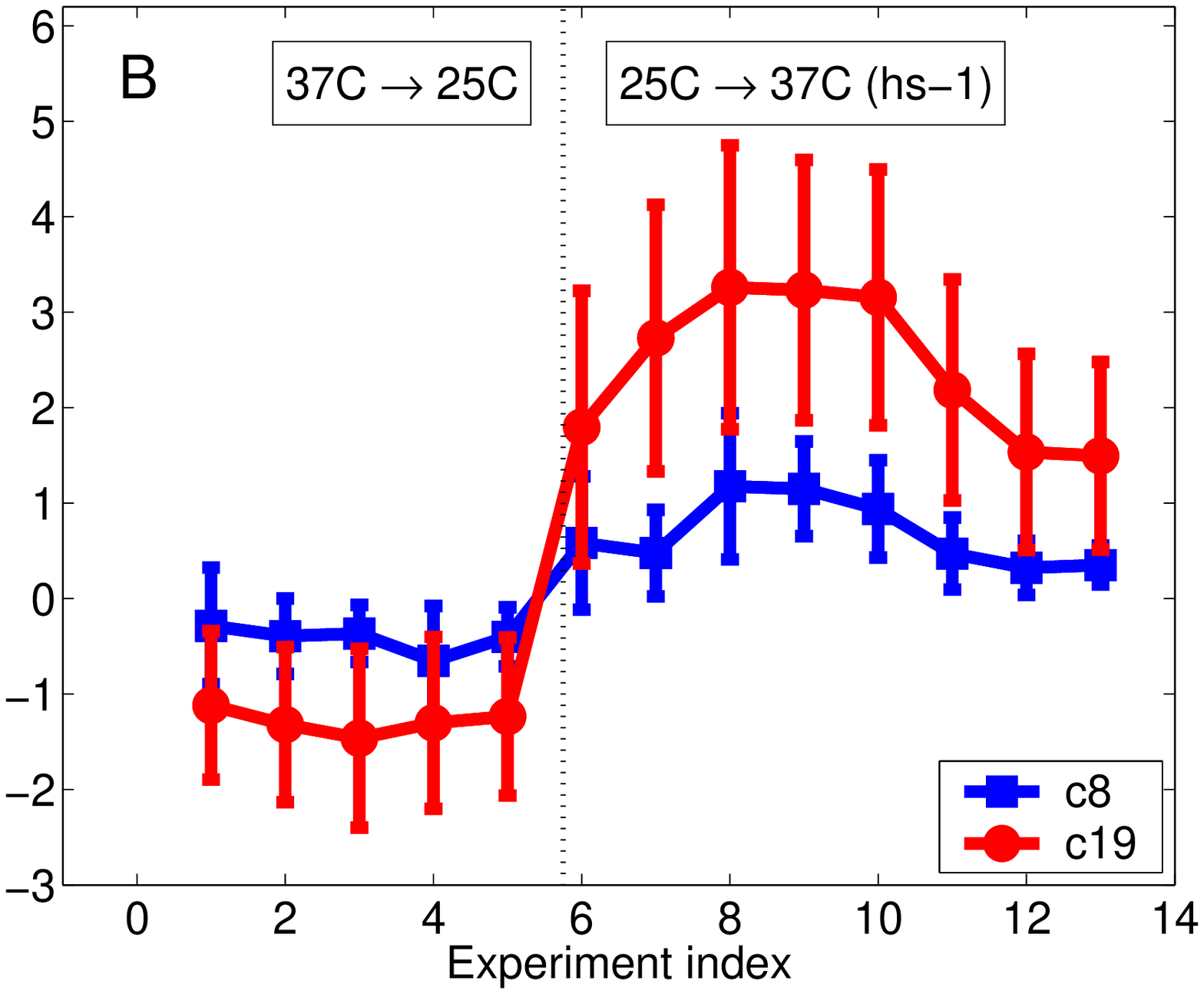,width=.31\columnwidth,height=1.8in} &
    \psfig{figure=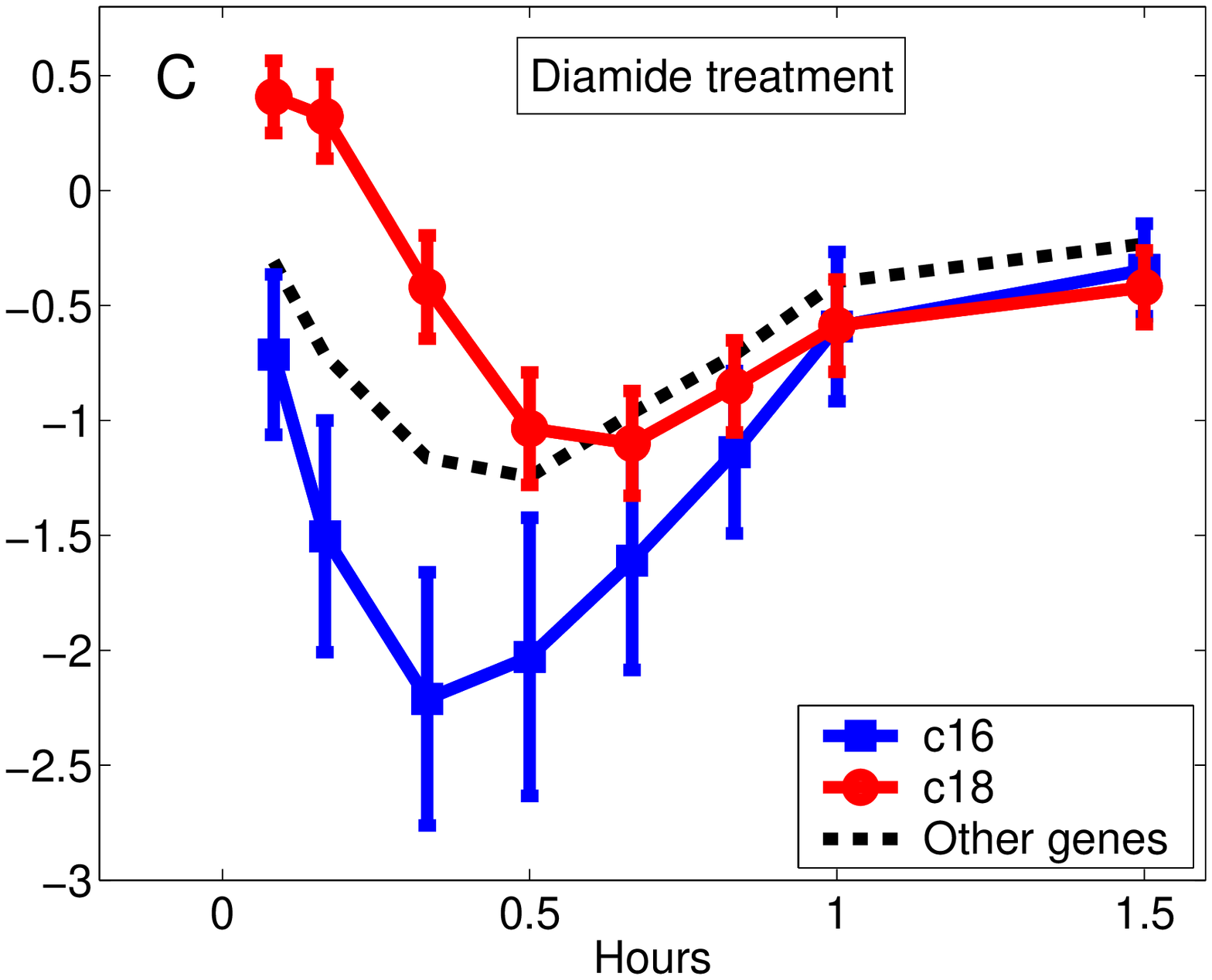,width=.31\columnwidth,height=1.8in} \\
    \psfig{figure=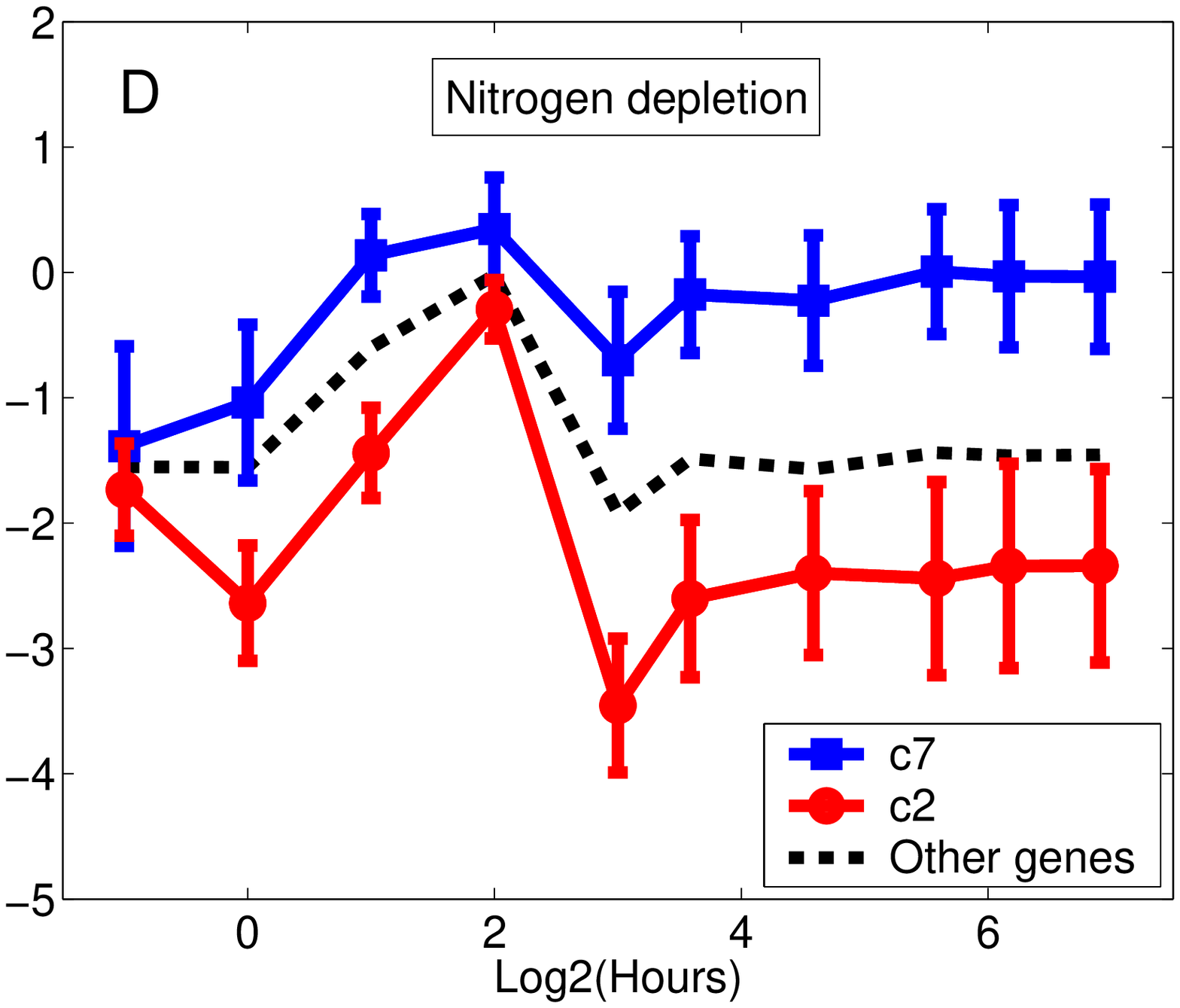,width=.31\columnwidth,height=1.8in} &
    \psfig{figure=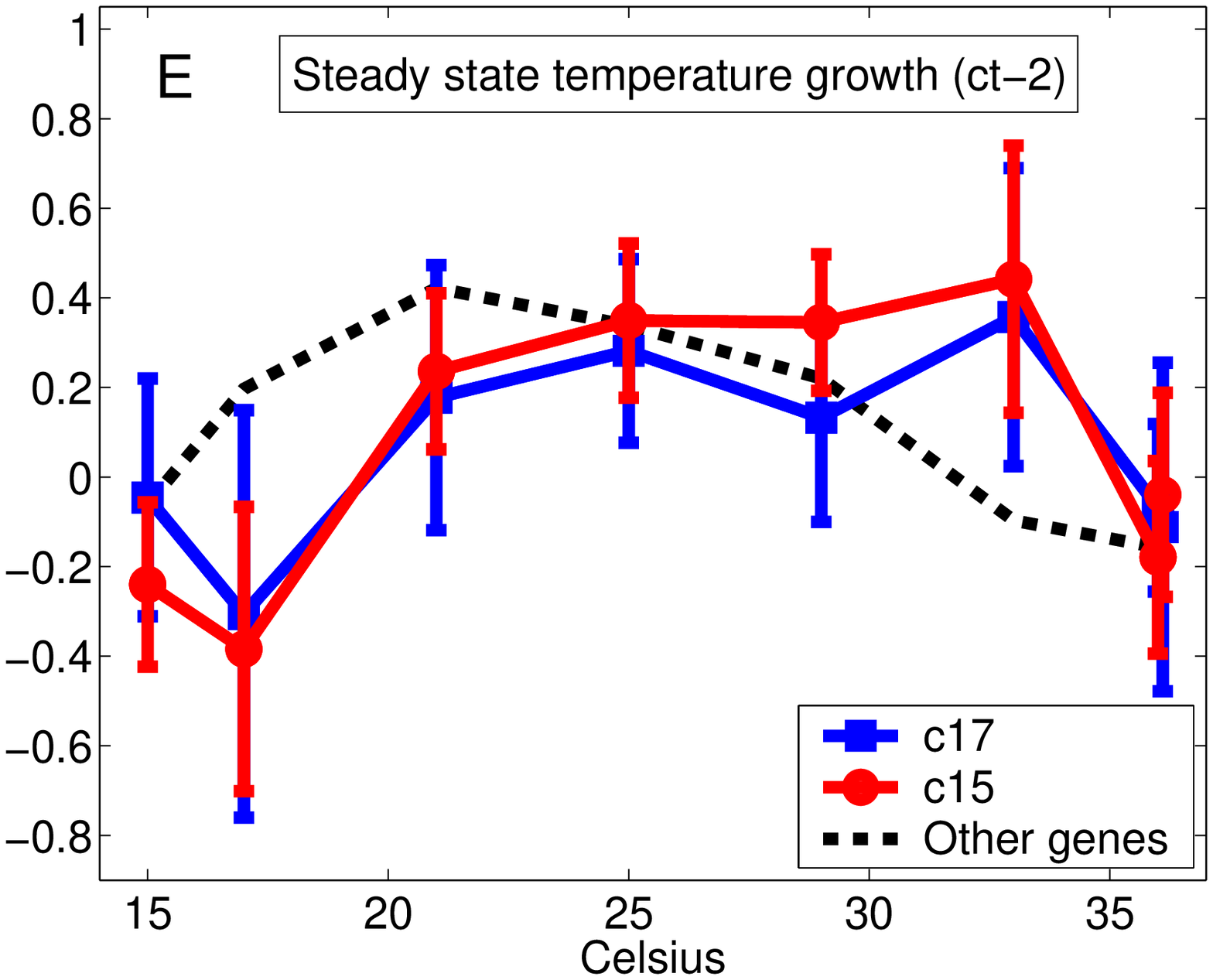,width=.31\columnwidth,height=1.8in} &
    \psfig{figure=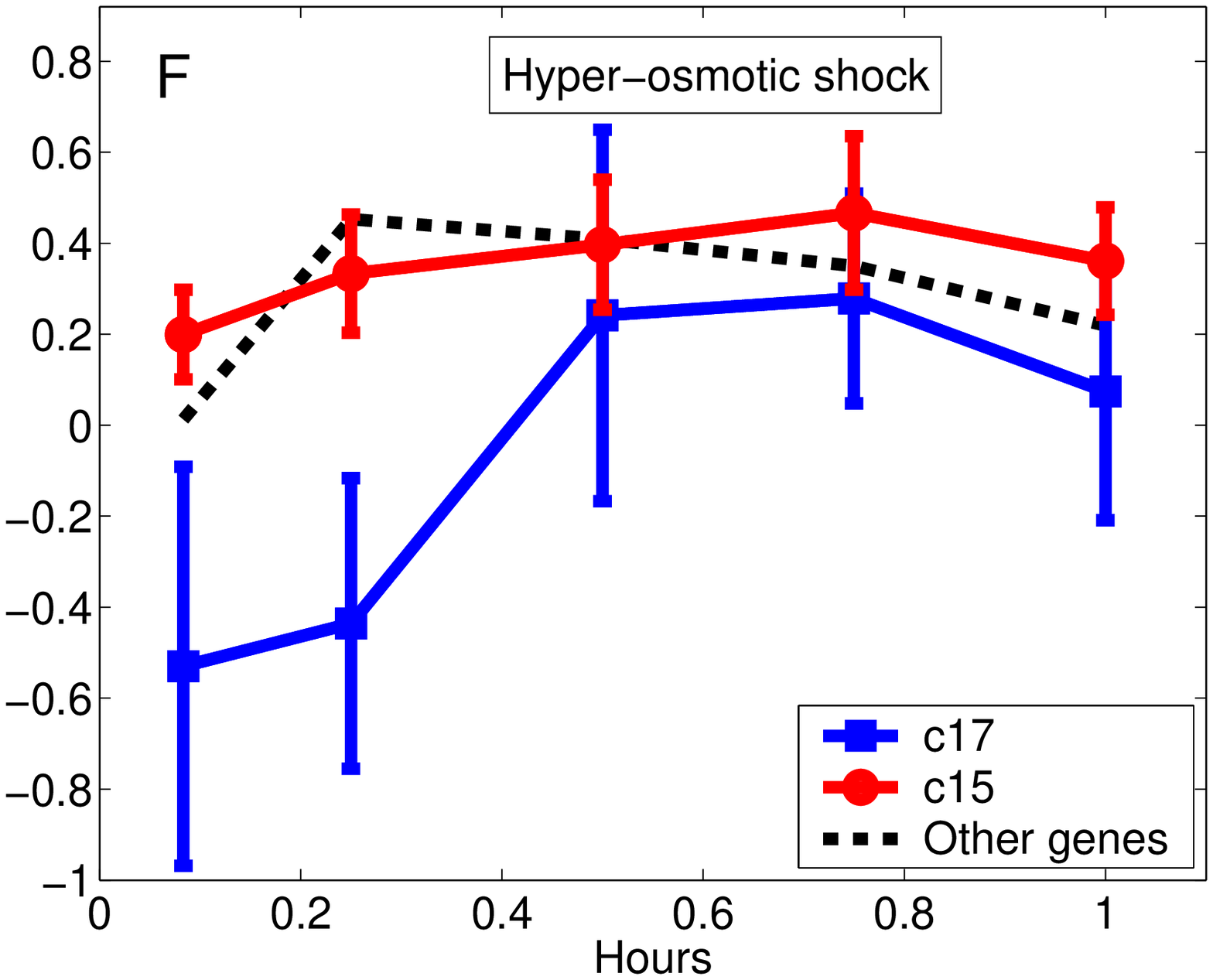,width=.31\columnwidth,height=1.8in} 
\end{tabular}
\end{center}
\caption{Examples of the average behavior of some of the clusters
obtained with $N_c=20$. 
Error-bars indicate standard deviation.
The vertical axis measures the $\log_2$ of expression ratio.
The dashed (``Other genes'') curve displays the
average behavior of the repressed genes, excluding those in the clusters
that are mentioned in the figure.
In panel A the upper dashed curve corresponds to the
average behavior of the induced genes, excluding those in $c5$.
{\bf (A)} $c5$ in Menadione exposure and stationary phase.
{\bf (B)} $c8$ and $c19$ in different temperatures shifts.
{\bf (C)} $c16$ and $c18$ in Diamide treatment.
{\bf (D)} $c7$ and $c2$ in Nitrogen depletion.
{\bf (E)} $c17$ and $c15$ in steady-state growth.
{\bf (F)} $c17$ and $c15$ in hyper--osmotic shock.}
\label{Fig:ESR_Centroids}
\end{figure}

\clearpage
\begin{figure}[] 
\begin{center}
\begin{tabular}{c}
    \psfig{figure=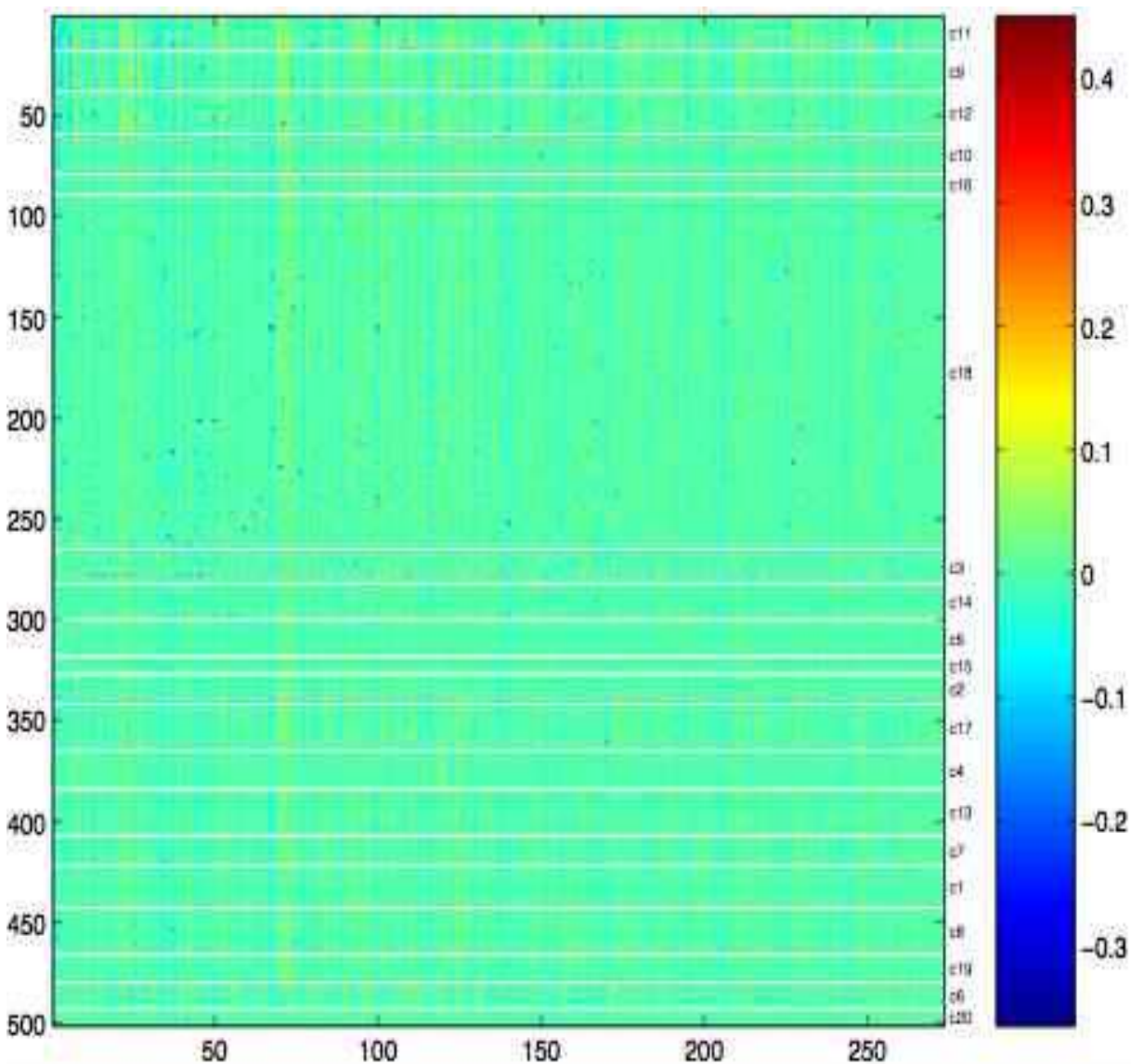,width=.9\columnwidth,height=6.0in} 
\end{tabular}
\end{center}
\caption{\small Fractional changes in stock price of the 
Standard and Poor's companies we considered during
the $273$ trading days of December 2002 -- December 2003.
The companies are sorted according to the clustering partition 
into $20$ clusters that we 
analyze in detail later on. Inside each cluster, 
companies are sorted according to the average mutual information relation
with other cluster members.}
\label{SP_data}
\end{figure}

\clearpage
\begin{figure}[] 
\begin{center}
\begin{tabular}{c}
    \psfig{figure=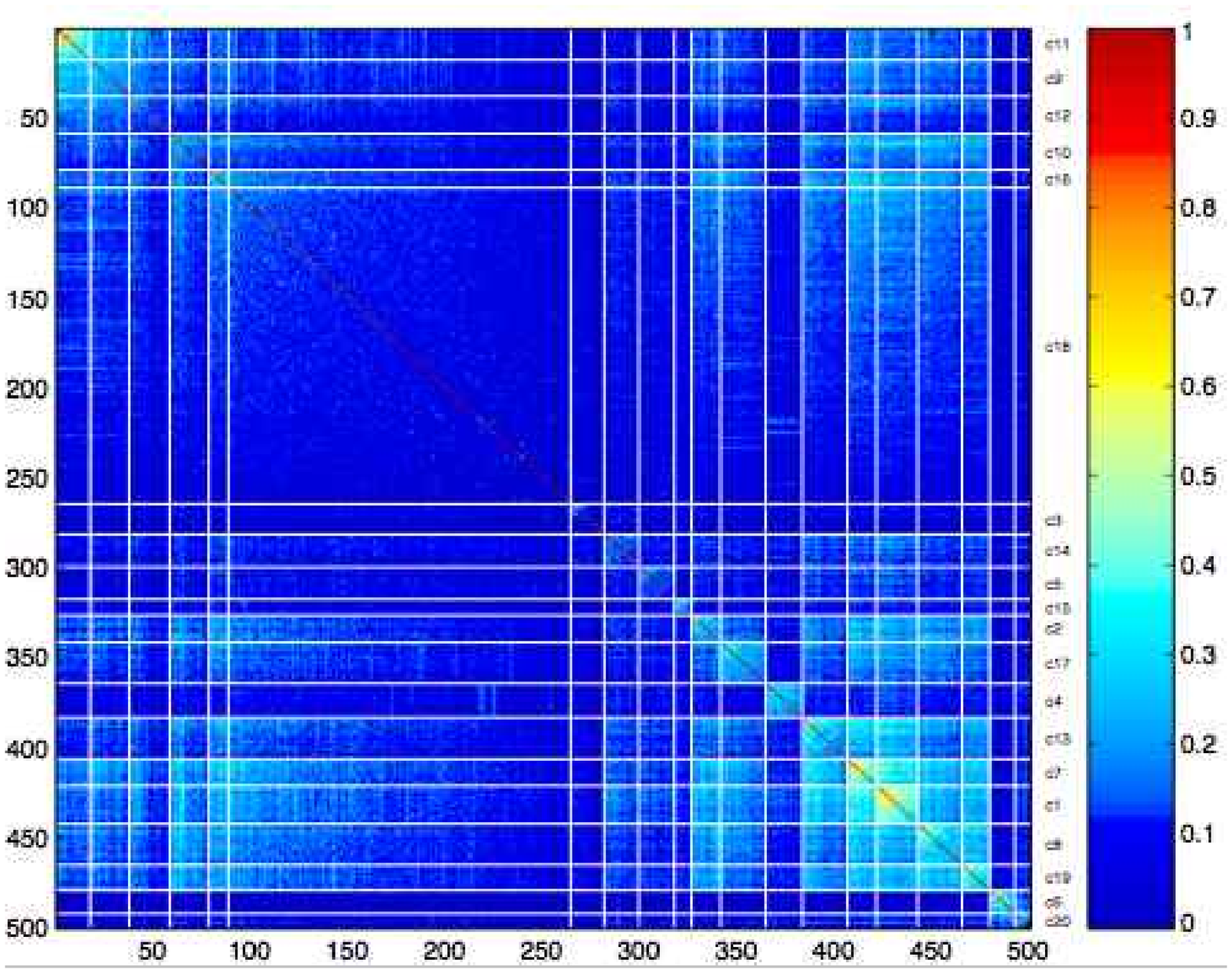,width=.9\columnwidth,height=6.0in} 
\end{tabular}
\end{center}
\caption{\small Pairwise mutual information relations for the SP500 data.
The companies are sorted according to the clustering partition 
into $20$ clusters that we analyze in detail later on. Inside each cluster, 
companies are sorted according to the average mutual information relation
with other cluster members.}
\label{SP_MIs}
\end{figure}

\clearpage
\begin{figure}[] 
\begin{center}
\begin{tabular}{c}
    \psfig{figure=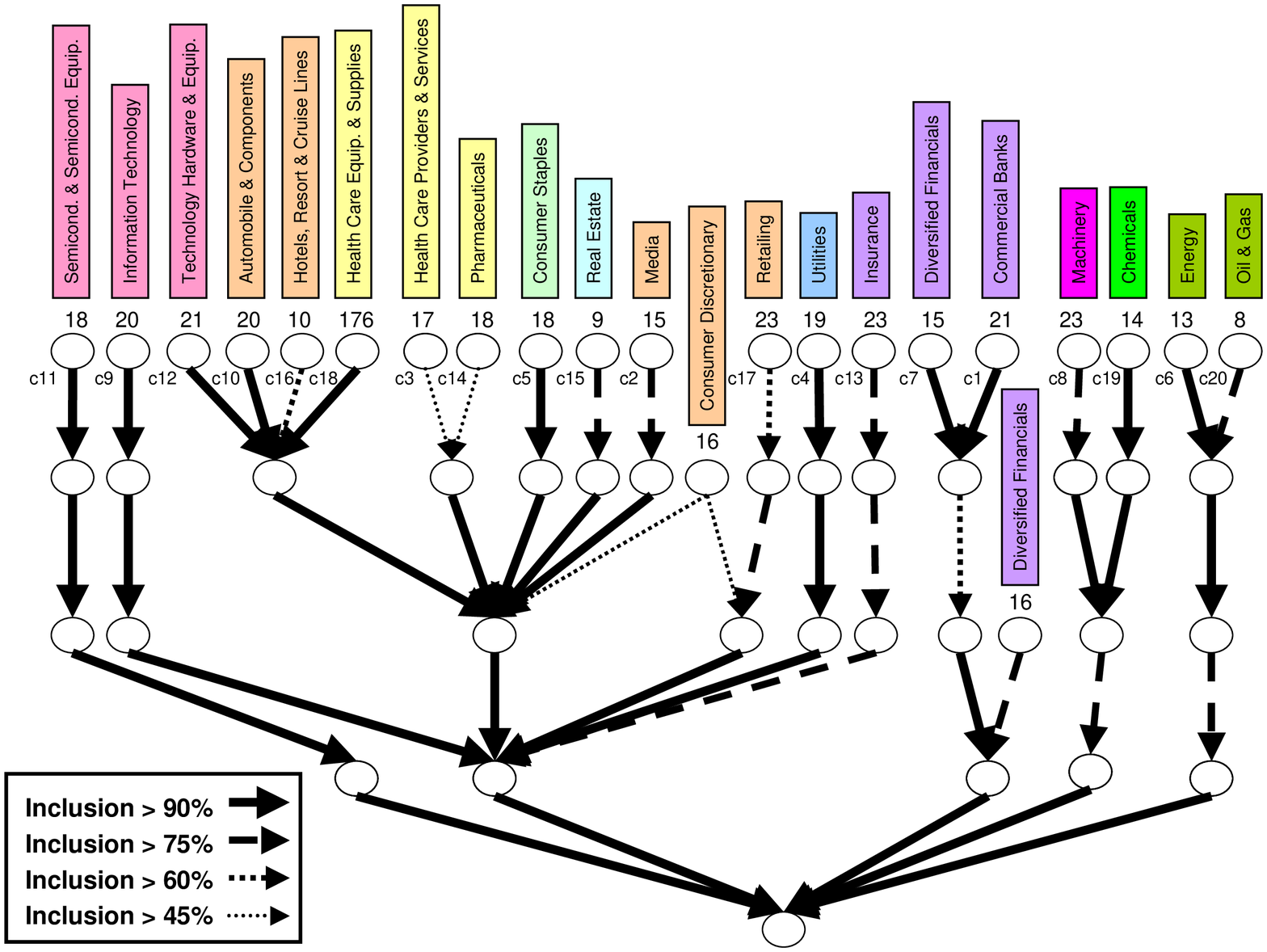,width=.9\columnwidth,height=5.5in} 
\end{tabular}
\end{center}
\caption{Relations between the optimal solutions with $N_c=\{5,10,15,20\}$
at $\frac{1}{T}=35$ for the SP500 data.
At the upper level, $N_c=20$ clusters, 
and the clusters are sorted as in \figref{SP_data} and \figref{SP_MIs}.
The numbers above every cluster indicate the number of companies in this cluster. 
The title of each cluster correspond to the most enriched annotation in the cluster, 
{\it i.e.\/}, to the annotation with the smallest $P$-value in the cluster. 
Similar color of text boxes indicate that the corresponding
annotations belong to the same major sector of economy
(see \secref{sec:SP500_annotation}).
Notice, that most clusters were enriched with more than
one annotation, hence the short titles might be too concise 
in some cases (see \secref{sec:SP500_fulldetails} for a detailed
description of every cluster).}
\label{Fig:SP500_Hierarchy}
\end{figure}

\clearpage
\begin{figure}[t] 
\begin{center}
\begin{tabular}{c}
    \psfig{figure=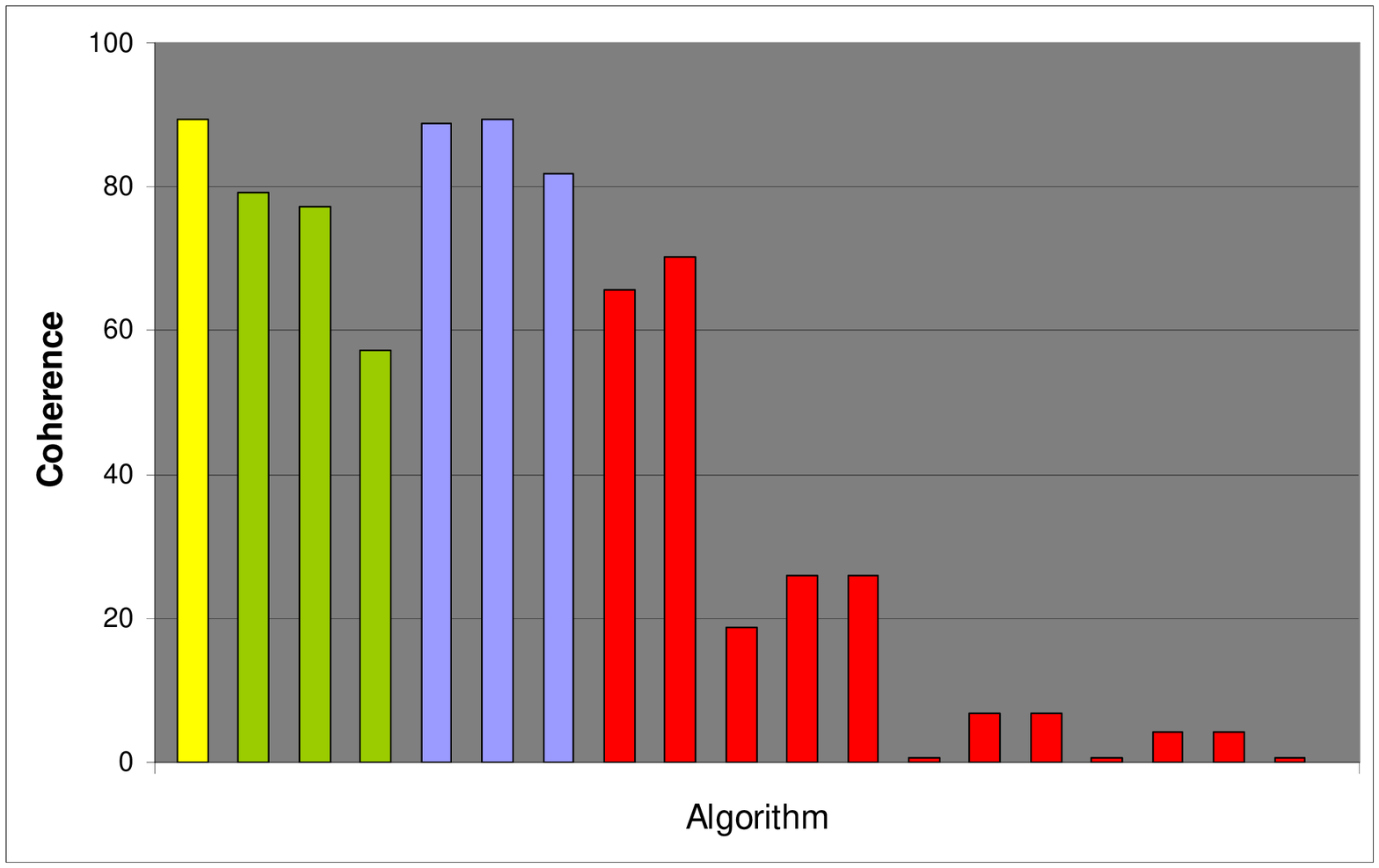,width=.9\columnwidth,height=3.0in} 
\end{tabular}
\end{center}
\caption{{\bf SP500 data}:
Comparison of average coherence results of the {\it Iclust\/} algorithm 
(yellow) with conventional clustering algorithms \cite{Cluster}:
$K$--means (green); $K$--medians (blue); Hierarchical (red).
For the hierarchical algorithms, four different variants are tried:
complete, average, centroid, and single linkage, respectively
from left to right.
For every algorithm, three different similarity measures are applied:
Pearson correlation (left); absolute value of Pearson correlation
(middle); Euclidean distance (right).
In all cases, the results are averaged over all the different numbers of clusters that 
we tried: $N_c=5,10,15,20$.}
\label{Fig:SP500_Coh}
\end{figure}

\clearpage
\begin{figure}[] 
\begin{center}
\begin{tabular}{c}
    \psfig{figure=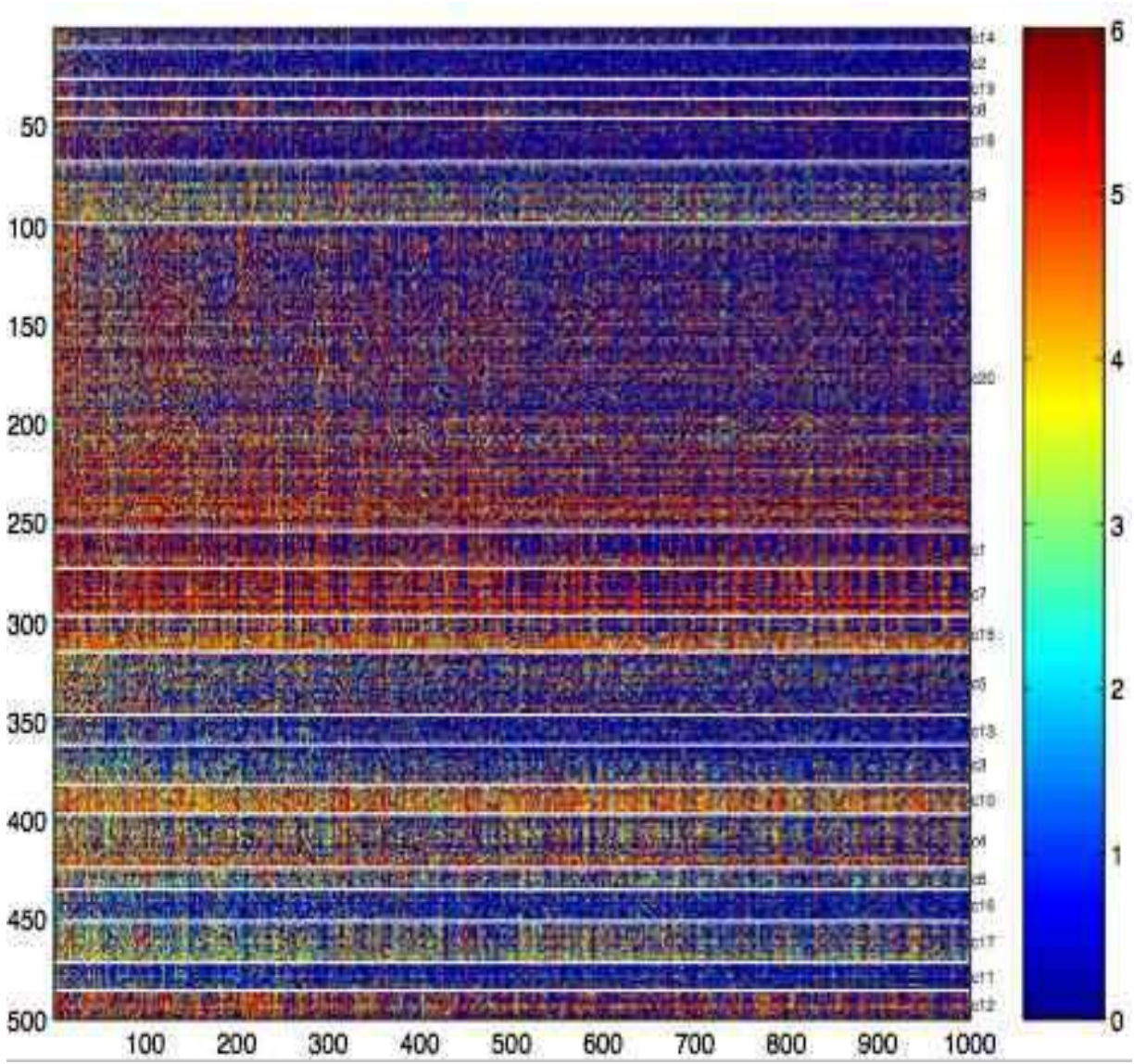,width=.9\columnwidth,height=6.0in} 
\end{tabular}
\end{center}
\caption{\small Discrete movie ratings for the $500$ movies with the maximal number of votes
in the EachMovie data. The ratings are presented only for the 
$1000$ viewers who rated the maximal number of movies.
Zeros represent missing values ({\it i.e.\/}, no vote).
The movies are sorted according to the clustering partition 
into $20$ clusters that we 
analyze in detail later on. Inside each cluster, 
movies are sorted according to the average mutual information relation
with other cluster members.}
\label{EachMovie_data}
\end{figure}

\clearpage
\begin{figure}[] 
\begin{center}
\begin{tabular}{c}
    \psfig{figure=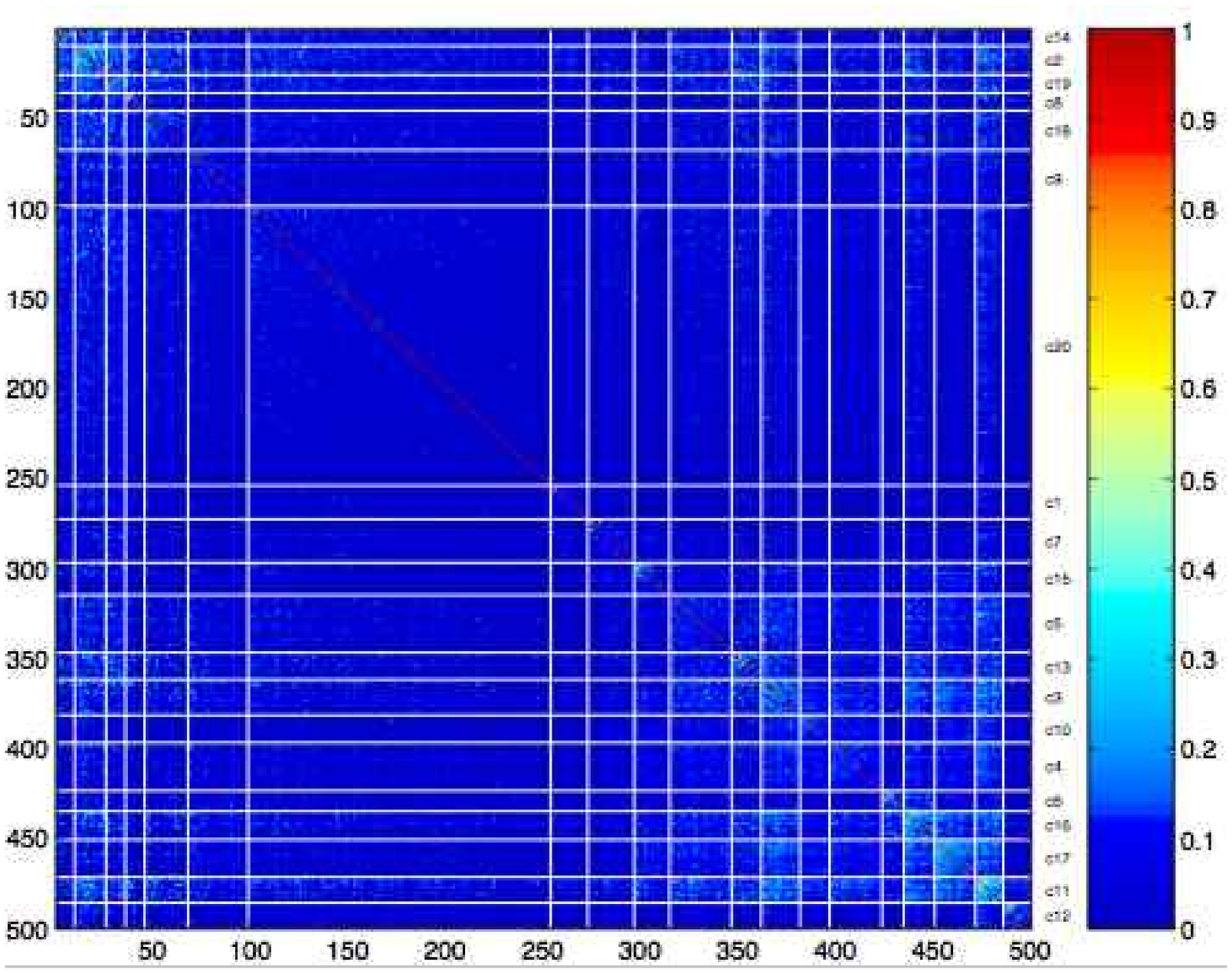,width=.9\columnwidth,height=6.0in} 
\end{tabular}
\end{center}
\caption{\small Pairwise mutual information relations for the EachMovie data. 
The movies are sorted according to the clustering partition 
into $20$ clusters that we analyze in detail later on. Inside each cluster, 
movies are sorted according to the average mutual information relation
with other cluster members.}
\label{EachMovie_MIs}
\end{figure}

\clearpage
\begin{figure}[] 
\begin{center}
\begin{tabular}{c}
    \psfig{figure=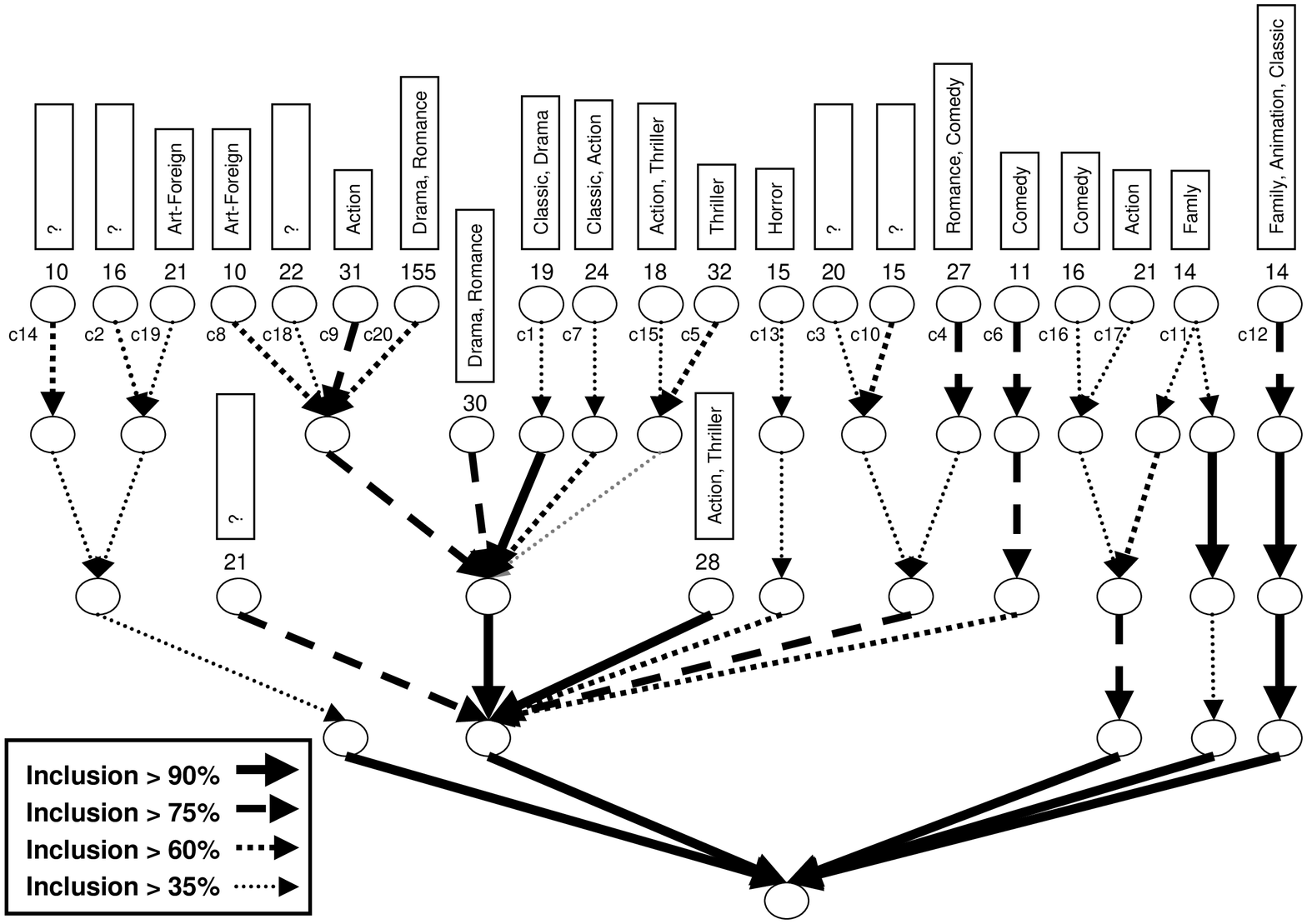,width=.9\columnwidth,height=5.5in} 
\end{tabular}
\end{center}
\caption{Relations between the optimal solutions with $N_c=\{5,10,15,20\}$
at $\frac{1}{T}=40$ for the EachMovie data.
At the upper level, $N_c=20$ clusters, 
and the clusters are sorted as in \figref{EachMovie_data} and \figref{EachMovie_MIs}.
The numbers above every cluster indicate the number of movies in this cluster. 
The title of each cluster corresponds to (all) enriched genre annotations 
in the cluster,
{\it i.e.\/}, to all annotations with a (Bonferroni corrected) $P$-value below $0.05$.
See \secref{sec:EachMovie_fulldetails} for a detailed
description of every cluster.}
\label{Fig:EachMovie_Hierarchy}
\end{figure}

\clearpage
\begin{figure}[t] 
\begin{center}
\begin{tabular}{c}
    \psfig{figure=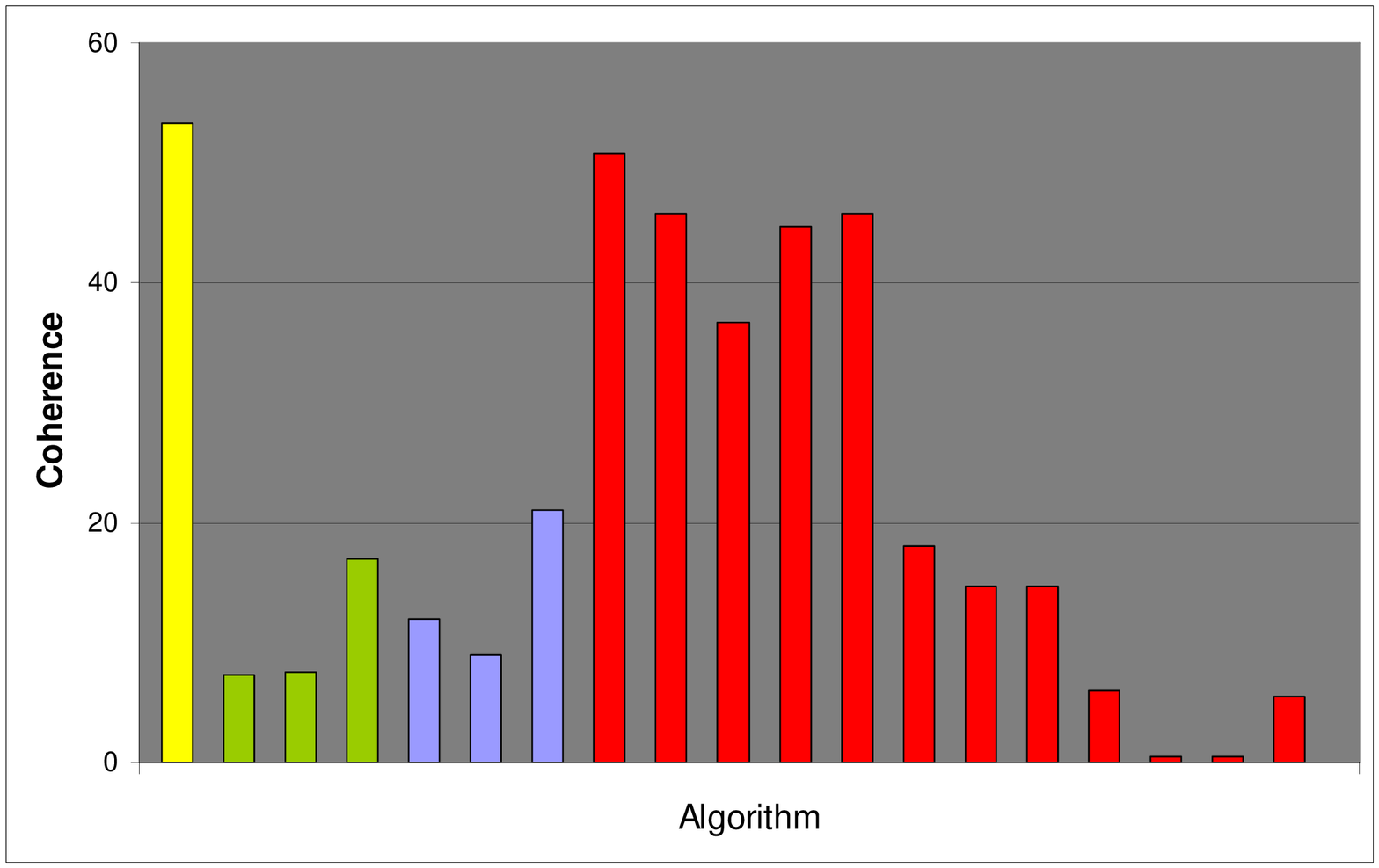,width=.9\columnwidth,height=3.0in} 
\end{tabular}
\end{center}
\caption{{\bf EachMovie data}:
Comparison of average coherence results of the {\it Iclust\/} algorithm 
(yellow) with conventional clustering algorithms \cite{Cluster}:
$K$--means (green); $K$--medians (blue); Hierarchical (red).
For the hierarchical algorithms, four different variants are tried:
complete, average, centroid, and single linkage, respectively
from left to right.
For every algorithm, three different similarity measures are applied:
Pearson correlation (left); absolute value of Pearson correlation
(middle); Euclidean distance (right).
In all cases, the results are averaged over all the different numbers of clusters that 
we tried: $N_c=5,10,15,20$.}
\label{Fig:EachMovie_Coh}
\end{figure}

\clearpage

\begin{table*}[b]
\caption{\small A simple example for an annotation matrix. 
Here, the total number of elements is $N=5$ and the total number of
distinct annotations is $R=4$. The first element is assigned
the second and third annotations, and so on.}
\label{Tbl:AnnotMatrix}
\small
\begin{center}
\begin{tabular}{|l||c|c||c|c||} \hline \hline 
{\bf Element index} & $a_1$ & $a_2$ & $a_3$ & $a_4$\\ \hline
$element_1$ & $0$ & $1$ & $1$ & $0$\\ \hline
$element_2$ & $1$ & $0$ & $1$ & $1$\\ \hline
$element_3$ & $1$ & $0$ & $0$ & $0$\\ \hline
$element_4$ & $0$ & $1$ & $1$ & $1$\\ \hline
$element_5$ & $1$ & $0$ & $1$ & $1$\\ \hline \hline
\end{tabular} 
\end{center}
\end{table*} 

\begin{table*}%[t]
\caption{\small Examples of $P$-values. When the annotation is
over-abundant in the cluster (with respect to its frequency 
in the entire population) the $P$-value is reduced accordingly.}
\label{Tbl:PvalDefinitions}
\small
\begin{center}
\begin{tabular}{|c|c|c|c|c||} \hline \hline 
$N$ & $K$ & $n$ & $x$ & $Pval$\\
(Population size) & (Annot. freq.) & (Cluster size) & (Annot. freq. in cluster) & \\ \hline
$1000$ & $100$ & $50$ & $5$ & $0.57$\\ \hline
$1000$ & $100$ & $50$ & $20$ & $10^{-8}$\\ \hline
$1000$ & $20$ &  $100$ & $2$ & $0.61$\\ \hline
$1000$ & $20$ &  $100$ & $20$ & $10^{-21}$\\ \hline
\end{tabular} 
\end{center}
\end{table*} 

\begin{table*}%[t]
\caption{\small A small subset of the $\bA_{BP}$ annotation matrix,
constructed for the ESR data out of the $GO_{BP}$ ontology.}
\label{Tbl:AnnotMatrixBP}
\small
\begin{center}
\begin{tabular}{|l||c|c||c|c|c||} \hline \hline 
{\bf ORF} & {\bf Metabolism} & {\bf Transcription} & {\bf RNA processing} & {\bf Ribosome biogenesis} & {\bf ...}\\ \hline
{\it YKL144C\/} & $1$ & $1$ & $0$ & $0$ & ...\\ \hline
{\it YML060W\/} & $1$ & $0$ & $0$ & $0$ & ...\\ \hline
{\it YGR251W\/} & $1$ & $1$ & $1$ & $1$ & ... \\ \hline
{\it YLL036C\/} & $1$ & $0$ & $1$ & $0$ & ...\\ \hline
{\it YNL163C\/} & $0$ & $0$ & $0$ & $1$ & ...\\ \hline 
{\it ...\/}   & ... & ... & ... & ... & ... \\ \hline\hline
\end{tabular} 
\end{center}
\end{table*} 

\clearpage
\begin{table*}%[t]
\caption{\small An example for a subset of genes from a single cluster
that are assigned different specific $GO_{BP}$ terms.
The functional relationship between these genes becomes 
statistically significant only if one considers the fact
that all these annotations have a common ancestor in the $GO_{BP}$ database,
the {\it tRNA aminoacylation for protein translation\/} term.}
\label{Tbl:c15example}
\small
\begin{center}
\begin{tabular}{|l|l||} \hline \hline 
{\bf ORF} & {\bf Direct $GO_{BP}$ annotation}\\ \hline\hline
{\it YDR037W\/} & lysyl-tRNA aminoacylation\\ \hline
{\it YGR094W\/} & valyl-tRNA aminoacylation\\ \hline
{\it YLR060W\/} & phenylalanyl-tRNA aminoacylation\\ \hline
{\it YNEuclidean47W\/} & cysteinyl-tRNA aminoacylation\\ \hline
{\it YPL160W\/} & leucyl-tRNA aminoacylation\\ \hline\hline
\end{tabular} 
\end{center}
\end{table*}

\begin{table*}[]
\caption{\small Details of the different annotation matrices
used for evaluating the statistical significance of the obtained
clusters for the yeast ESR genes.
{\small $^a$}Data source for constructing the annotation matrix.
{\small $^b$}Number of distinct annotations in the annotation matrix,
assigned at least two genes and thus participate in the analysis.
{\small $^c$}Number of genes assigned at least one annotation and
thus participate in the analysis. Notice that this number determines
the population size ($N$) for the $P$-value estimation. 
{\small $^d$}Average number of distinct annotations per gene. 
{\small $^e$}Maximal number of distinct annotations for a single gene.}
\label{Tbl:ESR_AnnotDetails}
\small
\begin{center}
\begin{tabular}{|l|c|c|c|c||} \hline \hline 
{\bf Data source}{\small $^a$} & {\bf $\#$ Annotations}{\small $^b$} & {\bf $\#$ Genes}{\small $^c$} & {\bf Avg. $\#$ Annot. per gene}{\small $^d$} & {\bf Maximal $\#$ Annot. per gene}{\small $^e$}\\ \hline\hline
$GO_{BP}$ \cite{GO} & $472$ & $614$ & $11.4$ & $63$ \\ \hline 
$GO_{MF}$ \cite{GO} & $215$ & $561$ & $4.6$ & $18$\\ \hline 
$GO_{CC}$ \cite{GO} & $94$ & $747$  & $5.4$ & $14$\\ \hline \hline
\end{tabular} 
\end{center}
\end{table*}

\clearpage
\begin{table*}%[t]
\caption{\small Coherence results for the ESR data
with respect to the three Gene Ontologies with $N_c=20$ clusters.
{\small $^a$}Clustering algorithm. 
In the $\langle$ $K$--means $\rangle$ row we present the average results
of all the six $K$--means variants. 
For each of these variants we performed $100$ runs from which
the best solution is chosen.
In the $\langle$ Hier. $\rangle$ row we present the average results
of all the $12$ Hierarchical clustering variants. 
In parenthesis we present the results
where the input are the $log_2$ of the expression ratio profiles.
{\small $^b$}Correlation measure used by the algorithm. 
{\it PC\/} stands for the (centered) Pearson Correlation. 
{\it $|$PC$|$\/} is the absolute value of this correlation. 
{\it Euclidean\/} stands for the Euclidean distance. 
{\small $^c$}Number of clusters with a positive coherence with respect to the 
$GO_{BP}$ ontology.
{\small $^d$}Average coherence of all $20$ clusters with respect to the $GO_{BP}$ ontology.
{\small $^e$}Number of clusters with a positive coherence with respect to the 
$GO_{MF}$ ontology.
{\small $^f$}Average coherence of all $20$ clusters with respect to the $GO_{MF}$ ontology.
{\small $^g$}Number of clusters with a positive coherence with respect to the 
$GO_{CC}$ ontology.
{\small $^h$}Average coherence of all $20$ clusters with respect to the $GO_{CC}$ ontology.}
\label{Tbl:ESR_Nc20}
\small
\begin{center}
\begin{tabular}{|l|c|c|c|c|c|c|c|} \hline \hline 
& & & & & & & \\
$N_c=20$ & & {\bf BP} & {\bf BP} & {\bf MF} & {\bf MF} & {\bf CC} & {\bf CC}\\
& & & & & & & \\ \hline
& & & & & & & \\
{\bf Algorithm}{\small $\;^a$} & 
{\bf Similarity}{\small $\;^b$} & 
{\bf $N^{pos}_c$}{\small $\;^c$} & 
{\bf $\langle Coh \rangle$}{\small $\;^d$} & 
{\bf $N^{pos}_c$}{\small $\;^e$} & 
{\bf $\langle Coh \rangle$}{\small $\;^f$} & 
{\bf $N^{pos}_c$}{\small $\;^g$} & 
{\bf $\langle Coh \rangle$}{\small $\;^h$}\\ 
& & & & & & & \\ \hline\hline
%%%%%%%%%%%%%%%%%%%%%%%%%%%%%%%%%%%%%%%%%%%%%%%%%%%%%%%%%%%%%%%%%%%%%%%%%
Iclust                & mutual information       & 17 & 51 & 16 & 41 & 14 & 33\\ \hline\hline
%%%%%%%%%%%%%%%%%%%%%%%%%%%%%%%%%%%%%%%%%%%%%%%%%%%%%%%%%%%%%%%%%%%%%%%%%
$K$--means	 & PC	 & 11 (13) & 30 (43) & 11 (11) & 31 (31) & 10 (12) & 19 (30)\\ \hline
$K$--means	 & $|$PC$|$	 & 9 (15) & 27 (50) & 8 (14) & 24 (40) & 8 (16) & 26 (42)\\ \hline
$K$--means	 & Euclidean	 & 7 (15) & 23 (52) & 9 (15) & 26 (39) & 5 (16) & 13 (51)\\ \hline
$K$--medians	 & PC	 & 11 (15) & 35 (51) & 13 (16) & 34 (48) & 10 (15) & 35 (46)\\ \hline
$K$--medians	 & $|$PC$|$	 & 12 (15) & 38 (41) & 16 (16) & 43 (39) & 13 (11) & 37 (35)\\ \hline
$K$--medians	 & Euclidean	 & 16 (18) & 49 (52) & 15 (14) & 39 (44) & 13 (16) & 43 (51)\\ \hline\hline
$\langle\;$ $K$--means $\;\rangle$ & & 11.0 (15.2) & 33.7 (48.2) & 12.0 (14.3) & 32.8 (40.2) & 9.8 (14.3) & 28.8 (42.5)\\ \hline\hline
Hier - Comp. linkage	 & PC	 & 9 (13) & 29 (41) & 10 (10) & 25 (30) & 7 (12) & 19 (34)\\ \hline
Hier - Comp. linkage	 & $|$PC$|$	 & 9 (10) & 25 (26) & 12 (9) & 31 (27) & 7 (10) & 17 (26)\\ \hline
Hier - Comp. linkage	 & Euclidean	 & 1 (13) & 2 (43) & 3 (11) & 8 (32) & 1 (8) & 2 (27)\\ \hline
Hier - Avg. linkage	 & PC	 & 5 (7) & 17 (20) & 5 (5) & 18 (17) & 4 (4) & 11 (12)\\ \hline
Hier - Avg. linkage	 & $|$PC$|$	 & 5 (4) & 17 (10) & 5 (2) & 18 (8) & 4 (2) & 10 (4)\\ \hline
Hier - Avg. linkage	 & Euclidean	 & 1 (9) & 2 (29) & 1 (4) & 1 (17) & 2 (6) & 6 (16)\\ \hline
Hier - Centr. linkage	 & PC	 & 4 (3) & 12 (10) & 4 (3) & 12 (10) & 4 (2) & 11 (8)\\ \hline
Hier - Centr. linkage	 & $|$PC$|$	 & 4 (4) & 12 (12) & 3 (4) & 7 (11) & 4 (2) & 9 (4)\\ \hline
Hier - Centr. linkage	 & Euclidean	 & 0 (4) & 0 (13) & 0 (4) & 0 (12) & 1 (2) & 1 (8)\\ \hline
Hier - Sing. linkage	 & PC	 & 2 (2) & 8 (8) & 2 (2) & 7 (7) & 2 (2) & 8 (8)\\ \hline
Hier - Sing. linkage	 & $|$PC$|$	 & 2 (0) & 6 (0) & 1 (0) & 5 (0) & 0 (0) & 0 (0)\\ \hline
Hier - Sing. linkage	 & Euclidean	 & 0 (0) & 0 (0) & 0 (0) & 0 (0) & 0 (0) & 0 (0)\\ \hline\hline
$\langle\;$ Hier. $\;\rangle$ & & 3.5 (5.8) & 10.8 (17.7) & 3.8 (4.5) & 11.0 (14.2) & 3.0 (4.2) & 7.8 (12.2)\\ \hline\hline
\end{tabular} 
\end{center}
\end{table*} 

\clearpage
\begin{table*}%[t]
\caption{\small Coherence results for the ESR data
with respect to the three Gene Ontologies with $N_c=15$ clusters.
The column and row definitions are as in \tblref{Tbl:ESR_Nc20}.
Again, in parenthesis we present the results
where the input are the $log_2$ of the expression ratio profiles.}
\label{Tbl:ESR_Nc15}
\small
\begin{center}
\begin{tabular}{|l|c|c|c|c|c|c|c|} \hline \hline 
& & & & & & & \\
$N_c=15$ & & {\bf BP} & {\bf BP} & {\bf MF} & {\bf MF} & {\bf CC} & {\bf CC}\\
& & & & & & & \\ \hline
& & & & & & & \\
{\bf Algorithm}{\small $\;^a$} & 
{\bf Similarity}{\small $\;^b$} & 
{\bf $N^{pos}_c$}{\small $\;^c$} & 
{\bf $\langle Coh \rangle$}{\small $\;^d$} & 
{\bf $N^{pos}_c$}{\small $\;^e$} & 
{\bf $\langle Coh \rangle$}{\small $\;^f$} & 
{\bf $N^{pos}_c$}{\small $\;^g$} & 
{\bf $\langle Coh \rangle$}{\small $\;^h$}\\ 
& & & & & & & \\ \hline\hline
Iclust                   & mutual information       & 12 & 51 & 14 & 54 & 14 & 52\\ \hline\hline
$K$--means	 & PC	 & 7 (14) & 29 (55) & 8 (13) & 32 (49) & 7 (10) & 18 (38)\\ \hline
$K$--means	 & $|$PC$|$	 & 10 (14) & 40 (47) & 9 (11) & 32 (37) & 8 (12) & 27 (38)\\ \hline
$K$--means	 & Euclidean	 & 10 (12) & 33 (50) & 8 (13) & 36 (46) & 3 (11) & 14 (44)\\ \hline
$K$--medians	 & PC	 & 11 (13) & 40 (46) & 11 (13) & 41 (49) & 10 (14) & 41 (47)\\ \hline
$K$--medians	 & $|$PC$|$	 & 11 (14) & 42 (50) & 11 (13) & 31 (44) & 10 (11) & 35 (38)\\ \hline
$K$--medians	 & Euclidean	 & 11 (14) & 50 (58) & 12 (13) & 42 (43) & 11 (13) & 46 (61)\\ \hline\hline
$\langle\;$ $K$--means $\;\rangle$ & & 10.0 (13.5) & 39.0 (51.0) & 9.8 (12.7) & 35.7 (44.7) & 8.2 (11.8) & 30.2 (44.3)\\ \hline\hline
Hier - Comp. linkage	 & PC	 & 8 (11) & 32 (43) & 9 (8) & 31 (31) & 6 (9) & 20 (44)\\ \hline
Hier - Comp. linkage	 & $|$PC$|$	 & 4 (8) & 17 (29) & 7 (7) & 21 (29) & 5 (8) & 15 (32)\\ \hline
Hier - Comp. linkage	 & Euclidean	 & 0 (8) & 0 (33) & 1 (8) & 2 (29) & 1 (6) & 2 (27)\\ \hline
Hier - Avg. linkage	 & PC	 & 5 (5) & 21 (22) & 4 (5) & 18 (21) & 4 (3) & 13 (12)\\ \hline
Hier - Avg. linkage	 & $|$PC$|$	 & 4 (4) & 15 (13) & 3 (3) & 11 (12) & 3 (2) & 5 (5)\\ \hline
Hier - Avg. linkage	 & Euclidean	 & 2 (7) & 8 (36) & 1 (4) & 1 (22) & 2 (4) & 8 (14)\\ \hline
Hier - Centr. linkage	 & PC	 & 4 (3) & 16 (13) & 4 (3) & 16 (15) & 4 (3) & 14 (12)\\ \hline
Hier - Centr. linkage	 & $|$PC$|$	 & 4 (3) & 16 (11) & 3 (3) & 7 (11) & 4 (3) & 11 (6)\\ \hline
Hier - Centr. linkage	 & Euclidean	 & 0 (3) & 0 (15) & 0 (3) & 0 (11) & 0 (2) & 0 (11)\\ \hline
Hier - Sing. linkage	 & PC	 & 2 (2) & 11 (11) & 2 (2) & 9 (9) & 2 (2) & 11 (11)\\ \hline
Hier - Sing. linkage	 & $|$PC$|$	 & 0 (0) & 0 (0) & 0 (0) & 0 (0) & 0 (0) & 0 (0)\\ \hline
Hier - Sing. linkage	 & Euclidean	 & 0 (0) & 0 (0) & 0 (0) & 0 (0) & 1 (0) & 6 (0)\\ \hline\hline
$\langle\;$ Hier. $\;\rangle$ & & 2.8 (4.5) & 11.3 (18.8) & 2.8 (3.8) & 9.7 (15.8) & 2.7 (3.5) & 8.8 (14.5)\\ \hline\hline
\end{tabular} 
\end{center}
\end{table*} 

\clearpage
\begin{table*}%[t]
\caption{\small Coherence results for the ESR data
with respect to the three Gene Ontologies with $N_c=10$ clusters.
The column and row definitions are as in \tblref{Tbl:ESR_Nc20}.
Again, in parenthesis we present the results
where the input are the $log_2$ of the expression ratio profiles.}
\label{Tbl:ESR_Nc10}
\small
\begin{center}
\begin{tabular}{|l|c|c|c|c|c|c|c|} \hline \hline 
& & & & & & & \\
$N_c=10$ & & {\bf BP} & {\bf BP} & {\bf MF} & {\bf MF} & {\bf CC} & {\bf CC}\\
& & & & & & & \\ \hline
& & & & & & & \\
{\bf Algorithm}{\small $\;^a$} & 
{\bf Similarity}{\small $\;^b$} & 
{\bf $N^{pos}_c$}{\small $\;^c$} & 
{\bf $\langle Coh \rangle$}{\small $\;^d$} & 
{\bf $N^{pos}_c$}{\small $\;^e$} & 
{\bf $\langle Coh \rangle$}{\small $\;^f$} & 
{\bf $N^{pos}_c$}{\small $\;^g$} & 
{\bf $\langle Coh \rangle$}{\small $\;^h$}\\ 
& & & & & & & \\ \hline\hline
Iclust                   & mutual information       & 7  & 50 & 7  & 43 & 9  & 53\\ \hline\hline
$K$--means	 & PC	 & 8 (9) & 45 (52) & 8 (8) & 44 (47) & 7 (9) & 37 (56)\\ \hline
$K$--means	 & $|$PC$|$	 & 7 (9) & 41 (48) & 6 (8) & 42 (41) & 8 (8) & 39 (48)\\ \hline
$K$--means	 & Euclidean	 & 5 (10) & 27 (62) & 6 (10) & 30 (57) & 3 (8) & 22 (55)\\ \hline
$K$--medians	 & PC	 & 9 (10) & 51 (57) & 8 (9) & 45 (53) & 9 (10) & 49 (54)\\ \hline
$K$--medians	 & $|$PC$|$	 & 7 (9) & 45 (52) & 8 (9) & 50 (47) & 9 (8) & 41 (56)\\ \hline
$K$--medians	 & Euclidean	 & 7 (9) & 48 (62) & 8 (9) & 46 (60) & 7 (9) & 49 (58)\\ \hline\hline
$\langle\;$ $K$--means $\;\rangle$ & & 7.2 (9.3) & 42.8 (55.5) & 7.3 (8.8) & 42.8 (50.8) & 7.2 (8.7) & 39.5 (54.5)\\ \hline\hline
Hier - Comp. linkage	 & PC	 & 6 (8) & 33 (44) & 7 (5) & 43 (32) & 5 (7) & 26 (43)\\ \hline
Hier - Comp. linkage	 & $|$PC$|$	 & 4 (6) & 24 (33) & 6 (5) & 32 (37) & 5 (6) & 22 (30)\\ \hline
Hier - Comp. linkage	 & Euclidean	 & 2 (7) & 12 (41) & 2 (7) & 8 (39) & 2 (5) & 7 (32)\\ \hline
Hier - Avg. linkage	 & PC	 & 3 (4) & 19 (30) & 3 (4) & 20 (30) & 3 (3) & 18 (19)\\ \hline
Hier - Avg. linkage	 & $|$PC$|$	 & 2 (4) & 8 (20) & 1 (3) & 7 (19) & 1 (2) & 1 (7)\\ \hline
Hier - Avg. linkage	 & Euclidean	 & 0 (4) & 0 (33) & 0 (5) & 0 (28) & 0 (4) & 0 (20)\\ \hline
Hier - Centr. linkage	 & PC	 & 3 (3) & 19 (19) & 3 (3) & 21 (20) & 3 (3) & 18 (18)\\ \hline
Hier - Centr. linkage	 & $|$PC$|$	 & 4 (3) & 19 (21) & 3 (3) & 11 (17) & 4 (3) & 16 (9)\\ \hline
Hier - Centr. linkage	 & Euclidean	 & 0 (3) & 0 (21) & 1 (3) & 2 (17) & 0 (3) & 0 (9)\\ \hline
Hier - Sing. linkage	 & PC	 & 2 (2) & 16 (16) & 2 (2) & 13 (14) & 2 (2) & 17 (17)\\ \hline
Hier - Sing. linkage	 & $|$PC$|$	 & 0 (0) & 0 (0) & 0 (0) & 0 (0) & 0 (0) & 0 (0)\\ \hline
Hier - Sing. linkage	 & Euclidean	 & 0 (0) & 0 (0) & 0 (0) & 0 (0) & 0 (0) & 0 (0)\\ \hline\hline
$\langle\;$ Hier. $\;\rangle$ & & 2.2 (3.7) & 12.5 (23.2) & 2.3 (3.3) & 13.1 (21.1) & 2.1 (3.2) & 10.4 (17.0)\\ \hline\hline
\end{tabular} 
\end{center}
\end{table*}

\clearpage
\begin{table*}%[t]
\caption{\small Coherence results for the ESR data
with respect to the three Gene Ontologies with $N_c=5$ clusters.
The column and row definitions are as in \tblref{Tbl:ESR_Nc20}.
Again, in parenthesis we present the results
where the input are the $log_2$ of the expression ratio profiles.}
\label{Tbl:ESR_Nc5}
\small
\begin{center}
\begin{tabular}{|l|c|c|c|c|c|c|c|} \hline \hline 
& & & & & & & \\
$N_c=5$ & & {\bf BP} & {\bf BP} & {\bf MF} & {\bf MF} & {\bf CC} & {\bf CC}\\
& & & & & & & \\ \hline
& & & & & & & \\
{\bf Algorithm}{\small $\;^a$} & 
{\bf Similarity}{\small $\;^b$} & 
{\bf $N^{pos}_c$}{\small $\;^c$} & 
{\bf $\langle Coh \rangle$}{\small $\;^d$} & 
{\bf $N^{pos}_c$}{\small $\;^e$} & 
{\bf $\langle Coh \rangle$}{\small $\;^f$} & 
{\bf $N^{pos}_c$}{\small $\;^g$} & 
{\bf $\langle Coh \rangle$}{\small $\;^h$}\\ 
& & & & & & & \\ \hline\hline
Iclust                & mutual information       & 5  & 75 & 5  & 77 & 5  & 86\\ \hline\hline
$K$--means	 & PC	 & 5 (5) & 62 (65) & 5 (5) & 63 (65) & 5 (5) & 75 (73)\\ \hline
$K$--means	 & $|$PC$|$	 & 5 (5) & 61 (70) & 5 (5) & 62 (67) & 5 (5) & 75 (70)\\ \hline
$K$--means	 & Euclidean	 & 3 (5) & 43 (71) & 3 (5) & 35 (56) & 3 (4) & 39 (65)\\ \hline
$K$--medians	 & PC	 & 5 (5) & 64 (62) & 5 (5) & 65 (63) & 5 (5) & 72 (72)\\ \hline
$K$--medians	 & $|$PC$|$	 & 5 (5) & 57 (59) & 5 (5) & 52 (58) & 5 (5) & 75 (69)\\ \hline
$K$--medians	 & Euclidean	 & 4 (5) & 52 (71) & 4 (5) & 59 (60) & 4 (4) & 68 (57)\\ \hline\hline
$\langle\;$ $K$--means $\;\rangle$ & & 4.5 (5.0) & 56.5 (66.3) & 4.5 (5.0) & 56.0 (61.5) & 4.5 (4.7) & 67.3 (67.7)\\ \hline\hline
Hier - Comp. linkage	 & PC	 & 4 (4) & 42 (44) & 5 (4) & 52 (46) & 4 (4) & 37 (57)\\ \hline
Hier - Comp. linkage	 & $|$PC$|$	 & 4 (4) & 47 (51) & 5 (3) & 34 (44) & 3 (4) & 30 (45)\\ \hline
Hier - Comp. linkage	 & Euclidean	 & 1 (3) & 11 (37) & 2 (4) & 13 (49) & 0 (4) & 0 (36)\\ \hline
Hier - Avg. linkage	 & PC	 & 3 (3) & 38 (39) & 3 (3) & 40 (47) & 3 (3) & 36 (37)\\ \hline
Hier - Avg. linkage	 & $|$PC$|$	 & 0 (1) & 0 (6) & 0 (1) & 0 (13) & 0 (2) & 0 (8)\\ \hline
Hier - Avg. linkage	 & Euclidean	 & 0 (2) & 0 (31) & 0 (3) & 0 (30) & 0 (2) & 0 (33)\\ \hline
Hier - Centr. linkage	 & PC	 & 3 (2) & 39 (32) & 3 (2) & 41 (27) & 3 (2) & 36 (33)\\ \hline
Hier - Centr. linkage	 & $|$PC$|$	 & 3 (1) & 21 (8) & 2 (0) & 19 (0) & 3 (1) & 13 (6)\\ \hline
Hier - Centr. linkage	 & Euclidean	 & 0 (1) & 0 (8) & 0 (0) & 0 (0) & 0 (1) & 0 (6)\\ \hline
Hier - Sing. linkage	 & PC	 & 2 (2) & 32 (32) & 2 (2) & 27 (27) & 2 (2) & 33 (33)\\ \hline
Hier - Sing. linkage	 & $|$PC$|$	 & 0 (0) & 0 (0) & 0 (0) & 0 (0) & 0 (0) & 0 (0)\\ \hline
Hier - Sing. linkage	 & Euclidean	 & 0 (0) & 0 (0) & 0 (0) & 0 (0) & 0 (0) & 0 (0)\\ \hline\hline
$\langle\;$ Hier. $\;\rangle$ & & 1.7 (1.9) & 19.2 (24.0) & 1.8 (1.8) & 18.8 (23.6) & 1.5 (2.1) & 15.4 (24.5)\\ \hline\hline
\end{tabular} 
\end{center}
\end{table*}

% Produced via CreateEnrichedAnnotTable2.m
\clearpage
\begin{table*}[]
\tiny
\caption{\small Enriched GO annotations in the {\it Iclust\/} solution with $N_c=20$ clusters.
Clusters are ordered as in \figref{ESR_data}, \figref{ESR_MIs},
and \figref{Fig:ESR_Hierarchy}.
Only annotations with a $P$-value below $0.05$ (Bonferroni corrected) are presented.
{\small $^a$}Cluster index, number of repressed genes in the cluster,
and number of induced genes in the cluster.
{\small $^b$}Cluster coherence (in percentage) in each of the three Gene Ontologies.
{\small $^c$}Enriched annotations for the $GO_{BP}$ ontology.
In parentheses: $(x/K,p)$ stands for 
the number of genes in the cluster to which this annotation is assigned,
the number of genes in the ESR module to which this annotation is assigned,
and the Bonferroni corrected $P$-value, respectively.
{\small $^d$}Enriched annotations for the $GO_{MF}$ ontology.
Parenthesis are as in the previous column.
{\small $^e$}Enriched annotations for the $GO_{CC}$ ontology.
Parenthesis are as in the previous column.}
\label{Tbl:ESR_Coh}
\begin{center}
\begin{tabular}{|l|l|l|l|l|} \hline \hline
& & & &\\ 
{\bf \normalsize C}{\small $^a$} & 
{\bf \normalsize Coh.}{\small $^b$} &  
{\bf \normalsize BP Enriched annot.}{\small $^c$} & 
{\bf \normalsize MF Enriched annot.}{\small $^d$} & 
{\bf \normalsize CC Enriched annot.}{\small $^e$}\\ 
& & & &\\ \hline
& & & &\\
c12 & BP : 18 &  nucleoside monophosphate metabolism (5/6,0.0032)  &  diphosphotransferase activity (4/4,0.0047)  &  --  \\ 
Rep : 57 & MF : 9 &  purine nucleoside monophosphate meta (5/6,0.0032)  &  --  &  --  \\ 
Ind : 3 & CC : 0 &  ribonucleoside monophosphate metabol (5/6,0.0032)  &  --  &  --  \\ 
& &  purine ribonucleoside monophosphate  (5/6,0.0032)  &  --  &  --  \\ 
& &  histidine biosynthesis (4/4,0.0076)  &  --  &  --  \\ 
& &  histidine metabolism (4/4,0.0076)  &  --  &  --  \\ 
& &  histidine family amino acid metaboli (4/4,0.0076)  &  --  &  --  \\ 
& &  histidine family amino acid biosynth (4/4,0.0076)  &  --  &  --  \\ 
& &  nucleoside monophosphate biosynthesi (4/4,0.0076)  &  --  &  --  \\ 
& &  purine nucleoside monophosphate bios (4/4,0.0076)  &  --  &  --  \\ 
& &  ribonucleoside monophosphate biosynt (4/4,0.0076)  &  --  &  --  \\ 
& &  purine ribonucleoside monophosphate  (4/4,0.0076)  &  --  &  --  \\ 
& &  biogenic amine biosynthesis (4/4,0.0076)  &  --  &  --  \\ 
& &  amine biosynthesis (7/16,0.0187)  &  --  &  --  \\ 
& &  amino acid derivative biosynthesis (4/5,0.0359)  &  --  &  --  \\ 
 & & & &\\ \hline 
 & & & &\\ 
c15 & BP : 67 &  tRNA aminoacylation for protein tran (8/14,0.00)  &  tRNA ligase activity (9/17,0.00)  &  --  \\ 
Rep : 12 & MF : 75 &  --  &  RNA ligase activity (9/17,0.00)  &  --  \\ 
Ind : 0 & CC : 0 &  --  &  ligase activity, forming carbon-oxyg (9/17,0.00)  &  --  \\ 
& &  --  &  ligase activity, forming aminoacyl-t (9/17,0.00)  &  --  \\ 
& &  --  &  ligase activity, forming phosphoric  (9/17,0.00)  &  --  \\ 
& &  --  &  ligase activity (9/23,0.00)  &  --  \\ 
 & & & &\\ \hline 
 & & & &\\ 
c14 & BP : 57 &  vacuolar acidification (3/3,0.0010)  &  hydrogen-transporting ATPase activit (3/3,0.0005)  &  chaper.-contain. T-compl. (3/3,0.0001)  \\ 
Rep : 15 & MF : 43 &  regulation of pH (3/4,0.0038)  &  ion transporter activity (4/8,0.0006)  &  hydrog.-transloc. ATPase (3/3,0.0001)  \\ 
Ind : 1 & CC : 53 &  monovalent inorganic cation homeosta (3/4,0.0038)  &  monovalent inorganic cation transpor (3/4,0.0020)  &  vacuolar membrane (3/8,0.0073)  \\ 
& &  hydrogen ion homeostasis (3/4,0.0038)  &  hydrogen ion transporter activity (3/4,0.0020)  &  cytoskeleton (3/8,0.0073)  \\ 
& &  organelle organization and biogenesi (6/38,0.0080)  &  ATPase activity, coupled to transmem (3/5,0.0048)  &  hydrog.-transp. ATPase V0 (2/2,0.0079)  \\ 
& &  vacuolar transport (3/5,0.0093)  &  cation transporter activity (3/6,0.0095)  &  membrane (5/47,0.0312)  \\ 
& &  cation homeostasis (3/7,0.0317)  &  asparagine synthase (glutamine-hydro) (2/2,0.0232)  &  --  \\ 
& &  asparagine biosynthesis (2/2,0.0488)  &  primary active transporter activity (3/9,0.0382)  &  --  \\ 
& &  aspartate family amino acid biosynth (2/2,0.0488)  &  P-P-bond-hydrolysis-driven transport (3/9,0.0382)  &  --  \\ 
& &  --  &  hydrolase activity, acting on acid a (3/9,0.0382)  &  --  \\ 
& &  --  &  ATPase activity, coupled to transmem (3/9,0.0382)  &  --  \\ 
 & & & &\\ \hline 
 & & & &\\ 
c10 & BP : 0 &  --  &  --  &  --  \\ 
Rep : 35 & MF : 0 &  --  &  --  &  --  \\ 
Ind : 1 & CC : 0 &  --  &  --  &  --  \\ 
 & & & &\\ \hline 
 & & & &\\ 
c2 & BP : 100 &  protein biosynthesis (15/189,0.00)  &  structural constituent of ribosome (12/127,0.0001)  &  ribosome (14/153,0.00)  \\ 
Rep : 17 & MF : 80 &  macromolecule biosynthesis (15/189,0.00)  &  structural molecule activity (12/128,0.0001)  &  ribonucleopro. complex (14/186,0.00)  \\ 
Ind : 0 & CC : 100 &  biosynthesis (15/236,0.00)  &  --  &  cytos. large ribos. subunit  (10/75,0.00)  \\ 
& &  translational elongation (5/10,0.00)  &  --  &  large ribosomal subunit (10/75,0.00)  \\ 
& &  protein metabolism (15/252,0.00)  &  --  &  cytos. ribos. (sens. Eukar.) (12/132,0.00)  \\ 
& &  --  &  --  &  cytosol (12/165,0.0001)  \\ 
& &  --  &  --  &  cytoplasm (16/525,0.0430)  \\ 
 & & & &\\ \hline 
 & & & &\\ 
c3 & BP : 17 &  transcription from Pol II promoter (5/11,0.0077)  &  ribonuclease activity (4/10,0.039)  &  DNA-direct. RNA polym. II-core (4/7,0.004)  \\ 
Rep : 29 & MF : 15 &  --  &  --  &  DNA-direct. RNA polym. II-holo (4/9,0.012)  \\ 
Ind : 6 & CC : 23 &  --  &  --  &  cytoplas. exosom. (RNase compl.) (3/5,0.028)  \\ 
 & & & &\\ \hline 
 & & & &\\ 
c20 & BP : 25 &  cell communication (5/18,0.0147)  &  --  &  vacuole (7/26,0.0014)  \\ 
Rep : 0 & MF : 0 &  signal transduction (4/12,0.0359)  &  --  &  storage vacuole (5/20,0.0288)  \\ 
Ind : 38 & CC : 22 &  --  &  --  &  lytic vacuole (5/20,0.0288)  \\ 
& &  --  &  --  &  vacuole (sensu Fungi) (5/20,0.0288)  \\ 
 & & & &\\ \hline 
\end{tabular}
\end{center}
\end{table*}

\clearpage
\begin{table*}[]
\tiny
\begin{center}
\begin{tabular}{|l|l|l|l|l|} \hline \hline
& & & &\\ 
{\bf \normalsize C}{\small $^a$} & 
{\bf \normalsize Coh.}{\small $^b$} &  
{\bf \normalsize BP Enriched annot.}{\small $^c$} & 
{\bf \normalsize MF Enriched annot.}{\small $^d$} & 
{\bf \normalsize CC Enriched annot.}{\small $^e$}\\ 
 & & & &\\ \hline
 & & & &\\ 
c6 & BP : 0 &  --  &  oxidoreductase activity (6/37,0.0255)  &  peroxisomal matrix (2/2,0.0353)  \\ 
Rep : 3 & MF : 33 &  --  &  --  &  --  \\ 
Ind : 36 & CC : 7 &  --  &  --  &  --  \\ 
 & & & &\\ \hline 
 & & & &\\ 
c8 & BP : 0 &  --  &  --  &  --  \\ 
Rep : 3 & MF : 0 &  --  &  --  &  --  \\ 
Ind : 22 & CC : 0 &  --  &  --  &  --  \\ 
 & & & &\\ \hline
 & & & &\\ 
c11 & BP : 63 &  carbohydrate biosynthesis (5/8,0.00)  &  protein kinase activity (5/12,0.00)  &  lipid particle (3/5,0.0029)  \\ 
Rep : 0 & MF : 58 &  gluconeogenesis (4/5,0.000)  &  phosphotransfer. activ., alcohol (5/19,0.000)  &  cAMP-depend. prot. kinase compl. (2/2,0.012)\\ 
Ind : 34 & CC : 100 &  hexose biosynthesis (4/5,0.0002)  &  kinase activity (5/27,0.0027)  &  cytoplasm (20/525,0.0139)  \\ 
& &  alcohol biosynthesis (4/5,0.0002)  &  protein serine/threonine kinase acti (3/6,0.0035)  &  --  \\ 
& &  monosaccharide biosynthesis (4/5,0.0002)  &  cyclic-nucleotide dependent protein  (2/2,0.0101)  &  --  \\ 
& &  regulation of carbohydrate metabolis (3/3,0.0015)  &  cAMP-depend prot kinase activi (2/2,0.01)  &  --  \\ 
& &  regulation of gluconeogenesis (3/3,0.0015)  &  transferase activity (7/103,0.0496)  &  --  \\ 
& &  regulation of biosynthesis (3/3,0.0015)  &  --  &  --  \\ 
& &  negative regulation of biosynthesis (3/3,0.0015)  &  --  &  --  \\ 
& &  negative regulation of metabolism (3/3,0.0015)  &  --  &  --  \\ 
& &  negative regulation of gluconeogenes (3/3,0.0015)  &  --  &  --  \\ 
& &  negative regulation of carbohydrate  (3/3,0.0015)  &  --  &  --  \\ 
& &  energy pathways (6/25,0.0015)  &  --  &  --  \\ 
& &  energy derivation by oxidation of or (6/25,0.0015)  &  --  &  --  \\ 
& &  protein amino acid phosphorylation (4/8,0.0021)  &  --  &  --  \\ 
& &  phosphorylation (4/9,0.0037)  &  --  &  --  \\ 
& &  main pathways of carbohydrate metabo (4/11,0.0093)  &  --  &  --  \\ 
& &  phosphorus metabolism (4/11,0.0093)  &  --  &  --  \\ 
& &  phosphate metabolism (4/11,0.0093)  &  --  &  --  \\ 
& &  carbohydrate metabolism (6/35,0.0120)  &  --  &  --  \\ 
& &  glucose metabolism (4/12,0.0138)  &  --  &  --  \\ 
& &  regulation of metabolism (3/6,0.0284)  &  --  &  --  \\ 
& &  hexose metabolism (4/15,0.0363)  &  --  &  --  \\ 
 & & & &\\ \hline 
 & & & &\\ 
c13 & BP : 40 &  pyruvate dehydrogenase bypass (3/3,0.0002)  &  pyruvate decarboxylase activity (2/3,0.0264)  &  --  \\ 
Rep : 16 & MF : 20 &  pyruvate metabolism (3/3,0.0002)  &  --  &  --  \\ 
Ind : 0 & CC : 0 &  fermentation (2/2,0.0189)  &  --  &  --  \\ 
& &  ethanol fermentation (2/2,0.0189)  &  --  &  --  \\ 
& &  glycolytic fermentation (2/2,0.0189)  &  --  &  --  \\ 
& &  alcohol metabolism (4/25,0.0303)  &  --  &  --  \\ 
 & & & &\\ \hline 
 & & & &\\ 
c1 & BP : 73 &  translational elongation (4/10,0.0004)  &  transl. elong. fact. activi (3/7,0.003)  &  nascent polypept.-associat. compl. (2/2,0.002)\\ 
Rep : 13 & MF : 36 &  nascent polypeptide association (2/2,0.0096)  &  translation factor activity, nucleic (4/29,0.0253)  &  --  \\ 
Ind : 0 & CC : 20 &  methionine metabolism (2/2,0.0096)  &  translation regulator activity (4/31,0.0328)  &  --  \\ 
& &  serine family amino acid metabolism (2/3,0.0286)  &  --  &  --  \\ 
 & & & &\\ \hline 
 & & & &\\ 
c17 & BP : 80 &  S phase of mitotic cell cycle (4/10,0.00)  &  $\alpha$ DNA polymer. activ. (2/2,0.003)  &  $\alpha$ DNA polymer.:primase complex (2/2,0.0010)  \\ 
Rep : 7 & MF : 33 &  DNA replication (4/10,0.00)  &  DNA-direct DNA polymer activ (2/4,0.015)  &  replication fork (2/4,0.0060)  \\ 
Ind : 3 & CC : 25 &  DNA replication and chromosome cycle (4/14,0.00)  &  --  &  --  \\ 
& &  mitotic cell cycle (4/14,0.00)  &  --  &  --  \\ 
& &  cell cycle (4/22,0.0002)  &  --  &  --  \\ 
& &  DNA metabolism (4/24,0.0003)  &  --  &  --  \\ 
& &  DNA dependent DNA replication (3/8,0.0004)  &  --  &  --  \\ 
& &  cell proliferation (4/27,0.0004)  &  --  &  --  \\ 
& &  DNA replication, priming (2/2,0.0016)  &  --  &  --  \\ 
& &  DNA replication initiation (2/3,0.0048)  &  --  &  --  \\ 
& &  lagging strand elongation (2/3,0.0048)  &  --  &  --  \\ 
& &  DNA strand elongation (2/4,0.0095)  &  --  &  --  \\ 
& &  DNA repair (2/8,0.0438)  &  --  &  --  \\ 
 & & & &\\ \hline 
 & & & &\\ 
c5 & BP : 24 &  regulation of redox homeostasis (3/5,0.0358)  &  oxidoreductase activity (9/37,0.00)  &  mitochondr. intermembr. space (3/6,0.017)  \\ 
Rep : 6 & MF : 45 &  cell redox homeostasis (3/5,0.0358)  &  --  &  --  \\ 
Ind : 26 & CC : 11 &  oxygen and reactive oxygen species m (4/12,0.0455)  &  --  &  --  \\ 
 & & & &\\ \hline 
\end{tabular}
\end{center}
\end{table*}

\clearpage
\begin{table*}[]
%\caption{\tblref{Tbl:ESR_Coh}: Continue.}
\tiny
\begin{center}
\begin{tabular}{|l|l|l|l|l|} \hline \hline
& & & &\\ 
{\bf \normalsize C}{\small $^a$} & 
{\bf \normalsize Coh.}{\small $^b$} &  
{\bf \normalsize BP Enriched annot.}{\small $^c$} & 
{\bf \normalsize MF Enriched annot.}{\small $^d$} & 
{\bf \normalsize CC Enriched annot.}{\small $^e$}\\ 
& & & &\\ \hline
& & & &\\ 
c4 & BP : 62 &  carbohydrate metabolism (11/35,0.00)  &  catalytic activity (24/280,0.0008)  &  $\alpha,\alpha$-trehalose-phosphate synt (3/3,0.005)\\ 
Rep : 1 & MF : 89 &  response to stress (10/50,0.0050)  &  hydrolase activity, acting on carbon (3/4,0.0201)  &  --  \\ 
Ind : 46 & CC : 7 &  --  &  --  &  --  \\ 
 & & & &\\ \hline 
 & & & &\\ 
c19 & BP : 45 &  response to stress (14/50,0.00)  &  heat shock protein activity (6/8,0.00)  &  --  \\ 
Rep : 0 & MF : 48 &  --  &  oxidoreductase activity (9/37,0.0031)  &  --  \\ 
Ind : 49 & CC : 0 &  --  &  chaperone activity (6/19,0.0139)  &  --  \\ 
 & & & &\\ \hline
 & & & &\\ 
c7 & BP : 89 &  nucleobase, nucleoside, nucleotide a (46/188,0.00)  &  --  &  nucleus (83/288,0.00)  \\ 
Rep : 114 & MF : 0 &  transcription, DNA-dependent (35/117,0.00)  &  --  &  nucleolus (40/118,0.00)  \\ 
Ind : 9 & CC : 78 &  transcription (35/119,0.00)  &  --  &  nucleoplasm (13/34,0.0125)  \\ 
& &  RNA metabolism (33/119,0.00)  &  --  &  DNA-direct. RNA polymer. III comp (7/13,0.03)  \\ 
& &  RNA processing (31/110,0.00)  &  --  &  --  \\ 
& &  ribosome biogenesis (30/110,0.0001)  &  --  &  --  \\ 
& &  transcription from Pol I promoter (28/102,0.0002)  &  --  &  --  \\ 
& &  rRNA processing (25/90,0.0008)  &  --  &  --  \\ 
& &  ribosome biogenesis and assembly (32/136,0.0013)  &  --  &  --  \\ 
& &  cellular process (48/263,0.0037)  &  --  &  --  \\ 
& &  cell growth and/or maintenance (47/256,0.0044)  &  --  &  --  \\ 
& &  cytoplasm organization and biogenesi (35/172,0.0142)  &  --  &  --  \\ 
& &  transcription from Pol III promoter (7/13,0.0392)  &  --  &  --  \\ 
& &  cell organization and biogenesis (37/196,0.0479)  &  --  &  --  \\ 
 & & & &\\ \hline 
& & & &\\ 
c16 & BP : 96 &  ribosome biogenesis (48/110,0.00)  &  snoRNA binding (14/25,0.00)  &  nucleolus (49/118,0.00)  \\ 
Rep : 87 & MF : 77 &  ribosome biogenesis and assembly (51/136,0.00)  &  RNA binding (21/70,0.00)  &  nucleus (68/288,0.00)  \\ 
Ind : 1 & CC : 88 &  RNA processing (43/110,0.00)  &  methyltransfer. activ. (10/20,0.000)  &  small nucleolar ribonucleoprot. co (11/27,0.001)  \\ 
& &  RNA metabolism (44/119,0.00)  &  transferase activ., transferr. o (10/20,0.000)  &  --  \\ 
& &  cytoplasm organization and biogenesi (52/172,0.00)  &  RNA helicase activity (9/17,0.0004)  &  --  \\ 
& &  transcription from Pol I promoter (40/102,0.00)  &  ATP depend. RNA helic. activ. (8/14,0.001)  &  --  \\ 
& &  rRNA processing (37/90,0.00)  &  ATP depend. helic. activ. (8/14,0.001)  &  --  \\ 
& &  transcription, DNA-dependent (41/117,0.00)  &  RNA depend. ATPase activ. (8/14,0.001)  &  --  \\ 
& &  transcription (41/119,0.00)  &  helicase activity (9/18,0.0008)  &  --  \\ 
& &  nucleobase, nucleoside, nucleotide a (51/188,0.00)  &  RNA methyltransfer. activ. (7/11,0.001)  &  --  \\ 
& &  cell organization and biogenesis (52/196,0.00)  &  nucleic acid binding (23/109,0.0041)  &  --  \\ 
& &  cellular process (55/263,0.00)  &  binding (25/129,0.0080)  &  --  \\ 
& &  cell growth and/or maintenance (54/256,0.00)  &  S-adenosylmethion.-depend. methy (6/13,0.048)  &  --  \\ 
& &  processing of 20S pre-rRNA (16/37,0.0001)  &  --  &  --  \\ 
& &  ribosomal large subunit biogenesis (9/13,0.0002)  &  --  &  --  \\ 
& &  ribosomal large subunit assembly and (10/26,0.0380)  &  --  &  --  \\ 
 & & & &\\ \hline 
 & & & &\\ 
c9 & BP : 62 &  RNA metabolism (17/119,0.0025)  &  binding (19/129,0.0006)  &  nucleolus (18/118,0.0336)  \\ 
Rep : 48 & MF : 58 &  RNA processing (16/110,0.0039)  &  nucleic acid binding (15/109,0.0251)  &  --  \\ 
Ind : 8 & CC : 33 &  nucleobase, nucleoside, nucleotide a (21/188,0.0083)  &  --  &  --  \\ 
& &  transcription, DNA-dependent (15/117,0.0369)  &  --  &  --  \\ 
& &  transcription (15/119,0.0451)  &  --  &  --  \\ 
 & & & &\\ \hline 
 & & & &\\ 
c18 & BP : 100 &  protein biosynthesis (112/189,0.00)  &  struct. constit. of riboso. (110/127,0.00)  &  cytos. ribos. (sensu Euka.) (110/132,0.00)  \\ 
Rep : 122 & MF : 98 &  macromolecule biosynthesis (112/189,0.00)  &  struct. molec. activ. (110/128,0.00)  &  ribosome (110/153,0.00)  \\ 
Ind : 0 & CC : 99 &  biosynthesis (112/236,0.00)  &  --  &  cytosol (110/165,0.00)  \\ 
& &  protein metabolism (112/252,0.00)  &  --  &  ribonucleoprotein complex (110/186,0.00)  \\ 
& &  metabolism (112/523,0.00)  &  --  &  cytosol. large ribos. subunit (62/75,0.00)  \\ 
& &  ribosomal small subunit assembly and (8/10,0.0013)  &  --  &  large ribosomal subunit (62/75,0.000)  \\ 
& &  regulation of translational fidelity (6/7,0.0082)  &  --  &  cytosol. small ribos. subunit (48/56,0.000)  \\ 
& &  regulation of translation (8/12,0.0107)  &  --  &  small ribosomal subunit (48/56,0.00)  \\ 
& &  ribosomal subunit assembly (15/36,0.0263)  &  --  &  eukaryotic 48S initiation complex (48/56,0.000)  \\ 
& &  --  &  --  &  eukaryotic 43S preinitiation complex (49/61,0.00)  \\ 
& &  --  &  --  &  cytoplasm (113/525,0.00)  \\ 
 & & & &\\ \hline 
\end{tabular}
\end{center}
\end{table*}

\clearpage
\begin{table*}%[t]
\caption{\small Coherence results for the SP500 data
with respect to the GICS companies' annotations
with $N_c=20$ clusters. 
{\small $^a$}Clustering algorithm. 
In the $\langle$ $K$--means $\rangle$ row we present the average results
of all the six $K$--means variants. 
For each of these variants we performed $100$ runs from which
the best solution is chosen.
In the $\langle$ Hier. $\rangle$ row we present the average results
of all the $12$ Hierarchical clustering variants. 
{\small $^b$}Correlation measure used by the algorithm. 
{\it PC\/} stands for the (centered) Pearson Correlation. 
{\it $|$PC$|$\/} is the absolute value of this correlation. 
{\it Euclidean\/} stands for the Euclidean distance. 
{\small $^c$}Number of clusters with a positive coherence. 
{\small $^d$}Average coherence of all $20$ clusters.}
\label{Tbl:SP_Nc20}
\small
\begin{center}
\begin{tabular}{|l|c|c|c|} \hline \hline 
$N_c=20$ & & &\\
{\bf Algorithm}{\small $\;^a$} & 
{\bf Similarity}{\small $\;^b$} & 
{\bf $N^{pos}_c$}{\small $\;^c$} & 
{\bf $\langle Coh \rangle$}{\small $\;^d$}\\
& & &\\ \hline\hline
Iclust                & mutual information       & 20 & 86\\ \hline\hline
$K$--means                & PC       & 19 & 74\\ \hline
$K$--means                & $|$PC$|$ & 17 & 69\\ \hline
$K$--means                & Euclidean       & 15 & 58\\ \hline
$K$--medians              & PC       & 20 & 85\\ \hline
$K$--medians              & $|$PC$|$ & 20 & 88\\ \hline
$K$--medians              & Euclidean       & 20 & 81\\ \hline\hline
$\langle\;$ $K$--means $\;\rangle$ & & 18.5 & 75.8\\ \hline\hline
Hier - Comp. linkage  & PC       & 16 & 70\\ \hline
Hier - Comp. linkage  & $|$PC$|$ & 16 & 70\\ \hline
Hier - Comp. linkage  & Euclidean       & 4  & 12\\ \hline
Hier - Avg. linkage   & PC       & 7  & 32\\ \hline
Hier - Avg. linkage   & $|$PC$|$ & 7  & 32\\ \hline
Hier - Avg. linkage   & Euclidean       & 0  & 0\\ \hline
Hier - Centr. linkage & PC       & 2  & 10\\ \hline
Hier - Centr. linkage & $|$PC$|$ & 2  & 10\\ \hline
Hier - Centr. linkage & Euclidean       & 0  & 0\\ \hline
Hier - Sing. linkage  & PC       & 2  & 10 \\ \hline
Hier - Sing. linkage  & $|$PC$|$ & 2  & 10 \\ \hline
Hier - Sing. linkage  & Euclidean       & 0  & 0 \\ \hline\hline
$\langle\;$ Hierarchical $\;\rangle$ & & 4.8 & 21.3\\ \hline\hline
\end{tabular} 
\end{center}
\end{table*} 

\clearpage
\begin{table*}%[t]
\caption{\small Coherence results for the SP500 data
with respect to the GICS companies' annotations with $N_c=15$ clusters.
The column and row definitions are as in \tblref{Tbl:SP_Nc20}.}
\label{Tbl:SP_Nc15}
\small
\begin{center}
\begin{tabular}{|l|c|c|c|} \hline \hline 
$N_c=15$ & & &\\
{\bf Algorithm}{\small $\;^a$} & 
{\bf Similarity}{\small $\;^b$} & 
{\bf $N^{pos}_c$}{\small $\;^c$} & 
{\bf $\langle Coh \rangle$}{\small $\;^d$}\\
& & &\\ \hline\hline
Iclust                   & mutual information       & 15 & 93\\ \hline\hline
$K$--means                & PC       & 12 & 69\\ \hline
$K$--means                & $|$PC$|$ & 13 & 68\\ \hline
$K$--means                & Euclidean       & 11 & 54\\ \hline
$K$--medians              & PC       & 15 & 90\\ \hline
$K$--medians              & $|$PC$|$ & 15 & 88\\ \hline
$K$--medians              & Euclidean       & 15 & 85\\ \hline\hline
$\langle\;$ $K$--means $\;\rangle$ & & 13.5 & 75.7\\ \hline\hline
Hier - Comp. linkage  & PC       & 11 & 63\\ \hline
Hier - Comp. linkage  & $|$PC$|$ & 11 & 63\\ \hline
Hier - Comp. linkage  & Euclidean       & 2  & 5\\ \hline
Hier - Avg. linkage   & PC       & 6  & 32\\ \hline
Hier - Avg. linkage   & $|$PC$|$ & 6  & 32\\ \hline
Hier - Avg. linkage   & Euclidean       & 0  & 0\\ \hline
Hier - Centr. linkage & PC       & 1  & 7 \\ \hline
Hier - Centr. linkage & $|$PC$|$ & 1  & 7 \\ \hline
Hier - Centr. linkage & Euclidean       & 0  & 0\\ \hline
Hier - Sing. linkage  & PC       & 1  & 7 \\ \hline
Hier - Sing. linkage  & $|$PC$|$ & 1  & 7 \\ \hline
Hier - Sing. linkage  & Euclidean       & 0  & 0 \\ \hline\hline
$\langle\;$ Hierarchical $\;\rangle$ & & 3.3 & 18.6\\ \hline\hline
\end{tabular} 
\end{center}
\end{table*}

\begin{table*}%[t]
\caption{\small Coherence results for the SP500 data
with respect to the GICS companies' annotations with $N_c=10$ clusters.
The column and row definitions are as in \tblref{Tbl:SP_Nc20}.}
\label{Tbl:SP_Nc10}
\small
\begin{center}
\begin{tabular}{|l|c|c|c|} \hline \hline 
$N_c=10$ & & &\\
{\bf Algorithm}{\small $\;^a$} & 
{\bf Similarity}{\small $\;^b$} & 
{\bf $N^{pos}_c$}{\small $\;^c$} & 
{\bf $\langle Coh \rangle$}{\small $\;^d$}\\
& & &\\ \hline\hline
Iclust                   & mutual information       & 10 & 91\\ \hline\hline
$K$--means                & PC       & 10 & 84\\ \hline
$K$--means                & $|$PC$|$ & 10 & 85\\ \hline
$K$--means                & Euclidean       & 8  & 63\\ \hline
$K$--medians              & PC       & 10 & 90\\ \hline
$K$--medians              & $|$PC$|$ & 10 & 90\\ \hline
$K$--medians              & Euclidean       & 10 & 77\\ \hline\hline
$\langle\;$ $K$--means $\;\rangle$ & & 9.7 & 81.5\\ \hline\hline
Hier - Comp. linkage  & PC       & 8  & 64\\ \hline
Hier - Comp. linkage  & $|$PC$|$ & 8  & 64\\ \hline
Hier - Comp. linkage  & Euclidean       & 4  & 22\\ \hline
Hier - Avg. linkage   & PC       & 2  & 20\\ \hline
Hier - Avg. linkage   & $|$PC$|$ & 2  & 20\\ \hline
Hier - Avg. linkage   & Euclidean       & 0  & 0\\ \hline
Hier - Centr. linkage & PC       & 1  & 10 \\ \hline
Hier - Centr. linkage & $|$PC$|$ & 1  & 10 \\ \hline
Hier - Centr. linkage & Euclidean       & 0  & 0\\ \hline
Hier - Sing. linkage  & PC       & 0  & 0 \\ \hline
Hier - Sing. linkage  & $|$PC$|$ & 0  & 0 \\ \hline
Hier - Sing. linkage  & Euclidean       & 0  & 0 \\ \hline\hline
$\langle\;$ Hierarchical $\;\rangle$ & & 2.2 & 17.5\\ \hline\hline
\end{tabular} 
\end{center}
\end{table*}

\begin{table*}%[t]
\caption{\small Coherence results for the SP500 data
with respect to the GICS companies' annotations with $N_c=5$ clusters.
The column and row definitions are as in \tblref{Tbl:SP_Nc20}.}
\label{Tbl:SP_Nc5}
\small
\begin{center}
\begin{tabular}{|l|c|c|c|} \hline \hline 
$N_c=5$ & & &\\
{\bf Algorithm}{\small $\;^a$} & 
{\bf Similarity}{\small $\;^b$} & 
{\bf $N^{pos}_c$}{\small $\;^c$} & 
{\bf $\langle Coh \rangle$}{\small $\;^d$}\\
& & &\\ \hline\hline
Iclust                   & mutual information       & 5  & 88\\ \hline\hline
$K$--means                & PC       & 5  & 90\\ \hline
$K$--means                & $|$PC$|$ & 5  & 87\\ \hline
$K$--means                & Euclidean       & 4  & 54\\ \hline
$K$--medians              & PC       & 5  & 90\\ \hline
$K$--medians              & $|$PC$|$ & 5  & 92\\ \hline
$K$--medians              & Euclidean       & 5 & 84\\ \hline\hline
$\langle\;$ $K$--means $\;\rangle$ & & 4.8 & 82.8\\ \hline\hline
Hier - Comp. linkage  & PC       & 4  & 66\\ \hline
Hier - Comp. linkage  & $|$PC$|$ & 5  & 84\\ \hline
Hier - Comp. linkage  & Euclidean       & 3  & 36\\ \hline
Hier - Avg. linkage   & PC       & 1  & 20\\ \hline
Hier - Avg. linkage   & $|$PC$|$ & 1  & 20\\ \hline
Hier - Avg. linkage   & Euclidean       & 0  & 0\\ \hline
Hier - Centr. linkage & PC       & 0  & 0 \\ \hline
Hier - Centr. linkage & $|$PC$|$ & 0  & 0 \\ \hline
Hier - Centr. linkage & Euclidean       & 0  & 0\\ \hline
Hier - Sing. linkage  & PC       & 0  & 0 \\ \hline
Hier - Sing. linkage  & $|$PC$|$ & 0  & 0 \\ \hline
Hier - Sing. linkage  & Euclidean       & 0  & 0 \\ \hline\hline
$\langle\;$ Hierarchical $\;\rangle$ & & 1.2 & 18.8\\ \hline\hline
\end{tabular} 
\end{center}
\end{table*} 

% Produced via CreateEnrichedAnnotTable2_SP.m
\begin{table*}%[h]
\tiny
\caption{\small Enriched GICS annotations in the {\it Iclust\/} solution with $N_c=20$ clusters
for the SP500 data. 
Clusters are ordered as in \figref{SP_data}, \figref{SP_MIs},
and \figref{Fig:SP500_Hierarchy}.
Only annotations with a $P$-value below $0.05$ (Bonferroni corrected) are presented.
{\small $^a$}Cluster index.
{\small $^c$}Cluster size.
{\small $^c$}Cluster coherence (in percentage).
{\small $^d$}Enriched annotations.
In parentheses: $(x/K,p)$ stands for 
the number of companies in the cluster to which this annotation is assigned,
the number of companies in the entire data to which this annotation is assigned,
and the Bonferroni corrected $P$-value, respectively.}
\label{Tbl:SP_Coh}
\begin{center}
\begin{tabular}{|l|l|l|l|} \hline \hline
 & & &\\ 
{\bf \normalsize C}{\small $^a$} & {\bf \normalsize C size}{\small $^b$} & {\bf \normalsize Coh.}{\small $^c$} & {\bf \normalsize Enriched annot.}{\small $^d$}\\ 
 & & &\\ \hline
 & & &\\ 
c11 & 18 & 100 & 4530 Semiconductors\& Semiconductor Equipment (16/19,0.000000)  \\ 
& & & 453010 Semiconductor\& Semiconductor Equipment (16/19,0.000000)  \\ 
& & & 45301020 Semiconductors (12/15,0.000000)  \\ 
& & & 45 Information Technology (18/81,0.000000)  \\ 
& & & 45301010 Semiconductor Equipment (4/4,0.000013)  \\ 
 & & &\\ \hline 
 & & &\\ 
c9 & 20 & 95 & 45 Information Technology (19/81,0.000000)  \\ 
& & & 4520 Technology Hardware\& Equipment (10/35,0.000002)  \\ 
& & & 451030 Software (6/15,0.000155)  \\ 
& & & 452030 Electronic Equipment\& Instruments (5/10,0.000276)  \\ 
& & & 4510 Software\& Services (7/27,0.000617)  \\ 
& & & 452010 Communications Equipment (5/14,0.001970)  \\ 
& & & 45201020 Communications Equipment (5/14,0.001970)  \\ 
& & & 45103010 Application Software (4/8,0.002368)  \\ 
& & & 45203020 Electronic Manufacturing Services (3/4,0.004177)  \\ 
 & & &\\ \hline 
 & & &\\ 
c12 & 21 & 95 & 4520 Technology Hardware\& Equipment (13/35,0.000000)  \\ 
& & & 45 Information Technology (17/81,0.000000)  \\ 
& & & 45202010 Computer Hardware (5/7,0.000035)  \\ 
& & & 452010 Communications Equipment (6/14,0.000132)  \\ 
& & & 45201020 Communications Equipment (6/14,0.000132)  \\ 
& & & 452020 Computers\& Peripherals (5/10,0.000383)  \\ 
& & & 501020 Wireless Telecommunication Services (2/2,0.040138)  \\ 
& & & 50102010 Wireless Telecommunication Services (2/2,0.040138)  \\ 
 & & &\\ \hline 
 & & &\\ 
c10 & 20 & 65 & 2510 Automobiles\& Components (6/9,0.000005)  \\ 
& & & 251010 Auto Components (4/6,0.000863)  \\ 
& & & 201010 Aerospace\& Defense (4/9,0.006685)  \\ 
& & & 20101010 Aerospace\& Defense (4/9,0.006685)  \\ 
& & & 25101010 Auto Parts\& Equipment (3/4,0.006730)  \\ 
& & & 2010 Capital Goods (7/37,0.008990)  \\ 
& & & 25102010 Automobile Manufacturers (2/2,0.046560)  \\ 
 & & &\\ \hline 
 & & &\\ 
c16 & 10 & 30 & 25301020 Hotels  Resorts\& Cruise Lines (3/4,0.000546)  \\ 
& & & 2530 Hotels Restaurants\& Leisure (3/11,0.020860)  \\ 
& & & 253010 Hotels Restaurants\& Leisure (3/11,0.020860)  \\ 
 & & &\\ \hline 
 & & &\\ 
c18 & 176 & 19 & 351010 Health Care Equipment\& Supplies (13/13,0.000133)  \\ 
& & & 35101010 Health Care Equipment (11/11,0.001186)  \\ 
& & & 2020 Commercial Services\& Supplies (11/12,0.009850)  \\ 
& & & 202010 Commercial Services\& Supplies (11/12,0.009850)  \\ 
& & & 2030 Transportation (9/9,0.010371)  \\ 
 & & &\\ \hline 
 & & &\\ 
c3 & 17 & 83 & 351020 Health Care Providers\& Services (9/16,0.000000)  \\ 
& & & 3510 Health Care Equipment\& Services (9/29,0.000000)  \\ 
& & & 35 Health Care (10/47,0.000000)  \\ 
& & & 35102030 Managed Health Care (4/5,0.000015)  \\ 
& & & 35102015 Health Care Services (2/4,0.045569)  \\ 
& & & 35102020 Health Care Facilities (2/4,0.045569)  \\ 
 & & &\\ \hline 
\end{tabular}
\end{center}
\end{table*}

\begin{table*}%[h]
\tiny
\begin{center}
\begin{tabular}{|l|l|l|l|} \hline \hline
 & & &\\ 
{\bf \normalsize C}{\small $^a$} & {\bf \normalsize C size}{\small $^b$} & {\bf \normalsize Coh.}{\small $^c$} & {\bf \normalsize Enriched annot.}{\small $^d$}\\ 
 & & &\\ \hline
 & & &\\ 
c14 & 18 & 94 & 352020 Pharmaceuticals (10/13,0.000000)  \\ 
& & & 35202010 Pharmaceuticals (10/13,0.000000)  \\ 
& & & 3520 Pharmaceuticals\& Biotechnology (10/18,0.000000)  \\ 
& & & 35 Health Care (12/47,0.000000)  \\ 
& & & 501010 Diversified Telecommunication Services (5/9,0.000066)  \\ 
& & & 50101020 Integrated Telecommunication Services (5/9,0.000066)  \\ 
& & & 50 Telecommunication Services (5/11,0.000232)  \\ 
& & & 5010 Telecommunication Services (5/11,0.000232)  \\ 
 & & &\\ \hline 
 & & &\\ 
c5 & 18 & 94 & 30 Consumer Staples (17/35,0.000000)  \\ 
& & & 3020 Food Beverage\& Tobacco (12/19,0.000000)  \\ 
& & & 302020 Food Products (9/10,0.000000)  \\ 
& & & 30202030 Packaged Foods\& Meats (8/9,0.000000)  \\ 
& & & 3030 Household\& Personal Products (4/6,0.000341)  \\ 
& & & 303010 Household Products (3/4,0.003000)  \\ 
& & & 30301010 Household Products (3/4,0.003000)  \\ 
& & & 302010 Beverages (3/6,0.014308)  \\ 
 & & &\\ \hline 
 & & &\\ 
c15 & 9 & 83 & 4040 Real Estate (5/6,0.000000)  \\ 
& & & 404010 Real Estate (5/6,0.000000)  \\ 
& & & 40401010 Real Estate Investment Trusts (5/6,0.000000)  \\ 
& & & 40 Financials (5/80,0.004491)  \\ 
 & & &\\ \hline 
 & & &\\ 
c2 & 15 & 93 & 2540 Media (10/14,0.000000)  \\ 
& & & 254010 Media (10/14,0.000000)  \\ 
& & & 25401040 Publishing (7/7,0.000000)  \\ 
& & & 25 Consumer Discretionary (14/83,0.000000)  \\ 
& & & 25401020 Broadcasting\& Cable TV (2/3,0.031371)  \\ 
& & & 252010 Household Durables (3/11,0.040516)  \\ 
 & & &\\ \hline 
 & & &\\ 
c17 & 23 & 100 & 2550 Retailing (19/30,0.000000)  \\ 
& & & 25 Consumer Discretionary (21/83,0.000000)  \\ 
& & & 255030 Multiline Retail (9/11,0.000000)  \\ 
& & & 255040 Specialty Retail (10/17,0.000000)  \\ 
& & & 25503010 Department Stores (5/7,0.000050)  \\ 
& & & 25503020 General Merchandise Stores (4/4,0.000065)  \\ 
& & & 25504010 Apparel Retail (3/3,0.001574)  \\ 
& & & 25504040 Specialty Stores (4/8,0.004005)  \\ 
& & & 30101040 HyperMarkets\& Super Centers (2/2,0.036344)  \\ 
 & & &\\ \hline 
 & & &\\ 
c4 & 19 & 100 & 55 Utilities (19/36,0.000000)  \\ 
& & & 5510 Utilities (19/36,0.000000)  \\ 
& & & 551010 Electric Utilities (14/22,0.000000)  \\ 
& & & 55101010 Electric Utilities (14/22,0.000000)  \\ 
& & & 551020 Gas Utilities (3/6,0.007516)  \\ 
& & & 55102010 Gas Utilities (3/6,0.007516)  \\ 
 & & &\\ \hline 
 & & &\\ 
c13 & 23 & 100 & 4030 Insurance (19/21,0.000000)  \\ 
& & & 403010 Insurance (19/21,0.000000)  \\ 
& & & 40 Financials (23/80,0.000000)  \\ 
& & & 40301040 Property\& Casualty Insurance (9/9,0.000000)  \\ 
& & & 40301020 Life\& Health Insurance (6/7,0.000000)  \\ 
& & & 40301030 Multi-line Insurance (3/3,0.001203)  \\ 
& & & 401020 Thrifts\& Mortgage Finance (3/6,0.021900)  \\ 
& & & 40102010 Thrifts\& Mortgage Finance (3/6,0.021900)  \\ 
 & & &\\ \hline 
 & & &\\ 
c7 & 15 & 100 & 4020 Diversified Financials (15/24,0.000000)  \\ 
& & & 402030 Capital Markets (13/16,0.000000)  \\ 
& & & 40 Financials (15/80,0.000000)  \\ 
& & & 40203020 Investment Banking\& Brokerage (6/7,0.000000)  \\ 
& & & 40203010 Asset Management\& Custody Banks (6/8,0.000000)  \\ 
 & & &\\ \hline 
 & & &\\ 
c1 & 21 & 100 & 401010 Commercial Banks (21/23,0.000000)  \\ 
& & & 4010 Banks (21/29,0.000000)  \\ 
& & & 40101015 Regional Banks (16/17,0.000000)  \\ 
& & & 40 Financials (21/80,0.000000)  \\ 
& & & 40101010 Diversified Banks (5/6,0.000003)  \\ 
 & & &\\ \hline 
\end{tabular}
\end{center}
\end{table*}

\begin{table*}%[h]
\tiny
\begin{center}
\begin{tabular}{|l|l|l|l|} \hline \hline
 & & &\\ 
{\bf \normalsize C}{\small $^a$} & {\bf \normalsize C size}{\small $^b$} & {\bf \normalsize Coh.}{\small $^c$} & {\bf \normalsize Enriched annot.}{\small $^d$}\\ 
 & & &\\ \hline
 & & &\\ 
c8 & 23 & 83 & 201060 Machinery (12/14,0.000000)  \\ 
& & & 2010 Capital Goods (16/37,0.000000)  \\ 
& & & 20 Industrials (16/58,0.000000)  \\ 
& & & 20106020 Industrial Machinery (7/9,0.000000)  \\ 
& & & 20106010 Construction\& Farm Machinery\& Heavy Trucks (5/5,0.000004)  \\ 
& & & 151050 Paper\& Forest Products (3/5,0.023469)  \\ 
 & & &\\ \hline 
 & & &\\ 
c19 & 14 & 93 & 151010 Chemicals (11/14,0.000000)  \\ 
& & & 15 Materials (13/33,0.000000)  \\ 
& & & 1510 Materials (13/33,0.000000)  \\ 
& & & 15101020 Diversified Chemicals (5/6,0.000001)  \\ 
& & & 15101050 Specialty Chemicals (4/5,0.000026)  \\ 
& & & 15101040 Industrial Gases (2/2,0.009228)  \\ 
& & & 15103020 Paper Packaging (2/3,0.027226)  \\ 
 & & &\\ \hline 
 & & &\\ 
c6 & 13 & 100 & 10 Energy (13/23,0.000000)  \\ 
& & & 1010 Energy (13/23,0.000000)  \\ 
& & & 101010 Energy Equipment\& Services (7/7,0.000000)  \\ 
& & & 10102020 Oil\& Gas Exploration\& Production (6/7,0.000000)  \\ 
& & & 10101020 Oil\& Gas Equipment\& Services (4/4,0.000002)  \\ 
& & & 101020 Oil\& Gas (6/16,0.000005)  \\ 
& & & 10101010 Oil\& Gas Drilling (3/3,0.000105)  \\ 
 & & &\\ \hline 
 & & &\\ 
c20 & 8 & 100 & 101020 Oil\& Gas (8/16,0.000000)  \\ 
& & & 10 Energy (8/23,0.000000)  \\ 
& & & 1010 Energy (8/23,0.000000)  \\ 
& & & 10102010 Integrated Oil\& Gas (5/6,0.000000)  \\ 
& & & 10102030 Oil\& Gas Refining\& Marketing\& Transportation (2/3,0.004224)  \\ 
 & & &\\ \hline 
\end{tabular}
\end{center}
\end{table*}

\begin{table*}%[t]
\caption{\small Coherence results for the EachMovie data
with respect to the movie genre annotations with $N_c=20$ clusters. 
{\small $^a$}Clustering algorithm. 
In the $\langle$ $K$--means $\rangle$ row we present the average results
of all the six $K$--means variants. 
For each of these variants we performed $100$ runs from which
the best solution is chosen.
In the $\langle$ Hier. $\rangle$ row we present the average results
of all the $12$ Hierarchical clustering variants. 
{\small $^b$}Correlation measure used by the algorithm. 
{\it PC\/} stands for the (centered) Pearson Correlation. 
{\it $|$PC$|$\/} is the absolute value of this correlation. 
{\it Euclidean\/} stands for the Euclidean distance. 
{\small $^c$}Number of clusters with a positive coherence. 
{\small $^d$}Average coherence of all $20$ clusters.}
\label{Tbl:EachMovie_Nc20}
\small
\begin{center}
\begin{tabular}{|l|c|c|c|} \hline \hline 
$N_c=20$ & & &\\
{\bf Algorithm}{\small $\;^a$} & 
{\bf Similarity}{\small $\;^b$} & 
{\bf $N^{pos}_c$}{\small $\;^c$} & 
{\bf $\langle Coh \rangle$}{\small $\;^d$}\\
& & &\\ \hline\hline
Iclust                & mutual information       & 15 & 54\\ \hline\hline
$K$--means                & PC       & 1  & 3\\ \hline
$K$--means                & $|$PC$|$ & 2  & 4\\ \hline
$K$--means                & Euclidean       & 5  & 12\\ \hline
$K$--medians              & PC       & 2  & 5\\ \hline
$K$--medians              & $|$PC$|$ & 4  & 8\\ \hline
$K$--medians              & Euclidean       & 2  & 6\\ \hline\hline
$\langle\;$ $K$--means $\;\rangle$ & & 2.7 & 6.3\\ \hline\hline
Hier - Comp. linkage  & PC       & 17 & 55\\ \hline
Hier - Comp. linkage  & $|$PC$|$ & 16 & 51\\ \hline
Hier - Comp. linkage  & Euclidean       & 10 & 34\\ \hline
Hier - Avg. linkage   & PC       & 12 & 43\\ \hline
Hier - Avg. linkage   & $|$PC$|$ & 12 & 43\\ \hline
Hier - Avg. linkage   & Euclidean       & 5  & 19\\ \hline
Hier - Centr. linkage & PC       & 4  & 16\\ \hline
Hier - Centr. linkage & $|$PC$|$ & 4  & 16\\ \hline
Hier - Centr. linkage & Euclidean       & 2  & 8\\ \hline
Hier - Sing. linkage  & PC       & 0  & 0 \\ \hline
Hier - Sing. linkage  & $|$PC$|$ & 0  & 0 \\ \hline
Hier - Sing. linkage  & Euclidean       & 1  & 5 \\ \hline\hline
$\langle\;$ Hierarchical $\;\rangle$ & & 6.9 & 24.2\\ \hline\hline
\end{tabular} 
\end{center}
\end{table*}

\begin{table*}%[t]
\caption{\small Coherence results for the EachMovie data
with respect to the movie genre annotations with $N_c=15$ clusters.
The column and row definitions are as in \tblref{Tbl:EachMovie_Nc20}.}
\label{Tbl:EachMovie_Nc15}
\small
\begin{center}
\begin{tabular}{|l|c|c|c|} \hline \hline 
$N_c=15$ & & &\\
{\bf Algorithm}{\small $\;^a$} & 
{\bf Similarity}{\small $\;^b$} & 
{\bf $N^{pos}_c$}{\small $\;^c$} & 
{\bf $\langle Coh \rangle$}{\small $\;^d$}\\
& & &\\ \hline\hline
Iclust                   & mutual information       & 11 & 54\\ \hline\hline
$K$--means                & PC       & 1  & 2\\ \hline
$K$--means                & $|$PC$|$ & 1  & 1\\ \hline
$K$--means                & Euclidean       & 2  & 6\\ \hline
$K$--medians              & PC       & 2  & 5\\ \hline
$K$--medians              & $|$PC$|$ & 1  & 3\\ \hline
$K$--medians              & Euclidean       & 4  & 14\\ \hline\hline
$\langle\;$ $K$--means $\;\rangle$ & & 21.8 & 5.2\\ \hline\hline
Hier - Comp. linkage  & PC       & 13 & 54\\ \hline
Hier - Comp. linkage  & $|$PC$|$ & 11 & 47\\ \hline
Hier - Comp. linkage  & Euclidean       & 6  & 29\\ \hline
Hier - Avg. linkage   & PC       & 10 & 47\\ \hline
Hier - Avg. linkage   & $|$PC$|$ & 10 & 46\\ \hline
Hier - Avg. linkage   & Euclidean       & 3  & 16\\ \hline
Hier - Centr. linkage & PC       & 2  & 8\\ \hline
Hier - Centr. linkage & $|$PC$|$ & 2  & 8\\ \hline
Hier - Centr. linkage & Euclidean       & 1  & 3\\ \hline
Hier - Sing. linkage  & PC       & 0  & 0 \\ \hline
Hier - Sing. linkage  & $|$PC$|$ & 0  & 0 \\ \hline
Hier - Sing. linkage  & Euclidean       & 1  & 7 \\ \hline\hline
$\langle\;$ Hierarchical $\;\rangle$ & & 4.9 & 22.1\\ \hline\hline
\end{tabular} 
\end{center}
\end{table*} 

\begin{table*}%[t]
\caption{\small Coherence results for the EachMovie data
with respect to the movie genre annotations with $N_c=10$ clusters.
The column and row definitions are as in \tblref{Tbl:EachMovie_Nc20}.}
\label{Tbl:EachMovie_Nc10}
\small
\begin{center}
\begin{tabular}{|l|c|c|c|} \hline \hline 
$N_c=10$ & & &\\
{\bf Algorithm}{\small $\;^a$} & 
{\bf Similarity}{\small $\;^b$} & 
{\bf $N^{pos}_c$}{\small $\;^c$} & 
{\bf $\langle Coh \rangle$}{\small $\;^d$}\\
& & &\\ \hline\hline
Iclust                   & mutual information       & 9  & 57\\ \hline\hline
$K$--means                & PC       & 1  & 3\\ \hline
$K$--means                & $|$PC$|$ & 1  & 6\\ \hline
$K$--means                & Euclidean       & 4  & 20\\ \hline
$K$--medians              & PC       & 2  & 6\\ \hline
$K$--medians              & $|$PC$|$ & 2  & 6\\ \hline
$K$--medians              & Euclidean       & 6  & 27\\ \hline\hline
$\langle\;$ $K$--means $\;\rangle$ & & 2.7 & 11.3\\ \hline\hline
Hier - Comp. linkage  & PC       & 8  & 43\\ \hline
Hier - Comp. linkage  & $|$PC$|$ & 8  & 44\\ \hline
Hier - Comp. linkage  & Euclidean       & 5  & 36\\ \hline
Hier - Avg. linkage   & PC       & 7  & 43\\ \hline
Hier - Avg. linkage   & $|$PC$|$ & 8  & 47\\ \hline
Hier - Avg. linkage   & Euclidean       & 2  & 16\\ \hline
Hier - Centr. linkage & PC       & 2  & 12\\ \hline
Hier - Centr. linkage & $|$PC$|$ & 2  & 12\\ \hline
Hier - Centr. linkage & Euclidean       & 1  & 4\\ \hline
Hier - Sing. linkage  & PC       & 0  & 0 \\ \hline
Hier - Sing. linkage  & $|$PC$|$ & 0  & 0 \\ \hline
Hier - Sing. linkage  & Euclidean       & 1  & 10 \\ \hline\hline
$\langle\;$ Hierarchical $\;\rangle$ & & 3.7 & 22.3\\ \hline\hline
\end{tabular} 
\end{center}
\end{table*} 

\begin{table*}%[t]
\caption{\small Coherence results for the EachMovie data
with respect to the movie genre annotations with $N_c=5$ clusters.
The column and row definitions are as in \tblref{Tbl:EachMovie_Nc20}.}
\label{Tbl:EachMovie_Nc5}
\small
\begin{center}
\begin{tabular}{|l|c|c|c|} \hline \hline 
$N_c=5$ & & &\\
{\bf Algorithm}{\small $\;^a$} & 
{\bf Similarity}{\small $\;^b$} & 
{\bf $N^{pos}_c$}{\small $\;^c$} & 
{\bf $\langle Coh \rangle$}{\small $\;^d$}\\
& & &\\ \hline\hline
Iclust                & mutual information       & 5  & 48\\ \hline\hline
$K$--means                & PC       & 2  & 21\\ \hline
$K$--means                & $|$PC$|$ & 2  & 19\\ \hline
$K$--means                & Euclidean       & 3  & 30\\ \hline
$K$--medians              & PC       & 4  & 32\\ \hline
$K$--medians              & $|$PC$|$ & 2  & 19\\ \hline
$K$--medians              & Euclidean       & 4  & 37\\ \hline\hline
$\langle\;$ $K$--means $\;\rangle$ & & 2.8 & 26.3\\ \hline\hline
Hier - Comp. linkage  & PC       & 5  & 51\\ \hline
Hier - Comp. linkage  & $|$PC$|$ & 5  & 41\\ \hline
Hier - Comp. linkage  & Euclidean       & 4  & 48\\ \hline
Hier - Avg. linkage   & PC       & 4  & 46\\ \hline
Hier - Avg. linkage   & $|$PC$|$ & 4  & 47\\ \hline
Hier - Avg. linkage   & Euclidean       & 2  & 21\\ \hline
Hier - Centr. linkage & PC       & 2  & 23\\ \hline
Hier - Centr. linkage & $|$PC$|$ & 2  & 23\\ \hline
Hier - Centr. linkage & Euclidean       & 1  & 9\\ \hline
Hier - Sing. linkage  & PC       & 0  & 0 \\ \hline
Hier - Sing. linkage  & $|$PC$|$ & 0  & 0 \\ \hline
Hier - Sing. linkage  & Euclidean       & 0  & 0 \\ \hline\hline
$\langle\;$ Hierarchical $\;\rangle$ & & 2.4 & 25.8\\ \hline\hline
\end{tabular} 
\end{center}
\end{table*} 

% Produced via CreateEnrichedAnnotTable2_EachMovie.m
\begin{table*}[]
\tiny
\caption{\small Enriched genre annotations in the {\it Iclust\/} solution with $N_c=20$ clusters
for the EachMovie data. 
Clusters are ordered as in \figref{EachMovie_data}, \figref{EachMovie_MIs},
and \figref{Fig:EachMovie_Hierarchy}.
Only annotations with a $P$-value below $0.05$ (Bonferroni corrected) are presented.
{\small $^a$}Cluster index.
{\small $^b$}Cluster size.
{\small $^c$}Cluster coherence (in percentage).
{\small $^d$}Enriched annotations.
In parentheses: $(x/K,p)$ stands for 
the number of movies in the cluster to which this annotation is assigned,
the number of movies in the entire data to which this annotation is assigned,
and the Bonferroni corrected $P$-value, respectively.}
\tiny
\label{Tbl:EachMovie_Coh}
\begin{center}
\begin{tabular}{|l|l|l|l|} \hline \hline
 & & &\\ 
{\bf \normalsize C}{\small$^a$} & {\bf \normalsize C size}{\small$^b$} & {\bf \normalsize Coh.}{\small$^c$} & {\bf \normalsize Enriched annot.}{\small$^d$}\\ 
 & & &\\ \hline 
 & & &\\ 
c14 & 10 & 0 & -- \\ 
 & & & \\ \hline 
 & & &\\ 
c2 & 16 & 0 & --  \\ 
 & & & \\ \hline 
 & & &\\ 
c19 & 10 & 50 & Art-Foreign (5/45,0.005254)  \\ 
 & & & \\ \hline 
 & & &\\ 
c8 & 10 & 70 & Art-Foreign (7/45,0.000019)  \\ 
 & & & \\ \hline 
 & & &\\ 
c18 & 22 & 0 & -- \\ 
 & & & \\ \hline 
 & & &\\ 
c9 & 31 & 55 & Action (17/110,0.000281)  \\ 
 & & & \\ \hline 
 & & &\\ 
c20 & 155 & 55 & Drama (68/160,0.001170)  \\ 
& & & Romance (30/61,0.011591)  \\ 
 & & & \\ \hline 
 & & &\\ 
c1 & 19 & 95 & Classic (10/44,0.000004)  \\ 
& & & Drama (15/160,0.000214)  \\ 
 & & & \\ \hline 
 & & &\\ 
c7 & 24 & 71 & Classic (10/44,0.000067)  \\ 
& & & Action (13/110,0.003526)  \\ 
 & & & \\ \hline 
 & & &\\ 
c15 & 18 & 94 & Action (16/110,0.000000)  \\ 
& & & Thriller (10/90,0.001778)  \\ 
 & & & \\ \hline 
 & & &\\ 
c5 & 32 & 39 & Thriller (12/90,0.034492)  \\ 
 & & & \\ \hline 
 & & &\\ 
c13 & 15 & 40 & Horror (6/33,0.001412)  \\ 
 & & & \\ \hline 
 & & &\\ 
c3 & 20 & 0 &  -- \\ 
 & & & \\ \hline 
 & & &\\ 
c10 & 15 & 0 &  -- \\ 
 & & & \\ \hline 
 & & &\\ 
c4 & 27 & 74 & Romance (12/61,0.000100)  \\ 
& & & Comedy (17/149,0.001613)  \\ 
 & & & \\ \hline 
 & & &\\ 
c6 & 11 & 100 & Comedy (11/149,0.000001)  \\ 
 & & & \\ \hline 
 & & &\\ 
c16 & 16 & 87 & Comedy (13/149,0.000012)  \\ 
 & & & \\ \hline 
 & & &\\ 
c17 & 21 & 76 & Action (16/110,0.000000)  \\ 
 & & & \\ \hline 
 & & &\\ 
c11 & 14 & 71 & Family (10/67,0.000003)  \\ 
 & & & \\ \hline 
 & & &\\ 
c12 & 14 & 100 & Family (13/67,0.000000)  \\ 
& & & Animation (8/25,0.000000)  \\ 
& & & Classic (5/44,0.019004)  \\ 
 & & & \\ \hline 
\end{tabular}
\end{center}
\end{table*}

\end{widetext}
\end{document}